\documentclass[aps,rmp,twocolumn,superscriptaddress,reprint,nofootinbib]{revtex4-1}

\usepackage{graphicx, epsfig, subfigure, tikz, color}

\usepackage{amsmath, bm, amssymb}
\usepackage{multirow, tabularx, dcolumn}

\definecolor{jlab_red}{RGB}{192,39,45}
\definecolor{jlab_orange}{RGB}{249,102,0}
\definecolor{jlab_blue}{RGB}{47,122,121}
\definecolor{jlab_green}{RGB}{65,125,10}
\definecolor{lightgray}{gray}{0.8}

\usepackage[colorlinks=true,
linkcolor=black,
breaklinks=true,
urlcolor=jlab_blue,
citecolor=jlab_green]{hyperref}

\newcommand{\beq}{\begin{equation}}
\newcommand{\eeq}{\end{equation}}
\newcommand{\bea}{\begin{eqnarray}}
\newcommand{\eea}{\end{eqnarray}}

\newcommand{\la}{\langle}
\newcommand{\ra}{\rangle}

\newcommand{\nn}{\nonumber}

\begin{document}


\vbox{ \hbox{JLAB-THY-17-2495} }
\vbox{ \hbox{ADP-17-28/T1034} }
\vspace{2mm}

\title[Hadron resonances from lattice QCD]{Scattering processes and resonances from lattice QCD}
\author{Ra\'ul~A.~Brice\~no} \email{rbriceno@jlab.org}
\affiliation{\mbox{Thomas Jefferson National Accelerator Facility, 12000 Jefferson Avenue, Newport News, Virginia 23606, USA}\vspace{2mm}}
\author{Jozef~J.~Dudek} \email{dudek@jlab.org}
\affiliation{\mbox{Thomas Jefferson National Accelerator Facility, 12000 Jefferson Avenue, Newport News, Virginia 23606, USA}\vspace{2mm}}
\affiliation{\mbox{Department of Physics, College of William and Mary, Williamsburg, Virginia 23187, USA}\vspace{2mm}}
\author{Ross~D.~Young\vspace{2mm}} \email{ross.young@adelaide.edu.au}
\affiliation{\mbox{Special Research Center for the Subatomic Structure of Matter (CSSM),
Department of Physics,} \mbox{University of Adelaide, Adelaide 5005, Australia}\vspace{2mm}}

\date{\today}

\begin{abstract}

The vast majority of hadrons observed in nature are not stable under the strong interaction, rather they are \emph{resonances} whose existence is deduced from enhancements in the energy dependence of scattering amplitudes. 
The study of hadron resonances offers a window into the workings of quantum chromodynamics (QCD) in the low-energy non-perturbative region, and in addition, many probes of the limits of the electroweak sector of the Standard Model consider processes which feature hadron resonances. 
From a theoretical standpoint, this is a challenging field: the same dynamics that binds quarks and gluons into hadron resonances also controls their decay into lighter hadrons, so a complete approach to QCD is required. Presently, lattice QCD is the only available tool that provides the required non-perturbative evaluation of hadron observables. 
In this article, we review progress in the study of few-hadron reactions in which resonances and bound-states appear using lattice QCD techniques. We describe the leading approach which takes advantage of the periodic finite spatial volume used in lattice QCD calculations to extract scattering amplitudes from the discrete spectrum of QCD eigenstates in a box. 
We explain how from explicit lattice QCD calculations, one can rigorously garner information about a variety of resonance properties, including their masses, widths, decay couplings, and form factors. 
The challenges which currently limit the field are discussed along with the steps being taken to resolve them.

\end{abstract}

\maketitle

\tableofcontents

\section{Introduction \label{Sec:introduction}}
\vspace{-4mm}

In the last couple of decades, the field of hadron spectroscopy has undergone a renaissance, spurred in part by the observation in experiments of a large number of states whose presence had not been anticipated. Many of these, known colloquially as the ``\emph{X,Y,Z}'' states, have been found in the heavy quark sector (see~\cite{Liu:2013waa, Chen:2016qju} for recent reviews on the topic), and a complete understanding of them remains elusive, with piecemeal proposals ranging from tetraquark constructions, to meson-meson molecules, to quark-gluon hybrid states etc... All these `pictures' strongly depend on the model chosen to locally approximate the low-energy behavior of the fundamental theory of the strong interaction, quantum chromodynamics (QCD). Ultimately it is QCD which builds these states, and it is to QCD that we should turn to understand them.

In practice, one does not have to go high in the experimental hadron spectrum to find states that put our understanding of QCD to a test. An iconic example is the $\sigma/f_0(500)$ resonance, which couples to the $\pi\pi$ scattering channel, and is both extremely light and unnaturally broad, having a decay width larger than its mass. Until recently even its existence was not a settled issue~\cite{Pelaez:2015qba, GarciaMartin:2011jx, Caprini:2005zr}, and at the QCD level, the origin of the $\sigma$ remains a mystery. On the other hand, there is also a spectrum of `conventional', mainly narrow, resonances, the lightest of which is the $\rho$, whose mass pattern is fairly well described by models that consider mesons to be constructed from a quark and an antiquark, having a relatively weak coupling to their decay channels~\cite{Shepherd:2016dni}. 

Motivated in part by the excitement in contemporary experimental exploration of the spectrum of hadrons, there is demand for a parallel theoretical program that can shed light on these states. Models, either at the quark level, or the hadron level, can help us to gain insight into how these states are constructed, but eventually we need to anchor our understanding in first-principles calculation within QCD. Historically, this has been extremely challenging, with one relevant complication being that the vast majority of hadrons observed in nature are \emph{unstable} under the strong interactions. We observe all but the lightest hadrons through their decay products as \emph{resonances}, whose lifetimes are of the same order as the time scale of QCD interactions. This forces us to consider theoretical techniques which can consistently encode the physics of both binding and decay. Presently, the only rigorous theoretical tool which has been shown to meet these demands is \emph{lattice QCD}. 

Lattice QCD considers  quark and gluon fields on a discrete grid of points of finite size, and by sampling possible configurations of these fields, with a probability dictated by the lagrangian of QCD, hadronic observables can be estimated, along with a measure of statistical uncertainty that can be reduced with increased computer time. Systematic errors arise from the choice of lattice spacing, the size of the box, and in many practical calculations, the values chosen for the quark masses, which for computational cost reasons may not be as low as the experimentally measured quark masses. These approximations are controlled, and the corresponding uncertainties can be systematically decreased, in principle.

The vast majority of lattice QCD calculations to date have focused on the properties of states that are \emph{stable} within QCD, and these calculations have matured to the level where they can be considered realistic, with up, down, strange and charm quark masses tuned to their correct physical values, and even some including the relatively small effects due to QED~\cite{Duncan:1996xy, Blum:2007cy, Borsanyi:2014jba, Horsley:2015eaa, Blum:2010ym, Aoki:2012st, deDivitiis:2013xla}.

The study of hadron \emph{resonances} using lattice QCD, and consequently the study of the QCD spectrum, is at a much earlier stage of its development, but rapid progress is being made that we will review here. An approach that relates the discrete spectrum of QCD eigenstates in the finite-volume defined by the lattice, to scattering amplitudes which may contain resonances, has been a powerful tool. Generically known as the ``L\"uscher method", it allows us access to hadron scattering amplitudes computed from first principles in QCD. By applying techniques similar to those used in the analysis of experimental scattering data, we are able to infer the resonance content of these amplitudes by analytically continuing into the complex energy plane, where resonances appear as pole singularities.

From a practical point of view, a first challenge has been to develop the algorithmic and computational techniques to make possible the determination of a tower of excited states, and recent years have seen tremendous advances in lattice QCD methods, such as those described in~\cite{Luscher:1986pf, Luscher:1990ck, Foley:2005ac, Peardon:2009gh, Blossier:2009kd, Dudek:2009qf, Morningstar:2011ka, Dudek:2010wm, Dudek:2013yja, Liu:2012ze, Dudek:2012ag, Dudek:2011bn, Edwards:2011jj, Dudek:2011tt}. It is now not unusual to observe lattice calculations determining as many as two dozen states in a single quantum number channel, with many methods being sufficiently flexible as to place almost no restriction on this number in future calculations.

The detailed discrete spectra of states extracted in lattice calculations have been used to constrain two-body scattering amplitudes in both elastic and coupled-channel cases, using a rigorous formalism that has been made rather general in derivations presented over the past few years. In this review we will consider all aspects of resonance physics within lattice QCD, including the coupling of resonances to external currents, and will outline challenges which still remain and possible avenues to overcome them.

\section{Resonances, composite particles, and scattering amplitudes  \label{Sec:resonances}}

The asymptotic states of QCD are QCD-stable hadrons, built from quarks and gluons, like the pion or the proton. Our interest in this review is in \emph{composite particles}, which we will consider through their dual identities as dynamical enhancements in hadron scattering amplitudes, and as objects built from quarks and gluons. Two categories of composite particle are \emph{resonances}, which are unstable, decaying into multiple stable hadrons, and \emph{bound-states}, which are stable against decay by virtue of being lighter than the relevant decay threshold. Bound-states can therefore be added to the list of possible asymptotic states of QCD.

Familiar bound-states in hadron physics are atomic nuclei, with the simplest being the deuteron, a composite particle having the quantum numbers of a proton and a neutron in a $S$-wave, with small fractional coupling to $D$-wave. A detailed spectrum of hadron resonances has been observed experimentally~\cite{Olive:2016xmw}, which includes both baryons and mesons with a wide range of angular momenta. Resonances with relatively long lifetimes, \emph{narrow resonances}, can often be observed as `bump-like' enhancements in the invariant mass distribution of their decay products. 

Bound-states and resonances may be considered more rigorously as being associated with \emph{pole singularities} of scattering amplitudes. We can illustrate this in the context of elastic scattering of two spinless particles of masses $m_1$, $m_2$. The total energy and momentum in any frame, $(E,\mathbf{P})$, the center-of-momentum frame energy, $E^\star$, and cm-frame momentum, $\mathbf{q}^\star$, are related to Mandelstam $s$ by
\begin{equation*}
 \sqrt{s} = \sqrt{E^2 - \mathbf{P}^2} = E^\star = \sqrt{m_1^2 + \mathbf{q}^{\star 2}} +  \sqrt{m_2^2 + \mathbf{q}^{\star 2}},
\end{equation*}
such that the magnitude of $\mathbf{q}^\star$ is determined,
\begin{equation}
q^\star = \frac{1}{2} \left[ s - 2(m_1^2+m_2^2) + \frac{\big( m_2^2-m_1^2 \big)^2}{s}\right]^{1/2}, \label{eqn:qstar}
\end{equation}
but the direction $\hat{\mathbf{q}}^\star$ is not, reflecting the possible angular dependence of the scattering amplitude, which can be expressed in terms of the angle $\theta_{\hat{\mathbf{q}}^\star}$ in the scattering plane, or via Mandelstam $t$.

The elastic scattering amplitude, $\mathcal{M}(s,t)$, can be decomposed in terms of \emph{partial-wave} amplitudes, ${\mathcal{M} = \frac{1}{4\pi} \sum_\ell P_\ell(\cos \theta_{\hat{\mathbf{q}}^\star}) \, \mathcal{M}_\ell(s)}$, and hadron resonances of definite angular momentum are expected to contribute to just one of the infinite set of partial-wave amplitudes. Conservation of probability above the kinematic threshold ($E^\star > E^\mathrm{thr.}=m_1 + m_2$) is enforced by the \emph{elastic unitarity} condition,
\begin{equation}
 \mathrm{Im} \frac{1}{\mathcal{M}_\ell} = - \frac{1}{16\pi} \frac{2 \, q^\star}{E^\star}\Theta( E^\star - E^\mathrm{thr.}),
 \label{eq:elastic_unitarity}
\end{equation}
and with the imaginary part specified by unitarity, it is common to express the real part of the elastic scattering amplitude as a function of a real variable, the \emph{phase-shift}, $\delta_\ell(E^\star)$,
\begin{equation}
\mathrm{Re} \frac{1}{\mathcal{M}_\ell} = \frac{1}{16\pi} \frac{2 \, q^\star}{E^\star} \cot \delta_\ell(E^\star).
 \label{eq:phaseshift}
\end{equation}

The presence of $q^\star$ in the unitarity relation indicates an important property of partial-wave amplitudes when they are considered to be functions of a \emph{complex} value of Mandelstam $s$. The square-root in Eq.(\ref{eqn:qstar}) means that $\mathcal{M}_\ell(s)$ features a branch cut beginning at threshold ${s=\big(m_1+m_2 \big)^2}$, and because of this there are two \emph{Riemann sheets}. The first, or \emph{physical} sheet, has $\mathrm{Im}(q^\star) > 0$, and is so named because it contains the real energy axis ($s+ i \epsilon$ with $\epsilon\to 0^+$), where physical scattering occurs. The second, or \emph{unphysical} sheet, has $\mathrm{Im}(q^\star) < 0$ and can be reached by moving down through the branch cut from the real axis.

\subsection{Pole singularities \label{sec:poles}}

The existence of a composite particle of angular momentum $\ell$ is indicated by the presence in $\mathcal{M}_\ell(s)$ of a \emph{pole singularity} --- in the vicinity of a pole at $s_0$, the elastic partial wave amplitude takes the form $\mathcal{M}_\ell \sim \frac{g^2}{s_0 - s}$. Causality forbids there to be poles off the real axis on the physical sheet~\cite{Gribov:1186219}, but poles \emph{on the real axis} below threshold on the physical sheet are allowed and are identified with bound states, with the bound-state mass being $\sqrt{s_0}$. Pole above threshold on the real axis violate unitarity, which one case see from Eq.~\ref{eq:elastic_unitarity}, which implies that,
\begin{equation}
 \mathrm{Im}\, \mathcal{M}_\ell^*\propto |\mathcal{M}_\ell|^2,
\end{equation}
which cannot be satisfied at a real-valued pole. 

Poles \emph{off the real axis} are allowed to appear if they are on the unphysical sheet --- they appear as complex conjugate pairs and can be identified with resonances. The real and imaginary parts of their position are often associated with a mass and a width for the resonance, $\sqrt{s_0} = m_R \pm i\frac{1}{2} \Gamma_R$. The proximity of the lower half-plane of the unphysical sheet to the real energy axis means that it is usually the pole at $\sqrt{s_0} = m_R - i\frac{1}{2} \Gamma_R$ which has the dominant effect on the measured scattering amplitude. Such a pole with a small value of $\Gamma_R$ lies close to the region of physical scattering and will give rise to a prominent `bump' in the energy region around $E^\star \approx m_R$. Resonance poles lying further from the real axis, or poles which lie close to the opening of a new threshold, do not necessarily have a simple `bump-like' signature on the real energy axis.

There is another possibility we have not yet considered: that a scattering amplitude has a pole singularity on the real energy axis below threshold, but on the \emph{unphysical} sheet. This case is known as a \emph{virtual bound-state}, and while the singularity will produce an enhancement at threshold, there is no asymptotic state possible. Virtual bound states can arise in cases when interactions are attractive, but not attractive enough to form a bound-state, with a famous experimental example being the di-neutron (spin-singlet $NN$ scattering).

\begin{figure}[t]
\includegraphics[width = \columnwidth]{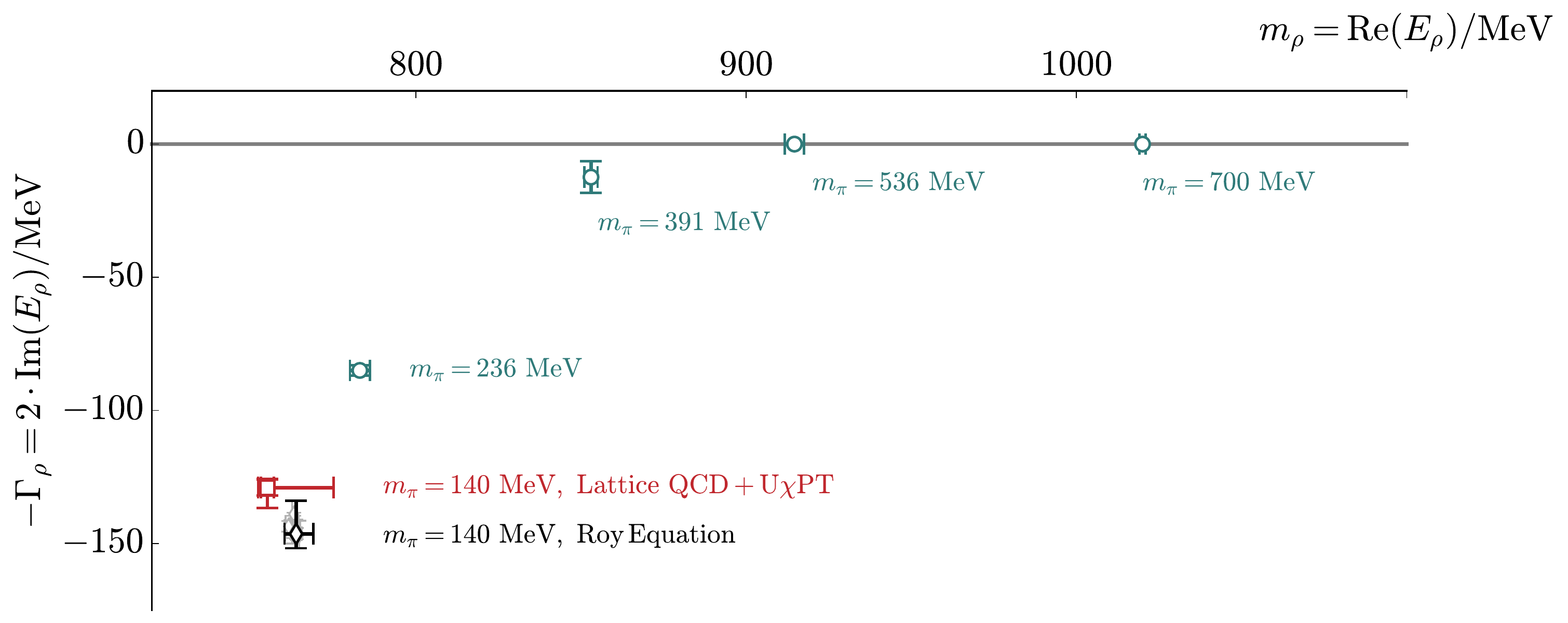}
\caption{The pole position of the $\rho$ resonance with varying light-quark mass (expressed in terms of the resulting pion mass) from lattice QCD calculations (blue, ~\cite{Lin:2008pr, Dudek:2013yja, Dudek:2012xn, Wilson:2015dqa}) and highly-constrained analysis of experimental data (black, ~\cite{Masjuan:2014psa, Ananthanarayan:2000ht, Colangelo:2001df, Zhou:2004ms, GarciaMartin:2011jx, Masjuan:2013jha}). Also shown (red), the result of extrapolating the the lattice QCD data down to the physical light quark mass using unitarized chiral perturbation theory (U$\chi$PT)~\cite{Bolton:2015psa} (U$\chi$PT is discussed in Sec.~\ref{sec:uchipt}). A transition from the $\rho$ being a stable bound state to being an unstable resonance is clearly visible.
}\label{fig:rho_mpi_0}
\end{figure}

We should be careful not to think of bound-states, virtual bound-states and resonances as necessarily having fundamentally different origins. For example, it has been observed in lattice QCD calculations that as the masses of the light $u,d$ quarks are increased from their physical values, the lightest resonance in $\pi\pi$ $P$-wave scattering, the $\rho$, has a width which decreases, until at a certain point the width becomes zero and the $\rho$ becomes a stable bound-state
~\cite{DeGrand:1990ip,Bernard:1993an,Leinweber:1993yw,Allton:1998gi,AliKhan:2001xoi,Bernard:2001av, Dudek:2013yja, Lin:2008pr, Feng:2010es, Feng:2014gba,  Dudek:2012xn, Wilson:2015dqa}
 (illustrated in Fig. \ref{fig:rho_mpi_0}). Another example is the di-neutron which, as presently understood,
evolves from a virtual bound-state when the $u,d$ quark masses take their physical value, to a true bound-state when the quarks are somewhat heavier~\cite{Berkowitz:2015eaa, Beane:2011iw, Beane:2012vq, Yamazaki:2012hi, Yamazaki:2015asa}.

\subsection{Coupled-channel scattering \label{sec:coupledchannel}}

Generalizing to the case where there are multiple two-body scattering channels kinematically accessible is relatively straightforward if we introduce the scattering \mbox{\emph{matrix} $\mathcal{M}$}, with matrix elements, $\mathcal{M}_{ab}$, where $a,b$ label channels (e.g. $\pi\pi$, $K\overline{K}$, $\eta\eta$ \ldots ). The symmetry of QCD under time-reversal ensures that this matrix is symmetric, and a constraint on the imaginary part of $\mathcal{M}$ is furnished by unitarity. The unitarity condition is most compactly expressed in terms of the matrix inverse of $\mathcal{M}$ in the partial wave basis,
\begin{equation}
\mathrm{Im} \, \left( \mathcal{M}_\ell^{-1}(E^\star) \right)_{ab} = - \delta_{ab}  \frac{1}{16\pi} \frac{2 \, q^\star_a}{E^\star} \, \Theta( E^\star - E^\mathrm{thr.}_a), \label{eq:unitarity}
\end{equation}
where we see that there are only imaginary pieces along the diagonal and only above the relevant kinematic threshold. The presence of the channel momentum, $q^\star_a$, indicates that $\mathcal{M}$ now has a branch cut starting at each kinematic threshold.\footnote{This complicates the Riemann sheet structure for complex $s$, leading to $2^{n_\mathrm{chan}}$ sheets for $n_\mathrm{chan}$ open channels. Typically only a small number of those sheets are close to physical scattering.}

A resonance pole will appear at the same position in each element of the matrix, $\left(\mathcal{M}_\ell(s) \right)_{ab} \sim \frac{r_{ab}}{s_0 -s}$. The residue can be factorized~\cite{Gribov:1186219}, $r_{ab} = g_a \cdot g_b$, such that we obtain \emph{couplings} describing the resonance's connection to each scattering channel.

Empirically one can measure only quantities which lead to estimates of $\left(\mathcal{M}_\ell(s) \right)_{ab}$ for \emph{real values} of $s=E^{\star 2}$, and the behavior in the complex plane must be obtained by \emph{analytically continuing} parameterized functions of $s$. Unitarity provides a strong constraint on the possible forms of such parameterizations, and a convenient way to implement this is by utilizing a $K$-matrix, where we write
\begin{equation}
\left(\mathcal{M_\ell}^{-1} \right)_{ab} = \left(\mathcal{K_\ell}^{-1}\right)_{ab} - i \,\delta_{ab}\,  \frac{1}{16\pi} \frac{2 \, q^\star_a}{E^\star}, \label{eq:kmat}
\end{equation}
and where by choosing $\mathcal{K}$ to be a matrix of \emph{real} functions of $s$ for real energies above kinematic threshold, we ensure the unitarity condition is satisfied.

\emph{Analyticity constraints} on scattering amplitudes in principle constrain the set of allowed forms for $\mathcal{K}$, but often these are not manifestly enforced, with forms typically chosen for their simplicity  used in fits to experimental data. For example, choosing the $K$-matrix to contain just a simple pole, $\mathcal{K}_{ab} = \frac{\gamma_a \gamma_b}{m^2 - s}$, leads to a scattering matrix of the Flatt\'e form~\cite{Flatte:1976xv, Flatte:1976xu}, or in the single-channel case, a Breit-Wigner. 

With an explicit choice of parameterization made, its free parameters can be varied to try to best describe experimental data for real energies. The amplitudes may then be analytically continued into the complex plane and examined for pole singularities. If they are found, the pole position can be interpreted in terms of a mass and width, and the pole residue can be factorized into couplings.

\subsection{Diagrammatic representation \label{sec:diagram_rep}}

We will later explore the behavior of scattering amplitudes when our theory is placed in a finite box, and the derivations that we will present are most easily performed using a diagrammatic representation. While QCD is a theory whose fundamental degrees of freedom are quark and gluons, these fields are confined, and the relevant asymptotic states for scattering processes feature only color-singlet hadrons. The Feynman diagrams we will be discussing in the following feature only these asymptotically allowed hadrons.\footnote{One could always envision that QCD can be mapped onto an all-encompassing low-energy effective field theory, which dictates all of the dynamics of hadrons.}

The diagrammatic representation of the scattering amplitude describing $n$ incoming and $n'$ outgoing hadrons, $\mathcal{M}(n \to n')$, is \emph{the sum over all diagrams with $n$ incoming and $n'$ outgoing legs that have been amputated and put on-shell. All intermediate propagators are evaluated using the $i\epsilon$-prescription and all intermediate loop momenta are integrated}. We make the fairly standard $i\epsilon$ prescription and momentum integration explicit in this statement because later we will have cause to adjust them.

\begin{figure}[t]
\begin{center}
\subfigure[]{ \label{fig:scat_ampa} \includegraphics[width = \columnwidth]{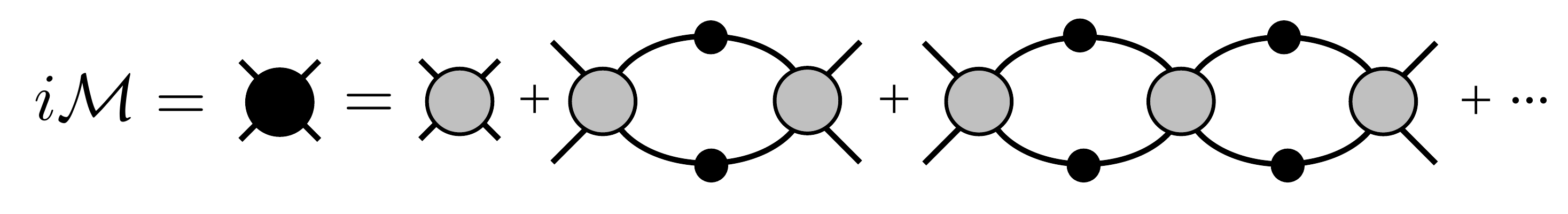}}\\
\subfigure[]{ \label{fig:kernel}    \includegraphics[width = 0.8\columnwidth]{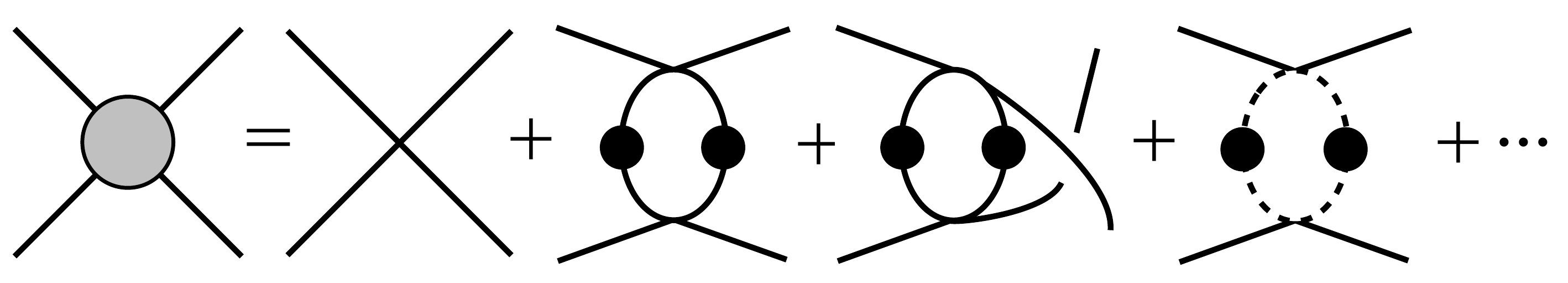}}\\
\subfigure[]{ \label{fig:1bodyprop} \includegraphics[width = 0.8\columnwidth]{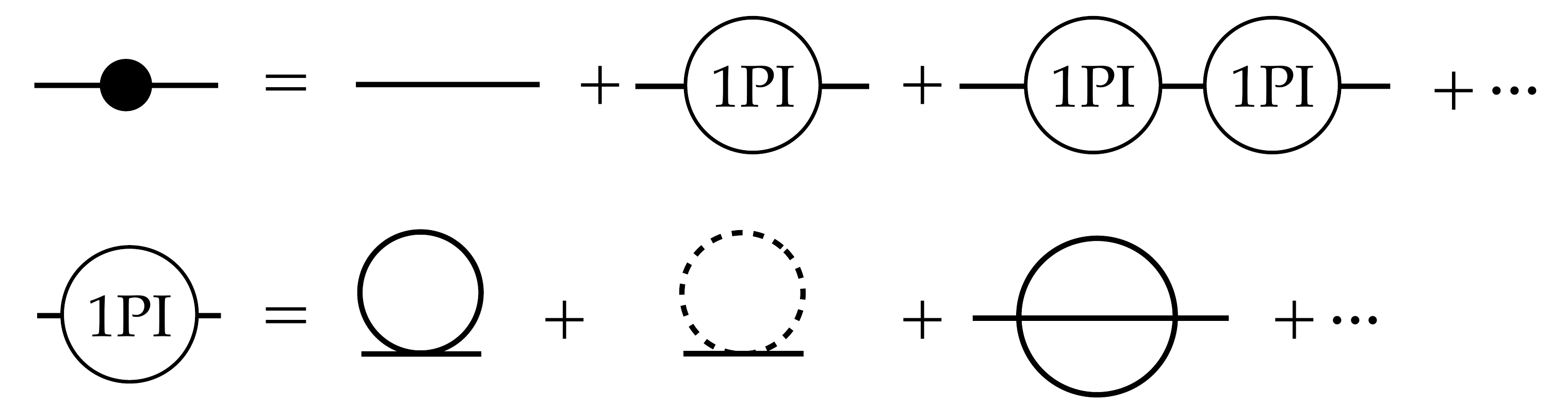}}
\caption{(a) The scattering amplitude, $\mathcal{M}$ as the sum over all on-shell amputated four-point diagrams in terms of the Bethe-Salpeter kernel (b) and the fully-dressed single particle propagator (c). Solid lines correspond to the particles in the primary channel being considered, while dashed lines denote fluctuations due to particles that cannot go on-shell. 
}\label{fig:scat_amp}
\end{center}
\end{figure}

For the simple case of elastic scattering of spinless hadrons, the sum of diagrams for $2 \to 2$ amplitudes can be presented immediately, as in Fig.~\ref{fig:scat_ampa}, where the diagrams can be drawn in a remarkably simple form by introducing the Bethe-Salpeter kernel, Fig.~\ref{fig:kernel}, and the fully-dressed single-particle propagator, Fig.~\ref{fig:1bodyprop}. The use of these two objects, which are themselves infinite sums, ensures that all diagrams, including those involving intermediate multi-particle states, are included in the definition of the scattering amplitude. This seemingly simple representation is in fact exact to all orders in the perturbative expansion. We will return to this diagrammatic representation of the scattering amplitude later when we consider correlation functions evaluated in a finite spatial volume.

\section{Lattice QCD  \label{Sec:lattice}}

Before proceeding to a discussion of the determination of resonance properties, we first present a basic overview of the numerical approach known as lattice QCD~(for a more detailed introduction to lattice QCD we point the reader to~\cite{gattringer2009quantum}).

Lattice QCD is a non-perturbative approach to QCD, where the quark and gluon fields are quantized on a discrete grid of spacetime points of finite size~\cite{Wilson:1974sk}. By transforming to Euclidean time, the path integral comes to feature a factor which can be treated as a probability according to which gauge-field configurations may be drawn in a Monte-Carlo approach. The gauge fields in lattice QCD are expressed in terms of $SU(3)$ matrices, $U$, one for each link of the lattice, and in terms of these and the quark fields, $\psi, \bar{\psi}$, the Euclidean partition function can be expressed as
\begin{eqnarray}
\mathcal{Z}_E=\int\! \mathcal DU \, \mathcal D\psi \, \mathcal D\bar{\psi} ~e^{-S_E(\psi, \bar{\psi}, U)},
\label{eq:ZE_IV}
\end{eqnarray}
where the integral is to be thought of as being over all possible configurations of the gauge and quark fields, and where the action $S_E$ is any suitable discretization of the QCD action. Correlation functions which feature fields at various space-time points can be similarly defined,
\begin{eqnarray}
\frac{1}{\mathcal{Z}_E} \int\! \mathcal DU \, \mathcal D\psi \, \mathcal D\bar{\psi} \; f(\psi, \bar{\psi}, U)  \;e^{-S_E(\psi, \bar{\psi}, U)},
\label{eq:Zcorr}
\end{eqnarray}
where for a suitable choice of function we may be able to relate the value of the correlation function to a physical observable.

Because the QCD lagrangian is bilinear in $\psi$, $\bar{\psi}$, the fermionic part of the integral can be done exactly leaving
\begin{eqnarray}
\mathcal{Z}_E=\int\! \mathcal DU  \, \det Q(U)\, e^{-S_g(U)} = \int\! \mathcal DU  \, e^{-\widetilde{S}_E(U)},
\label{eq:ZE_det}
\end{eqnarray}
where $Q(U)$ is the ``Dirac matrix'' which appears between $\bar{\psi}$ and $\psi$ in the discretization of the QCD lagrangian. In the case of correlation functions, all possible Wick contractions of $\psi$ and $\bar{\psi}$ in the function are replaced with ``propagators'', $Q^{-1}(U)$, when the fermion fields are integrated out.

The quantity $e^{-\widetilde{S}_E}$ is positive definite, and can be treated as a probabilistic measure of the importance of a given gauge field configuration, to be used in a Monte-Carlo generation of a finite number of possible gauge field configurations. With these configurations in hand, the value of a correlation function can be computed on each configuration, and by averaging over the ensemble of configurations, a statistical estimate obtained.

In this approach one discretizes and truncates spacetime, which introduces respectively an ultraviolet cutoff (via the lattice spacing, $a$) and an infrared cutoff (via the lattice volume, $L$), and QCD in its usual form can in principle be recovered as $L \to \infty$ and $a \to 0$. In this document we will not be overly concerned with the behavior of lattice QCD under changes in $a$, but the dependence on $L$ will prove to be precisely the tool needed to investigate scattering of hadrons in lattice QCD.

The introduction of a space-time boundary requires us to specify boundary conditions for the fields, and the most common choice is to have fermion(gauge) fields be anti-periodic(periodic) in the temporal direction, and to have all fields be periodic in spatial directions. Periodic boundary conditions on a cube, $L\times L \times L$, mean that free particles can only have three-momenta ${\mathbf{p} = \frac{2\pi}{L}(n_x, n_y, n_z)}$ for integer $n_i$. The finite size in the temporal direction, usually denoted $T$, effectively puts the system at a finite temperature, but provided $m_\pi T \gg 1$, where $m_\pi$ is the mass of the lightest asymptotic particle in QCD (the pion), this effect is usually negligibly small, being suppressed exponentially.

Truncation to a finite periodic spatial volume has several impacts on the theory, some of which we will take advantage of to determine scattering amplitudes --- these will be discussed in the next section. Other impacts are not so useful. An example is the change in the properties of stable hadrons due to them being able to `see' themselves around the periodic volume. These `polarization' effects can be shown to be exponentially suppressed with increasing volume~\cite{Luscher:1985dn}, scaling like $e^{- m_\pi L}$, and by working in large enough volumes, $m_\pi L \gg 1$, they can be reduced to a negligible level.

The parameters which must be specified in order to carry out a computation in lattice QCD include the masses of the quarks. Historically, most lattice QCD calculations have been carried out with mass values for the light \emph{up} and \emph{down} quarks somewhat larger than their physical masses, owing to the large computational cost of working with very light fermion fields. An obvious effect of this is that the masses of hadrons computed in these lattice QCD calculations come out larger than they would in the physical version of QCD (or in experiment). In particular it is usual to characterize the quark mass used by quoting the \emph{pion mass} calculated using that quark mass. In recent years we have seen an increasing number of lattice QCD calculations (of relatively simple quantities) that use a quark mass value that is rather close to the physical value.

\section{Scattering in a finite-volume  \label{Sec:finitevolume}}

In a finite volume, strictly speaking, we cannot introduce the asymptotic states we require to define a scattering system -- when we separate particles by large distances they begin to feel significant effects from the boundary, such that they are not truly free or asymptotic. Furthermore, the spectrum of eigenstates of QCD in a finite volume is qualitatively different from that in infinite volume. The continuous spectrum of multihadron states we observe in scattering experiments is present because the volume of the system is effectively infinite, allowing a particle to have any continuous value of momentum. In a finite volume, no such continuous distribution can exist, as application of the boundary conditions will quantize momenta, leading to a \emph{discrete} spectrum of states. 

In this chapter we will show that information about scattering amplitudes can be obtained from the discrete spectrum in a finite volume, and in particular from its dependence on the volume. We will illustrate the basic idea using the simplest possible system, non-relativistic quantum mechanics in one space dimension, before moving to the desired case, relevant to lattice QCD calculations, of scattering in quantum field theory in three space dimensions when the boundary is a cube with periodic boundary conditions. In the latter case, while conceptually the physics is the same, complications arise from the mismatch between the partial-wave expansion, which relies upon continuous rotational invariance, and the cubic geometry of the lattice boundary.

The final result, which we will refer to as the \emph{L\"uscher quantization condition}
\footnote{The idea of extracting scattering information from finite-volume spectra predates~\cite{DeWitt:1956be, Huang:1957im} L\"uscher's seminal work, but he was the first to find a \emph{non-perturbative} relation}
, is of the following form:
\begin{equation}
\det \big[ F^{-1}(E, \mathbf{P}; L) + \mathcal{M}(E) \big] = 0. \label{eq:QC_2body}
\end{equation}
This equation features the determinant of a sum of two complex, energy-dependent matrices, and is quite general, with a version of it being applicable to all possible $2\to 2$ scattering processes, be they elastic, coupled-channel, featuring spinless hadrons or hadrons with spin
~\cite{Luscher:1986pf, Luscher:1990ux, Rummukainen:1995vs, Feng:2004ua, He:2005ey, Bedaque:2004kc, Liu:2005kr, Kim:2005gf, Christ:2005gi, Lage:2009zv, Bernard:2010fp, Fu:2011xz, Leskovec:2012gb,  Briceno:2012yi, Hansen:2012tf, Guo:2012hv, Li:2012bi, Briceno:2013hya, Briceno:2014oea}. 
 The components are $\mathcal{M}$, which is a matrix, diagonal in total angular momentum, built out of the infinite volume scattering matrices introduced in Section~\ref{sec:coupledchannel}, and $F^{-1}$ which encodes the `kinematics' of the finite-volume. $F^{-1}$ is in general not diagonal in angular momentum, but is diagonal in the space of dynamically coupled channels, and it differs depending on the value of the total momentum of the two-body system, $\mathbf{P}$.

We can interpret Eq.(\ref{eq:QC_2body}) in the following way -- for a given total momentum, $\mathbf{P}$, and a set of scattering amplitudes, $\mathcal{M}_\ell(E)$ (which might be matrices in the space of kinematically open channels), in an $L\times L \times L$ volume, the discrete spectrum of states having a specified quantum number
\footnote{we will discuss later what these `quantum numbers' are with a cubic symmetry, given that the broken rotational symmetry indicates that they will not be values of angular momentum}
, $E_n(\mathbf{P},L)$, is given by all energies for which the determinant evaluates to zero. Techniques to determine the scattering amplitudes from values of $E_n(\mathbf{P},L)$ computed using lattice QCD will be discussed later, after we first motivate the finite-volume approach, and then sketch a derivation of Eq.(~\ref{eq:QC_2body}).

\vspace{20mm}
\subsection{Scattering in non-relativistic quantum mechanics in one space dimension \label{sec:1dqm}}
We can illustrate the essential relationship between the discrete spectrum of states in a finite periodic volume and the infinite volume scattering amplitudes using the simple case of elastic scattering of two identical spinless bosons in one spatial dimension in non-relativistic quantum mechanics~\cite{deGrand:deTar}.

We begin by considering the infinite-volume system where we suppose that the bosons, separated by a distance $|x|$, interact through a finite-range potential, $V(|x|)$, with $V(|x| > R) = 0$. Outside the potential, the wavefunction of the two-boson system will be of the form $\psi_p(|x|) \sim \cos \big( p |x| + \delta(p) \big)$, where all values of the momentum, $p \ge 0$, are allowed, and where the \emph{elastic scattering phase-shift}, $\delta(p)$, describes the scattering amplitude. In principle, if the potential is specified, one can solve for the positive energy eigenfunctions inside the well (thus accounting for the dynamics of the system), and match the wavefunctions at $|x|=R$ to determine the phase-shift. 

Now consider putting the system in a periodic box (a circle of circumference $L > R$). A single boson in this box would have a simple momentum spectrum, $p_n = \frac{2\pi}{L} n$ for integer $n$, which follows from applying the periodic boundary conditions to free-particle wavefunctions, $e^{ipx}$. For the \emph{interacting boson pair}, we apply the periodic boundary conditions (at $x = \pm L/2$) to the wavefunction, $\cos \big( p |x| + \delta(p) \big)$, and its derivative, which leads to the following condition on the momentum:
\begin{equation*}
p = \frac{2\pi}{L}n - \frac{2}{L} \delta(p).
\end{equation*}
This simple result illustrates most of the important features of the spectrum in a finite-volume: there will be only \emph{discrete} values of $p$ (and hence $E$) which solve this equation; if there is no scattering ($\delta(p)=0$), we recover the free-particle spectrum; when the particles interact, the discrete spectrum depends on the infinite-volume scattering amplitude (via $\delta(p)$) and the volume of the `box' ($L$). 

The result for quantum field theory in a three-dimensional periodic cubic box (a torus) shares all these qualitative features -- let us now sketch a derivation.

\vspace{9mm}
\subsection{Scattering in a periodic cubic volume \label{sec:luscher}}

To arrive at Eqn.~\ref{eq:QC_2body}, we equate two different but equivalent representations of the \emph{two-point correlation function} in a finite volume: the dispersive representation which expresses the correlation function in terms of the discrete spectrum of eigenstates, and an all-orders diagrammatic representation. The derivation sketched here follows the approach first presented in~\cite{Kim:2005gf, Hansen:2012tf}, using a notation similar to that in~\cite{Briceno:2015tza}.

A two-point correlation function in a finite Euclidean spacetime of spatial dimension $L \times L \times L$ can be written\footnote{on a lattice, the integral $\int_L \! d \mathbf x$ would be replaced by a finite sum over lattice sites},
%
%
\begin{align}
C_L(x_4-y_4, \mathbf P)   \equiv \int_L &\! d \mathbf x \int_L \! d \mathbf y \, e^{- i \mathbf P \cdot (\mathbf x - \mathbf y)} \nonumber \\
&\times \Big [ \langle 0 \vert T \mathcal A(x) \mathcal B^\dagger(y) \vert 0 \rangle \Big ]_L \,,
\label{eq:two-point}
\end{align}
where $B^\dagger$ and $\mathcal A$ are creation and annihilation operators having the quantum numbers of the hadron-hadron scattering channel we wish to study, and where we have projected into definite three-momentum, ${\mathbf{P} = \frac{2\pi}{L}\big( n_x, n_y, n_z \big)}$. The dispersive representation follows by inserting a complete set of discrete eigenstates of the finite-volume Hamiltonian, $C_L(x_4-y_4, \mathbf P) = $
\begin{equation}
  L^6 \sum_n e^{ - E_n(x_4 - y_4)}   \big\la 0 \big| \mathcal{A}(0) \big| E_n, \mathbf{P}; L\big\ra  \,  \big\la E_n, \mathbf{P}; L \big| \mathcal{B}^\dag(0) \big| 0 \big\ra, \label{eqn:dispersive}
\end{equation}
%
%
where the eigenstates in a finite volume are normalized according to ${ \big\la E_n', \mathbf{P}'; L \big| E_n, \mathbf{P}; L \big\ra = \delta_{n,n'} \, \delta_{\mathbf{P}',\mathbf{P}} }$.  

\begin{figure}
\begin{center}
\centering
\subfigure[]{\includegraphics[width =\columnwidth]{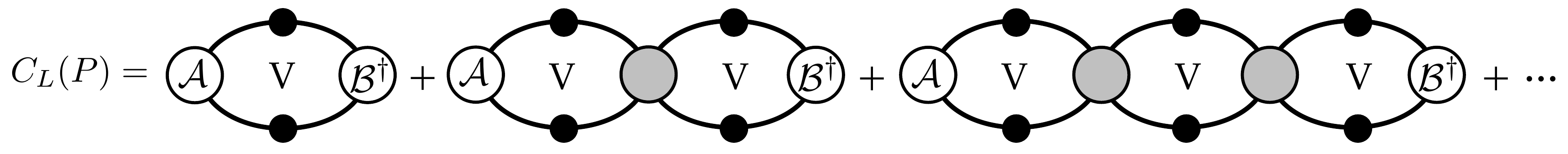}  \label{fig:FVcorr}}
\subfigure[]{\includegraphics[width =\columnwidth]{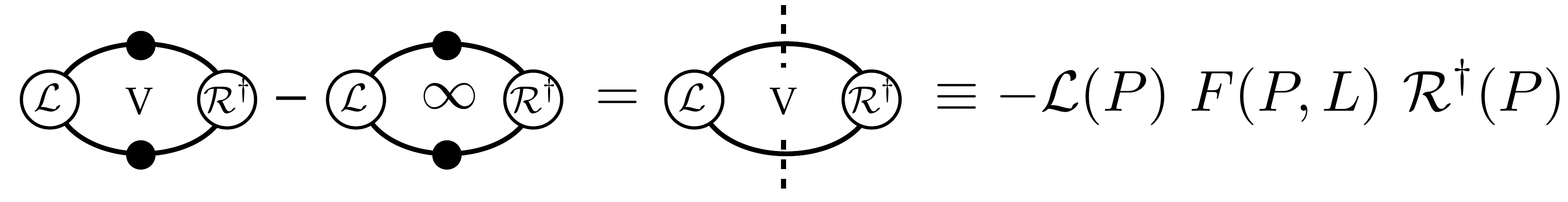} \label{fig:F_function}}
\caption{(a) The diagrammatic representation of the finite-volume two-point correlation function for energies where the two-particle states can go on-shell. (b) The finite-volume function, $F(P,L)$, defined in Eq.~(\ref{eq:Fscdef}), expressed in terms of the difference between finite and infinite volume two-particle loops.
}\label{fig:FVcorr_F_function}
\end{center}
\end{figure}

The diagrammatic representation of the same finite-volume correlation function, for energies below any three-particle threshold, can be constructed according to the approach laid out in~\cite{Kim:2005gf}. As illustrated in Fig.~\ref{fig:FVcorr_F_function} those diagrams where an intermediate two-particle state can go on-shell play the dominant role in determining the dependence of the correlation function on the finite-volume.  Qualitatively this can be understood by recognizing that on-shell particles can propagate over arbitrary distances and hence sample the boundaries of the volume, and the quantitative manifestation of this will be pole singularities at energies corresponding to allowed free two-particle states. Diagrams in which the intermediate two-particle state cannot go on shell can be shown to contribute at a level which is exponentially suppressed $\sim e^{-m_\pi L}$, and these can be neglected for volumes $L \gg m_\pi^{-1}$.

The core object in the diagrammatic representation is the \emph{difference} between the two-particle loop in finite and infinite volume as shown in Fig.~\ref{fig:F_function}. We will illustrate this in the simplest case of two identical spinless particles -- the extension to cases of unequal masses and particles having non-zero spin can be found in the literature~\cite{Briceno:2012yi, Hansen:2012tf, Briceno:2014oea, Briceno:2015tza}. Inside the loop, one particle carries momentum $\textbf k$ and the other $\textbf P-\textbf k$, and their on-shell energies are ${\omega_k=\sqrt{k^2+m^2} }$ and ${ \omega_{Pk}=\sqrt{(\textbf P-\textbf k)^2+m^2} }$ respectively. In a finite-volume, the integral over continuous particle momentum is replaced by a sum over the allowed discrete momenta $ \textbf k = (2 \pi/L) \textbf n$ with $\textbf n \in\mathbb Z^3$,
\begin{align*}
\int\!\! \frac{d\textbf k}{(2 \pi)^3}\longrightarrow&
\sum_{\mathbf k } \frac{\Delta k_x \Delta k_y \Delta k_z}{(2 \pi)^3}\\
&=
\sum_{\mathbf k } \frac{\Delta n_x \Delta n_y \Delta n_z}{L^3}
=
\frac{1}{L^3}\sum_{\mathbf k }.
\end{align*}
The difference between finite and infinite-volume loop functions, having arbitrary smooth functions $\mathcal L(P-k,  k)$ and $\mathcal R^\dag(P-k,  k)$ at the vertices, takes the form
\footnote{For identical particles, one should include a symmetry factor of $\frac{1}{2}$. Through this review we ignore such factors, both in finite- and infinite-volume quantities, in such a way that the relationship between these is consistent.}
\begin{align*}
\mathcal{F}_L&=\left[\frac{1}{L^3} \sum_\mathbf{k} - \int\!\!\frac{d\textbf{k}}{(2\pi)^3}\right]\int\frac{d k_4}{2\pi}\\
&\quad\times\mathcal{L}(P-k,k)   \,   \Delta(k)   \,   \Delta(P-k)   \,   \mathcal{R}^\dagger(P-k,k),
\end{align*}
with $\Delta$ being the fully-dressed Euclidean single-particle propagator, defined with unit residue at the poles. Performing the contour integration over $k_4$, dropping terms which do not have singularities for physical $E$ (i.e. terms which vanish exponentially in the volume), and approximating the non-singular part of the function by the value at the pole, we arrive at
%
%
\begin{align*}
\mathcal F_L  &=  -     \bigg[ \frac{1}{L^3} \sum_{\mathbf k } - \int\!\! \frac{d\textbf k}{(2 \pi)^3} \bigg ] \frac{1}{2 \omega_k \, 2 \omega_{Pk}} \nonumber \\ 
& \times \mathcal L(P - k,k)  
\frac{1}{E -   \omega_k - \omega_{Pk} + i \epsilon}    \mathcal R^\dagger(P-k,k )  \bigg \vert_{k_4=i\omega_k}
\end{align*}
%
%
The presence of the pole at $E=  \omega_k - \omega_{Pk} + i \epsilon$ will ensure the dominance of the on-shell values of $\mathcal L$ and $\mathcal R^\dagger$. These have a familiar decomposition when expressed in the \mbox{cm-frame}. The on-shell condition written in terms of the cm-frame relative momentum, $\mathbf{q}^\star$, is $E^\star = 2 \omega_q^\star$, which determines the magnitude of $\mathbf{q}^\star$, but not its direction, which coincides with $\hat {\textbf{k}}$ boosted into the cm-frame. We can decompose $\mathcal L$ and $\mathcal R^\dagger$ in terms of spherical harmonics, as
\begin{align}
\mathcal L_{\mathrm{on}}(P, \textbf k^\star ) &\equiv \sqrt{4 \pi} \,  Y_{\ell m }(\hat {\textbf k}^\star)\mathcal \, \mathcal{L}_{ \ell m }(P), \nn \\
\mathcal R_{\mathrm{on}}^\dagger(P, \textbf k^\star )
&\equiv 
 \sqrt{4 \pi} \,  Y^*_{\ell m }(\hat {\textbf k}^\star)\mathcal \, \mathcal{R}^\dagger_{ \ell m}(P) \,,
\,\,\,
\label{eq:finalonfunc}
\end{align}
and the factors $\mathcal{L}_{\ell m}, \mathcal{R}^\dag_{\ell m}$, being independent of the loop momentum, can be taken outside the integral and sum to give
\begin{equation}
\label{eq:Fsum_final}
\mathcal F_L= - \mathcal L_{ \ell m }(P)  \, F_{ \ell m ;  \ell' m'} (P,L)  \, \mathcal R^\dagger_{\ell' m'}(P)
,
\end{equation}
where
\begin{widetext}
\begin{align}
\label{eq:Fscdef}
&F_{ \ell m ;  \ell' m'}(P,L)  \equiv 
   \bigg[ \frac{1}{L^3} \sum_{\mathbf k } - \int\!\! \frac{d\textbf k}{(2 \pi)^3} \bigg ]
   \quad\frac{ 4 \pi \,  Y^{}_{\ell m }(\hat {\textbf k}^\star)   \, Y^*_{\ell' m'}(\hat {\textbf k}^\star)  }{2 \omega_k \, 2 \omega_{Pk}  \, (E -  \omega_k - \omega_{Pk} + i \epsilon )} \left (\frac{k^{\star}}{q^\star} \right)^{\ell+\ell'} \,.
\end{align}
\end{widetext}
The factors $\bigg ( \frac{k^\star}{q^\star} \bigg )^{\ell}$ are introduced to cancel the ambiguity in defining the spherical harmonics at threshold.

In summary, the difference between a finite-volume two-particle loop and its infinite-volume counterpart can be written as a product of matrices in angular momentum space featuring only \emph{on-shell quantities}. We may apply this result to the diagrams appearing in Fig.~\ref{fig:FVcorr} --- there the functions appearing inside the loops are either Bethe-Salpeter kernels or overlaps with the source and sink operators. For each diagram we replace it with its infinite volume expression plus a finite-volume correction which is functionally identical to Eq.(\ref{eq:Fsum_final}) with $\mathcal L$ and $\mathcal R^\dagger$ replaced with the appropriate functions. The sum of all $2 \to 2$ on-shell amputated diagrams is replaced with the scattering amplitude
\footnote{whose partial-wave expansion we generalize to ${\mathcal{M}(P, \hat{\mathbf{q}}'^\star, \hat{\mathbf{q}}^\star) = Y^*_{\ell m}(\hat{\mathbf{q}}^\star) \, \mathcal{M}_{\ell m; \ell' m'}(E^\star)\, Y^*_{\ell' m'}(\hat{\mathbf{q}}'^\star)}$ for arbitrary directions of initial and final-state particles, realizing that rotational invariance in infinite volume will make the matrix $\mathcal{M}_{\ell m; \ell' m'}$ diagonal in angular momentum}
$\mathcal M$. 
After a substantial amount of algebra, one can show that the correlation function can be written in terms of the infinite-volume correlation function, $C_{\infty}$, and a geometric series in $-\mathcal M\,F$ which appears between the fully dressed overlaps to the source and sink operators $B^\star$ and $A^\star$,
\begin{widetext}
\begin{eqnarray}
C_L(x_4-y_4, \textbf P)
&=& L^3 \!\int \!\frac{d P_4}{2 \pi} e^{i P_4(x_4-y_4)} \Bigg[ C_{\infty}(P) - A^\star(P) \,  F(P,L)
\sum_{n=0}\Big(\!-\mathcal M(P) \, F(P,L) \Big)^n \, B^\star(P) \Bigg ] \,, 
 \nn\\
&=& L^3 \!\int \!\frac{d P_4}{2 \pi} e^{i P_4(x_4-y_4)} \left[ C_{\infty}(P) - A^\star(P) \Big[ F^{-1}(P,L) + \mathcal M(P)\Big]^{-1} B^\star(P) \right] \,. \label{eq:steptwo} 
\end{eqnarray}
\end{widetext}
In order to complete the derivation by comparing to the dispersive representation, Eq.(\ref{eqn:dispersive}), we must evaluate the integral over $P_4$, which can be done by considering the analytic properties of the integrand. In order for Eq.(\ref{eqn:dispersive}) to result, the integrand must include a series of poles of the form $\frac{1}{P_4 - i E_n(L)}$. Since the position of the poles corresponds to the spectrum in a finite volume, they are surely volume dependent, and as such they must arise from the factor $\big[F^{-1}(P,L) + \mathcal M(P)\big]^{-1}$ in Eq.(\ref{eq:steptwo}). This factor will diverge appropriately if the matrix $\big[ F^{-1}(P,L) + \mathcal M(P)\big]$ is singular, that is if $\det \big[F^{-1}(P,L) + \mathcal M(P)\big] = 0$, which we recognize as Eq.(\ref{eq:QC_2body}).

The relationship between the finite-volume spectrum and the infinite-volume scattering amplitudes, Eq.(\ref{eq:QC_2body}) has thus been derived, but one additional useful result concerning matrix-elements follows if we evaluate the $dP_4$ integral, giving,
\begin{align}
C_L(x_4-y_4, \textbf P)   &= \sum_n e^{- E_n(\mathbf{P},L) (x_4-y_4)} \nonumber \\
  &\times L^3  
A^\star(E_{n}, \textbf P) \, \mathcal R(E_{n}, \textbf P)   \, B^\star(E_{n}, \textbf P)
\label{eq:CL_diag}  
\end{align}
where $\mathcal R$ is the residue of the finite-volume two-particle propagator,
\begin{align}
\label{eq:Rdef}
\mathcal R(E_{n}, \textbf P) &\equiv 
\nn\\
&\hspace{-1.5cm} \lim_{P_4 \rightarrow i E_{n}} \left( - (i P_4 + E_{n}) \Big[ F^{-1}(P,L) + \mathcal M(P)\Big]^{-1}    \right) \,.
\end{align}
%
%
We will consider the importance of this result later when we investigate the computation of resonant matrix elements in QCD.

\newpage
\subsection{Relating scattering amplitudes to finite-volume spectra \label{sec:decomposing}}

Eq.(\ref{eq:QC_2body}), which we repeat here,
\begin{equation*}
\det \big[ F^{-1}(E, \mathbf{P}; L) + \mathcal{M}(E) \big] = 0,
\end{equation*}
describes the relationship between the finite-volume energy spectrum, $E_n(\mathbf{P}, L)$, and the infinite-volume scattering amplitudes. As the derivation in the previous section indicates, the matrices $F^{-1}$ and $\mathcal{M}$ are, even in the simplest elastic spinless case, formally of infinite size, featuring all possible integer values of $\ell \ge 0$. How then can we hope to use this equation in practice? To make progress we first note that at low energies, higher partial-wave amplitudes typically reduce rapidly in magnitude as $\ell$ is increased. This is expected on the grounds of angular momentum conservation which leads to a behavior at threshold which much fall at least as fast as $\mathcal{M}_\ell \sim (q^\star)^{\ell}$. It follows that, in practice, the contribution of higher partial waves to Eq.(\ref{eq:QC_2body}) is numerically negligible at low energies, and one is well justified truncating the angular momentum space to the lowest few $\ell$ values.

An additional simplification comes from considering the fact that, although the use of a cubic symmetry for the lattice boundary has broken the continuous rotational symmetry needed for angular momentum to be a good quantum number, there is a smaller residual symmetry group still present. In the case of the rest frame ($\mathbf{P} = \mathbf{0}$), this is the group of rotations which leave a cube invariant, the \emph{cubic} or \emph{octahedral} group. This group has a finite number of irreducible representations, or \emph{irreps}. For example, for bosonic systems the irreps are labelled $A_1, A_2, T_1, T_2$ and $E$ (plus an additional parity label), and these can be though of as the `quantum numbers' carried by finite-volume eigenstates. When projected into one of these irreps (\emph{subduced} is the term-of-art used), the dense matrix $F^{-1}$ (whose rows and columns are labelled by $\ell, m$ values) becomes block diagonal, reflecting the fact that only certain $\ell$ values subduce into each irrep, see Table~\ref{tab:subduction}. 

\begin{table}[b]
\begin{tabular}{c|l}
$\Lambda$(dim) & $\ell$ \\
\hline\\[-1.4ex]
$A_1(1)$ & $0,4,\ldots$ \\
$T_1(3)$ & $1,3,4,\ldots$ \\
$T_2(3)$ & $2,3,4,\ldots$ \\
$E(2)$ & $2,4,\ldots$ \\
$A_2(1)$ & $3,\ldots$ \\
\end{tabular}
\caption{The lowest partial waves, $\ell$, that are subduced into the $\Lambda$ irrep for the cubic group for bosonic systems -- the dimension of each irrep is given in parentheses. An irrep of dimension $d$ has $d$ equivalent ``rows'' -- analogous to the $2\ell +1$ $m$-values in the rotationally invariant case. 
\label{tab:subduction}}
\end{table}

For non-zero momenta $\mathbf{P}$, the symmetry is reduced still further, to the appropriate \emph{little group}, each of which has its own set of irreducible representations. We will not go into detail here (see e.g.~\cite{Thomas:2011rh}) except to note that in general, these \emph{in-flight} irreps are typically more dense in $\ell$ content and usually feature both parities.

Although subduction simplifies the behavior of $F$ in angular momentum space somewhat, the dependence on energy remains complicated, featuring many singularities --- some illustration is provided in Appendix~\ref{Sec:F_funct}.

\subsubsection{Dominance of the lowest partial-wave \label{sec:one_pw}}

An extremely important case, the one most commonly considered in the literature to date, has only the lowest partial-wave subduced into an irrep having a non-negligible amplitude, with all higher partial-waves assumed small enough to ignore. In the case of elastic scattering this assumption leads to $F^{-1}$ and $\mathcal{M}$ being $1\times 1$ matrices, and Eq.(\ref{eq:QC_2body}) reducing to a simple equation, $\mathcal{M} = - F^{-1}$, which is often expressed in terms of the elastic scattering phase-shift as
\begin{equation}
\cot \delta_\ell(E) = - \cot \phi^\Lambda_\ell(\mathbf{P}, L, E)\label{eq:cotd}
\end{equation}
where the right-hand-side is related to the finite-volume function we have previously discussed, ${\cot \phi^\Lambda_\ell(\mathbf{P}, L, E) = 16 \pi \, \frac{E^\star}{2q^\star} \,\mathrm{Re}\, F^\Lambda_\ell}$, where the energy dependence of $F$ is illustrated in Appendix~\ref{Sec:F_funct}.

Under these conditions, determining the scattering amplitude from values of $E_n(\mathbf{P},L)$ computed using lattice QCD is straightforward -- one simply inserts the lattice energy value into the function on the right-hand-side of Eq.(\ref{eq:cotd}), to yield a value of $\delta_\ell$ at that energy. The more energy levels are determined, the more discrete points one will have on the phase-shift curve.

One might wonder how we can know, in any particular calculation, whether we are justified in neglecting higher partial waves. The simplest approach is to directly determine the corresponding higher partial-wave phase-shift using a relevant irrep. For example, if we are considering $\pi\pi$ scattering with $I=1$, we might be concerned about the contribution of the $\ell=3$ partial-wave to the $T_1^-$ irrep, which should be dominated by $\ell=1$ at low energies. We can use energy levels in the $T_2^-$ and $A_2^-$ irreps, which have $\ell = 3$ as their lowest partial wave\footnote{ $\ell = 2$ cannot contribute to $I=1$ in this case due to Bose symmetry. We can also make use of levels in moving-frame irreps which have $\ell = 3$ as their leading contribution.}, assuming that still higher partial-waves are negligible. 

Going beyond this simplest case of a single partial wave dominating, Eq.(\ref{eq:QC_2body}), suitably subduced and truncated to include only the lowest few relevant partial-waves, in principle has each finite-volume energy level as a function of the phase-shift for all included partial waves. Of course, level--by--level this is just one equation in multiple unknowns, and it cannot be solved. However, in practice, we will determine many energy levels with each one providing a constraint. If the scattering amplitudes are smooth functions of the energy, they can be suitably parameterized, and we can attempt to describe the spectrum as a whole, by varying a small number of parameters. A similar situation arises in the case of coupled-channel scattering, and indeed is more acute there, as there is usually no sense in which one channel can be considered `weak' relative to a `dominant' channel, so we will illustrate the \emph{parameterization of amplitudes} approach there.

\subsubsection{Coupled-channel scattering and parameterization of scattering amplitudes \label{sec:fv_coupled}}

We can illustrate the approach most easily if we initially assume a system dominated by $S$-wave scattering of two coupled channels of spinless particles that we will label $\pi\pi$ and $K\overline{K}$. If $m_\pi < m_K$ there is a region of elastic $\pi\pi$ scattering before the $K\overline{K}$ threshold opens. \mbox{$S$-wave} scattering is described by the matrix of scattering amplitudes,
\begin{equation*}
\mathcal{M} = \left(
		\begin{array}{cc}
				\mathcal{M}_{\pi\pi; \pi\pi}         &  \mathcal{M}_{\pi\pi; K\overline{K}} \\
				\mathcal{M}_{\pi\pi; K\overline{K}}  &  \mathcal{M}_{K\overline{K}; K\overline{K}}  
		\end{array}
		\right),
\end{equation*}
where each entry is a complex number at each value of cm-energy. Multichannel unitarity, Eq.(\ref{eq:unitarity}), specifies the imaginary parts so that just three real numbers are needed at each value of energy to completely specify the scattering. This matrix appears in a version of Eq.(\ref{eq:QC_2body}),
\begin{equation}
\det \left[   
	\left(
		\begin{array}{cc}
				F^{-1}_{\pi\pi}         &  0 \\
				0  &  F^{-1}_{K\overline{K}}  
		\end{array}
	\right)
	+
	\left(
		\begin{array}{cc}
				\mathcal{M}_{\pi\pi; \pi\pi}         &  \mathcal{M}_{\pi\pi; K\overline{K}} \\
				\mathcal{M}_{\pi\pi; K\overline{K}}  &  \mathcal{M}_{K\overline{K}; K\overline{K}}  
		\end{array}
	\right)	
\right] = 0, \label{eq:2chan}
\end{equation}
where $F^{-1}_{\pi\pi}$ and $F^{-1}_{K\overline{K}}$ are known functions of $E$ and $L$ whose imaginary parts are such that this equation is real for matrices $\mathcal{M}$ satisfying the multichannel unitarity condition (see Appendix~\ref{Sec:F_funct}). It follows that for any particular finite-volume energy level $E_n(L)$, we have one real equation and three unknowns.

An approach that has proven successful in a number of explicit calculations (which will be discussed in Section~\ref{Sec:examples}) was outlined in~\cite{Guo:2012hv}. In it the energy dependence of the scattering matrix is \emph{parameterized} by some explicit analytic form featuring a small number of free parameters, $\{a_i\}$. For any particular set of values of these free parameters, the explicit parameterized form for $\mathcal{M}$ can be substituted into Eq.(\ref{eq:2chan}) and that equation solved for a discrete spectrum of finite-volume energies in an $L\times L \times L$ box, $E^\mathrm{par.}_n(L; \{a_i\})$. This spectrum can be compared to the spectrum obtained in a lattice calculation, and by varying $\{a_i\}$ a $\chi^2$ can be minimized to find a best description. 

Of course this approach requires us to propose particular parameterization forms for $\mathcal{M}$. Since they must satisfy multichannel unitarity if Eq.(\ref{eq:2chan}) is to have solutions, the question of how to parametrize  $\mathcal{M}$ is equivalent to to choosing a parametrization for the $K$-matrix introduced in Eq.(\ref{eq:kmat}). One  might be correctly concerned that by having choosen a particular parametrization for the $K$-matrix one might introduce possible systematic bias. This can and should be tested by varying the choice made in the parametrization. It is empirically observed in explicit calculations presented in Section~\ref{Sec:examples} that in cases where a large density of energy levels are determined, strongly constraining the free parameters in the parameterization, any sufficiently flexible form for the \mbox{$K$-matrix} will give compatible amplitudes. Furthermore, if the channel contains a fairly narrow resonance, the analytic continuation of all parametrization forms into the complex plane will show pole singularities at consistent locations.

\subsubsection{Examples of finite-volume spectra for simple scattering amplitudes \label{sec:examples}}
In this subsection we will present some illustrative examples of the finite-volume spectra obtained by solving Eq.(\ref{eq:QC_2body}) for some simple scattering amplitudes. The very simplest case is that of no scattering at all, and in this case the finite-volume spectrum will simply be the discrete spectrum of non-interacting hadron pairs. For example for a pair of spinless particles, which we might call $\pi\pi$, in the rest frame, $E(L) = 2 \sqrt{ m_\pi^2 + k^2 }$ with $\mathbf{k} = \frac{2\pi}{L}(n_x,n_y,n_z)$.

\begin{figure*}
\includegraphics[width = 0.8\textwidth]{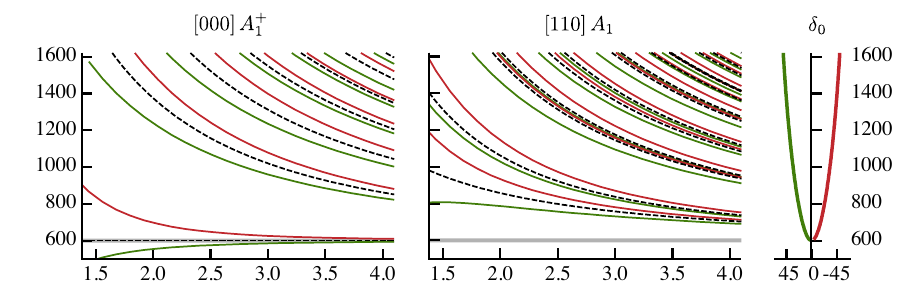}
\caption{Finite volume spectrum in a rest-frame irrep and a moving-frame irrep for weak attractive (green) and repulsive (red) elastic scattering. Non-interacting energy levels are indicated by the dashed black lines and the gray band shows the kinematic threshold ($2m_\pi$). Plotted is cm-frame energy, $E^\star$ in MeV, against $L$ in fm. Scattering particles have mass $300$ MeV and the scattering length is $|a| = 0.32$ fm. Rightmost panel shows the corresponding elastic phase-shift in degrees.}\label{fig:fv_weak} 
\end{figure*}

The simplest non-trivial case we will consider is weakly interacting elastic $S$-wave scattering, described by $\tan \delta(q^\star) = a \, q^\star$, with \emph{scattering length} $ a < 0$ being repulsive and $a > 0$ being attractive. Figure~\ref{fig:fv_weak} shows the finite volume spectra in the rest frame $\mathbf{P} = [000]$ \mbox{($A_1^+$ irrep)} and, for illustration, one possible moving frame, $\mathbf{P} = [110]$ ($A_1$ irrep). We observe that in the weak attractive case ($a >0$, green curves), the energies lie close to, but systematically lower than, the non-interacting energy levels (black dashed curves), while in the repulsive case ($a<0$, red curves) they lie systematically higher.

\begin{figure*}
\includegraphics[width = 0.8\textwidth]{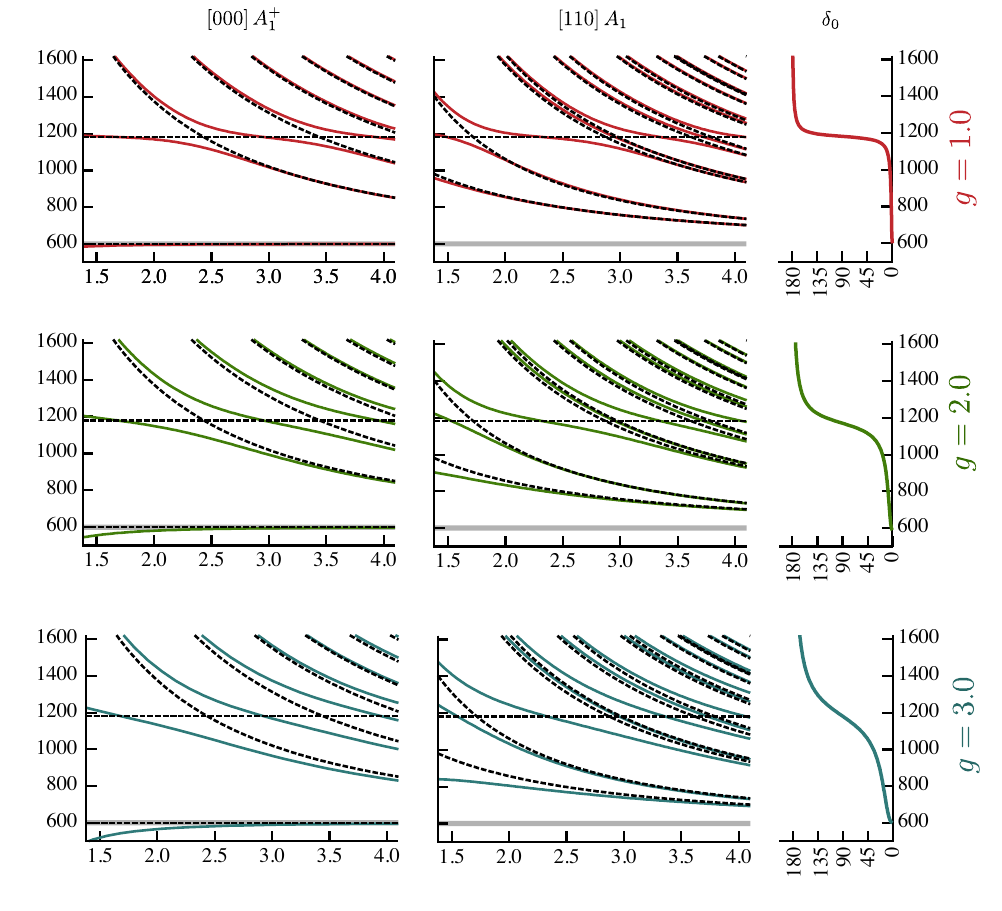}
\caption{Finite volume spectrum in two irreps for a Breit-Wigner resonance with three values of decay coupling. Plotted is cm-frame energy, $E^\star$ in MeV, against $L$ in fm. Scattering particles have mass $300$ MeV and Breit-Wigner mass is $m=1182$~MeV. Dashed black curves show non-interacting energy levels, and the gray band at 600 MeV indicates the kinematic threshold. Rightmost panel shows the elastic phase-shift in degrees.
}\label{fig:fv_elastic_res} 
\end{figure*}

A more interesting case is presented in Figure~\ref{fig:fv_elastic_res}, where the elastic amplitude is a relativistic $S$-wave Breit-Wigner,
\begin{equation}
\tan \delta_0(E^\star) = \frac{E^\star \, \Gamma(E^\star)}{m^2 - E^{\star 2}} 
\;\;\; \mathrm{with} \;\;\; \Gamma(E^\star) = \frac{g^2}{6\pi} \frac{m^2}{E^{\star 2} } q^\star
\end{equation}
with $m$ chosen to be 1182 MeV and three increasing values of coupling, $g$. For the narrowest resonance ($g=1.0$, upper panel), it is clear that at every value of $L$ there is an `extra' level (beyond those expected in the non-interacting case) in the vicinity of $1182$ MeV, and in those locations where a non-interacting curve (black dashed curves) crosses this energy, there is an `avoided level crossing'. Levels at energies far from $1182$ MeV are observed to lie very close to the non-interacting curves, as we might expect given that at those energies, $\delta_0 \approx 0^\circ$ or $180^\circ$, either of which correspond to no scattering. As the coupling is increased (lower panels), the avoided level crossings become broader and the effect of the resonance is effectively spread over a larger energy region, where it is no long possible identify any single level as `belonging' to the resonance.

\begin{figure*}
\includegraphics[width = 0.8\textwidth]{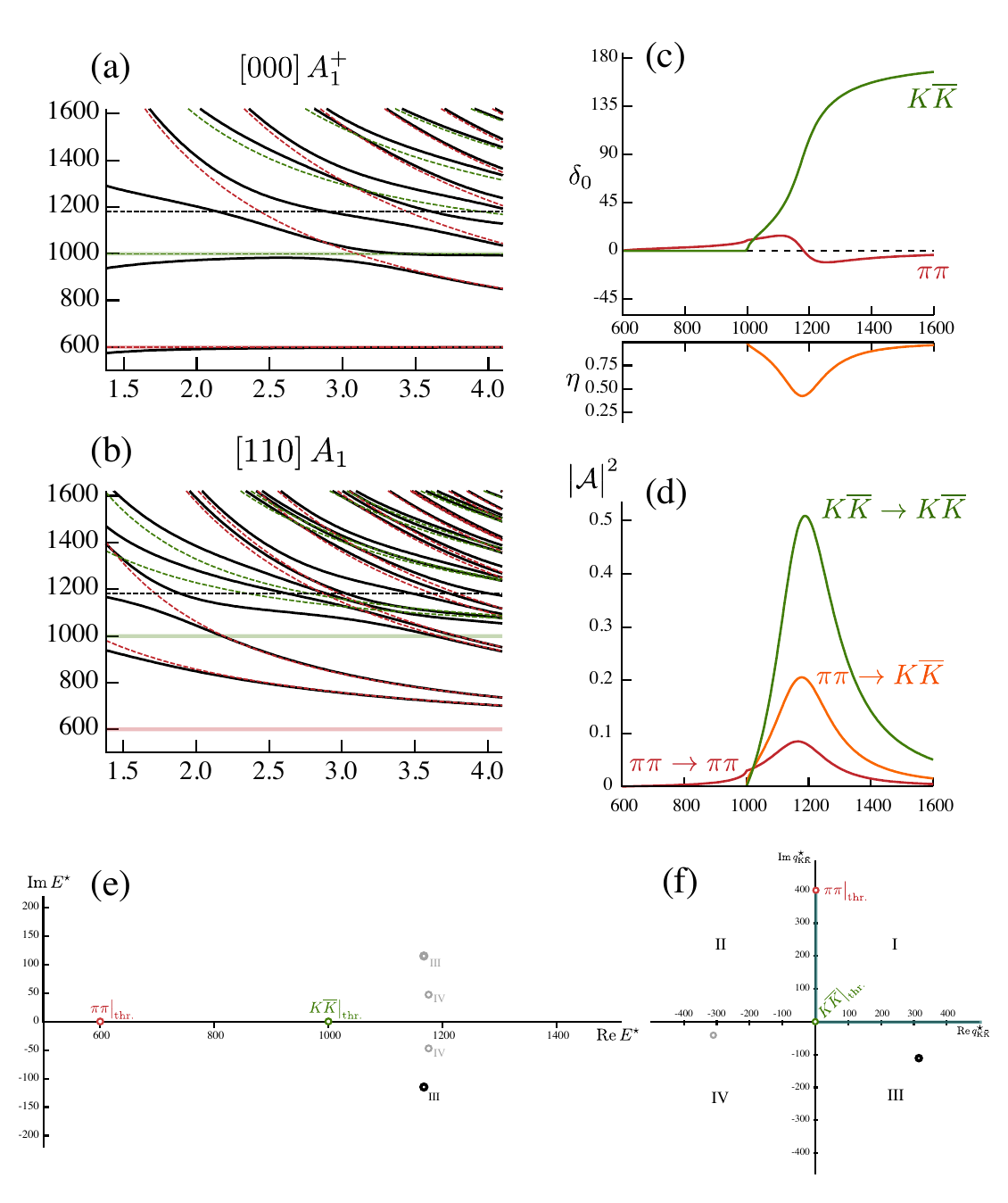}
\caption{Narrow resonance in two channel scattering modeled by a Flatt\'e amplitude. Channels are labelled ``$\pi\pi$'' and ``$K\overline{K}$'' with $m_\pi = 300$~MeV and $m_K = 500$~MeV.\\
(a) Finite-volume spectrum in rest-frame: interacting theory shown by black curves, non-interacting $\pi\pi$ and $K\overline{K}$ energies shown by red, green dashed curves respectively, kinematic thresholds ($2m_\pi$, $2m_K$) shown by the horizontal red, green bands. Horizontal dashed black line indicates the Flatt\'e mass, $m = 1182$ MeV.\\
(b) Same as (a) for the frame moving with momentum $[110]$ -- energy is the corresponding cm-frame energy, $E^\star$.\\
(c) Phase-shifts and inelasticity, defined in two-channel scattering via 
${ \mathcal{M}_{\pi\pi; \pi\pi} =  \frac{4\pi E^\star}{i q^\star_{\pi\pi}}  \big( \eta \, e^{2i \delta_{\pi\pi}} - 1 \big) }$,
 ${ \mathcal{M}_{\pi\pi; K\overline{K}} =  \frac{4\pi E^\star}{ \sqrt{q^\star_{\pi\pi} q^\star_{K\overline{K}} }} \,  \sqrt{1-\eta^2} \, e^{i ( \delta_{\pi\pi} + \delta_{K\overline{K}}  )}  }$,
 ${ \mathcal{M}_{K\overline{K}; K\overline{K}} =  \frac{4\pi E^\star}{i q^\star_{K\overline{K}}}   \big( \eta \, e^{2i \delta_{K\overline{K}}} - 1 \big)}$. \\
(d) Square of the amplitude modulus, $\big| \mathcal{A} \big| = \frac{1}{16\pi} \frac{2 \, q^\star}{E^\star} \big| \mathcal{M}_{\ell = 0} \big|$.\\
(e) Poles of the amplitude plotted in the complex energy plane -- above the $K\overline{K}$ threshold, the lower half plane of sheet III is closest to physical scattering, and the pole highlighted in black is the cause of the prominent bump in the amplitudes near 1182 MeV.\\
(f) Poles of the amplitude plotted in the complex $q^\star_{K\overline{K}}$ plane -- physical scattering runs down the positive imaginary axis below $K\overline{K}$ threshold, then along the positive real axis above. That the IV sheet pole is always rather distant from physical scattering is apparent.}\label{fig:fv_flatte} 
\end{figure*}

Figure~\ref{fig:fv_flatte} shows an example of an $S$-wave \emph{two-coupled-channel} process constructed to feature a simple narrow resonance coupled to both channels, which we label $\pi\pi$ and $K\overline{K}$, where these channels are otherwise uncoupled. A simple Flatt\'e form achieves this
\footnote{ in $S$-wave, $\mathcal{M}_{ij}(E^{\star}) = 16 \pi \frac{g_i g_j}{m^2 - E^{\star 2} - i \sum_k g_k^2 (2 q^\star_k / E^\star) }$ } 
, and gives us our canonical view of a multichannel narrow resonance: it appears as a bump in the channels to which it couples, and the height of the bump is simply related to the coupling to that decay channel. It has a nearby pole on the nearest unphysical sheet (since we are above both threshold in two-channel scattering, this is sheet III, $\mathrm{Im} \, q^\star_{\pi\pi} <0$, $\mathrm{Im}  \, q^\star_{K\overline{K}} <0$) that is partnered by an `image' pole~\cite{Au:1986vs} on a more distant Riemann sheet (in this case, sheet IV, $\mathrm{Im} \, q^\star_{\pi\pi} >0$, $\mathrm{Im}  \, q^\star_{K\overline{K}} <0$)
.

Focussing on the corresponding finite-volume spectrum we notice that, as was the case in elastic scattering, there is a typically an `extra' state in the vicinity of the resonance mass. The characteristic `avoided level crossing' behavior is again present. It is clear that computing in moving frames (such as $\mathbf{P} = [110]$ shown in the figure) provides a high density of levels which can be used to constrain the scattering matrix.

\begin{figure*}
\includegraphics[width = 0.8\textwidth]{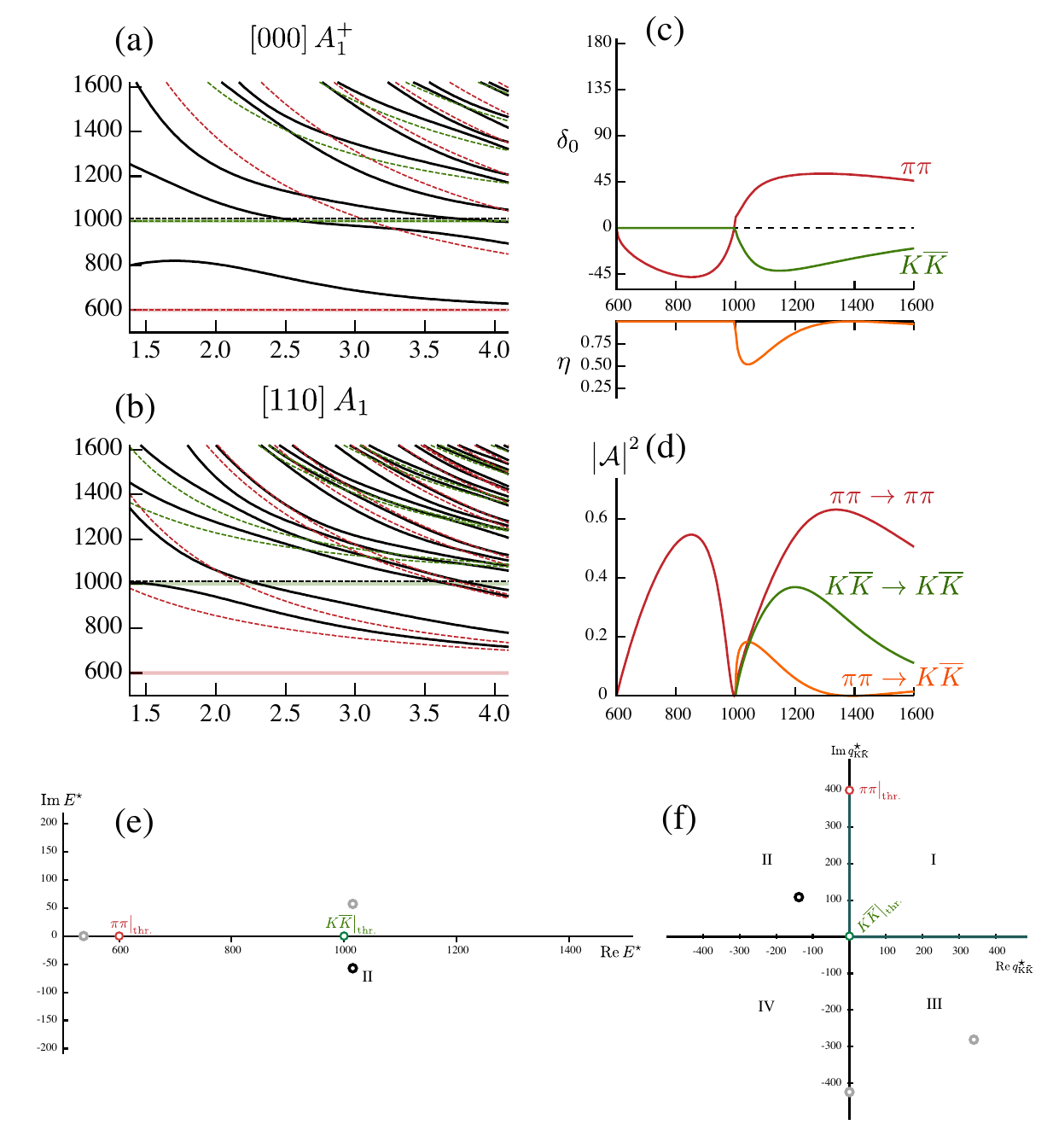}
\caption{As Figure~\ref{fig:fv_flatte} for an amplitude featuring a nearby sheet II pole. 
}\label{fig:f0980} 
\end{figure*}

We end this section by pointing out that the physics of multichannel resonances in infinite and finite volumes is not always as simple as was implied by the previous example. Figure~\ref{fig:f0980} shows a two-channel amplitude which features a resonant pole causing a rapid energy variation at the opening of the $K\overline{K}$ threshold
\footnote{The scattering matrix here is constructed from a $K$-matrix parameterized via
\begin{equation}
\mathcal{K}^{-1} = \left( \begin{array}{cc}
a & b + c (E^\star)^2 \\
b + c (E^\star)^2 		& d + e (E^\star)^2
\end{array} \right),
\end{equation}
where $a,b,c,$ and $d$ are constants, and we have replaced the phase space factor in Eq.(\ref{eq:kmat}) with the so-called  Chew-Mandelstam phase space, first introduced in~\cite{Chew:1960iv} and described in detail in~\cite{Wilson:2014cna}.}. 
It is clear that this resonance does not manifest itself as a bump, but rather as a \emph{dip} in the magnitude of the $\pi\pi \to \pi\pi$ amplitude, and as a rapid turn-on of the $\pi\pi \to K\overline{K}$ and $K\overline{K} \to K\overline{K}$ amplitudes at threshold. There is a pole on sheet II whose presence is being felt at the $K\overline{K}$ threshold -- a very distant pole on sheet III has little impact on the behavior at $K\overline{K}$ threshold. We will not discuss here what physics might cause such a pole distribution, only point out that this is a possible scattering amplitude, one which may be somewhat reminiscent of the experimental $f_0(980)$ resonance.
 
Now, examining the finite-volume spectra corresponding to this amplitude, we see that, while there can be large departures from the non-interacting spectrum in the region around the resonance position (which is very close to the $K\overline{K}$ threshold), there is no obvious `extra' level present. This example illustrates the case that in multichannel scattering, a presence of an additional level might suggest the existence of a resonance in the theory, but the converse, that the absence of an `extra' level implies the absence of a resonance, is certainly not true.  

\section{Determining the finite-volume spectrum  \label{Sec:spectrum}}
It should be clear from the results of Section~\ref{Sec:finitevolume} that one can learn about scattering amplitudes if one has determined the discrete spectrum of states in one or more volumes, in one or more frames. In this section we present a discussion of how the reliable extraction of a discrete excited state spectrum can be achieved in practical lattice QCD calculations.

\subsection{Variational analysis of correlation matrices \label{sec:variational}}
From the form of Eq.(\ref{eqn:dispersive}),
\begin{align*}
C(t)   &\equiv \int_L \! d \mathbf x \int_L \! d \mathbf y \ e^{- i \mathbf P \cdot (\mathbf x - \mathbf y)} \Big [ \langle 0 \vert  \mathcal A(\mathbf{x}, t) \mathcal B^\dagger(\mathbf{y},0) \vert 0 \rangle \Big ]_L  \nn \\
&= L^6 \sum_n e^{ - E_n t} \, \big\la 0 \big| \mathcal{A}(0) \big| E_n, \mathbf{P}; L\big\ra  \,  \big\la E_n, \mathbf{P}; L \big| \mathcal{B}^\dag(0) \big| 0 \big\ra, \nn
\end{align*}
it would seem that by decomposing the time-dependence of a single correlation function in terms of a sum of exponentials one could get access to the entire spectrum of states. Unfortunately this approach is not practical for the determination of anything beyond the energy of the lightest state with the quantum numbers of $\mathcal{A}$, $\mathcal{B}^\dag$. A technique that has proven extremely effective is to compute a \emph{matrix} of correlation functions using a basis of operators,
\begin{equation*}
C_{ij}(t) \equiv \int_L \! d \mathbf x \int_L \! d \mathbf y \ e^{- i \mathbf P \cdot (\mathbf x - \mathbf y)} \Big [ \langle 0 \vert\mathcal{A}^{}_i(\mathbf{x}, t) \mathcal{A}_j^\dagger(\mathbf{y}, 0) \vert 0 \rangle \Big ]_L,
\end{equation*}
which can be analyzed \emph{variationally}~\cite{Michael:1985ne, Luscher:1990ck,Blossier:2009kd}.
The set of $N$ operators $\{ \mathcal{A}_i \}_{i=1\ldots N}$ all have the same quantum numbers but will ideally have differing magnitudes of overlap $\la 0 | \mathcal{A}_i | E_n \ra$, onto each state in the spectrum. It follows that there will be a certain linear superposition of operators, $v_i^{(n)} \mathcal{A}_i$, that will optimally (in the variational sense) interpolate each state $n$ in the spectrum, and these superposition weights can be shown to be the eigenvectors which solve a generalized eigenvalue problem,
\begin{equation}
 C_{ij}(t)\,  v^{(n)}_j(t,t_0) = \lambda_n(t,t_0) \,  C_{ij}(t_0)\,  v^{(n)}_j(t,t_0). 
 \label{eq:gen_eval_prob}
\end{equation}
In this expression, $t_0$ is a suitably chosen reference time, which ideally is selected to be such that only $N$ states are required to saturate $C_{ij}(t_0)$ to a good approximation~\cite{Dudek:2007wv}. The eigenvalues feature the relevant state energies, ${\lambda_n(t,t_0) \sim e^{-E_n(t-t_0)} }$.
Fitting the time dependence of each of these quantities, also known as \emph{principal correlators}, therefore enables a determination of the discrete spectrum. There are various approaches to handling the time-dependence of the solution of Eq.(\ref{eq:gen_eval_prob}), see e.g.~\cite{Dudek:2007wv, Bulava:2009jb, Mahbub:2013ala, Kiratidis:2015vpa, Kiratidis:2016hda}.
%

\subsection{Operator construction \label{sec:operators}}
\begin{figure*}
\includegraphics[width = 0.8\textwidth]{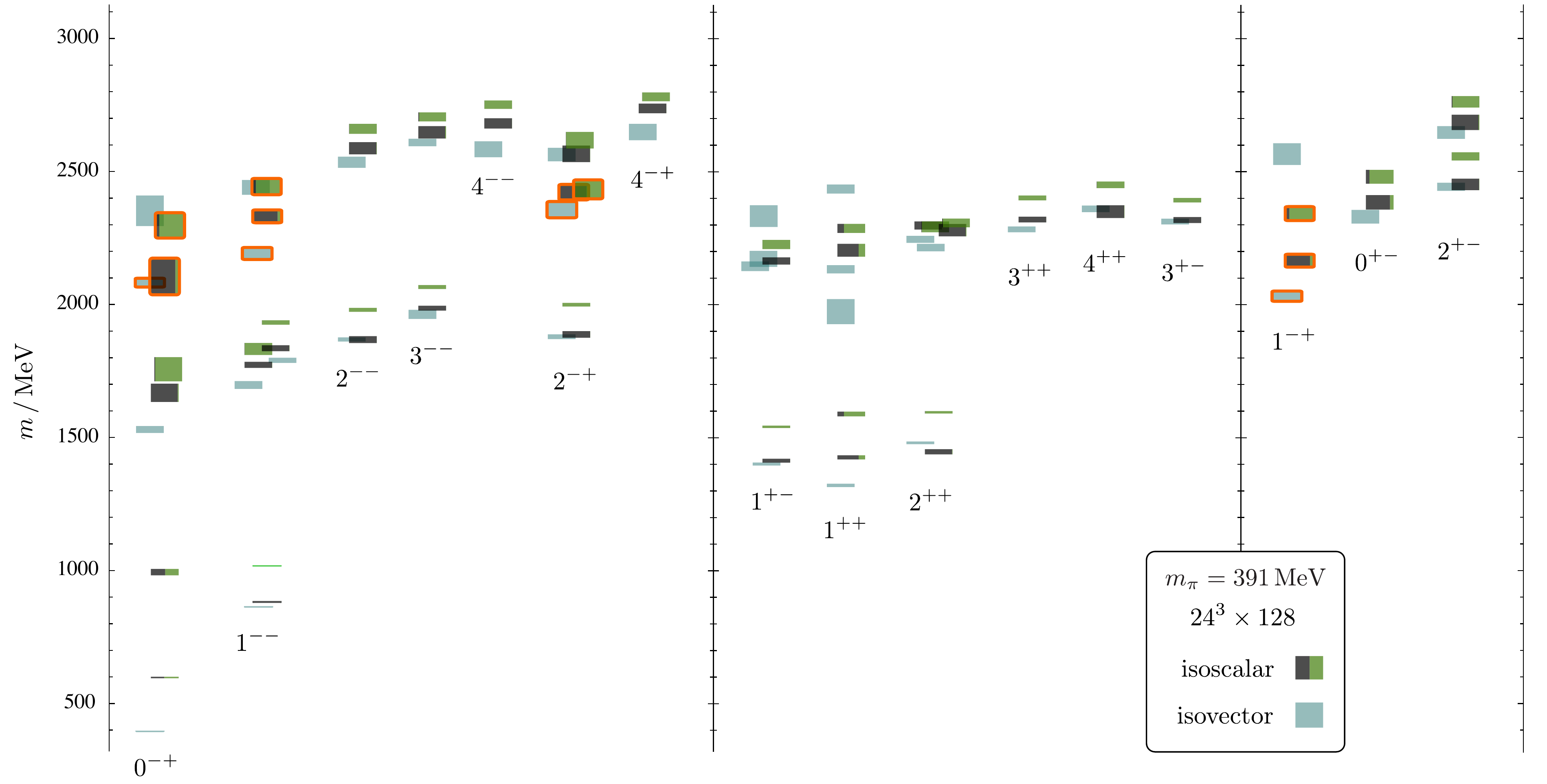}
\caption{Spectrum of isoscalar(black/green boxes) and isovector(blue boxes) mesons extracted from variational analysis of large matrices of correlation functions in~\cite{Dudek:2013yja}. Degree of black/green indicates the hidden light ($u\bar{u}+d\bar{d}$) versus hidden strange ($s\bar{s}$) content of each state determined from relative sizes of operator overlap. Orange outlines indicate lightest set of states having significant overlap with operators featuring a chromomagnetic gluonic construction, suggested to be a signal for them being \emph{hybrid mesons}.}\label{fig:isoscalar} 
\end{figure*}

There still remains the question of what form the operators, which must have the quantum numbers of the hadronic system of interest, but which need to be constructed from the basic quark and gluon fields of QCD, should take. A longstanding approach for the case of systems with meson quantum numbers is to make use of fermion bilinear operators, $\overline{\psi} \boldsymbol{\Gamma} \psi$, where the quark fields may be smeared over space, and where the $\boldsymbol{\Gamma}$ object controls the spin-structure and may, if desired, have spatial dependence. A recent application of this approach~\cite{Dudek:2009qf,Dudek:2010wm,Dudek:2011tt,Thomas:2011rh,Dudek:2013yja} considers a large basis of operators constructed out of the following basic form,
\begin{equation}
\overline{\psi} \boldsymbol{\Gamma} \psi = \overline{\psi} \Gamma D_i \ldots D_k \psi,
\end{equation}
where the $D_i$ are gauge-covariant derivatives in the spatial directions. Using up to three derivatives, a very large basis of operators can be constructed, and with these operators, large matrices of correlation functions computed. When these are analyzed variationally, spectra like those shown in Figure~\ref{fig:isoscalar} can be obtained~\cite{Dudek:2013yja}. The set of states observed closely matches the expectations of a model where mesons are quark-antiquark constructions, with the addition of some states that also have gluonic content~\cite{Dudek:2011bn}. But this spectrum is surely incomplete --- in particular when the volume is changed, the extracted spectrum varies relatively little, in contrast to the expectations of the previous section (where for example at least some levels approximately track the volume dependence of non-interacting levels). Subsequent calculations using an augmented operator basis have shown that this spectrum is indeed incomplete, and it is to the form of the required additional operators that we now turn.

\subsubsection{The importance of ``multi-hadron'' operators \label{sec:meson-meson}}

In order to resolve the complete low-lying spectrum of states through variational analysis of correlation matrices, it proves to be necessary to include in the operator basis some operators which `resemble' the expected multi-hadron states in that energy region. This was illustrated in~\cite{Wilson:2015dqa} for the case of the $\rho$ resonance appearing in $\pi\pi$ scattering. The operator basis used (in the $\mathbf{P} = \mathbf{0}$ frame) included a large number of operators of the `single-meson' $\overline{\psi} \bf{\Gamma} \psi$ type described in the previous section, \emph{and in addition} several operators of form $\sum_{\hat{\mathbf{k}}} c_{\hat{\mathbf{k}}}\, \pi(\mathbf{k}) \, \pi(-\mathbf{k})$, where $\pi(\mathbf{k})$ is a shorthand notation for a superposition of $\overline{\psi} \bf{\Gamma} \psi$ operators which optimally
\footnote{Optimal operator determined through variational analysis of a matrix of correlators for the irrep in which the moving pion sits.}. 
interpolates the lightest pseudoscalar with momentum $\mathbf{k}$
The product of two ``pion'' operators, in a fictitious world where pions do not interact, would closely resemble an eigenstate of the system with energy $2\sqrt{m_\pi^2 + \mathbf{k}^2}$. These `meson-meson' operators are qualitatively different from the `single-meson' operators --- each meson operator is projected into a definite momentum and this means that they effectively sample the entire lattice volume.

\begin{figure*}
\includegraphics[width = .7\textwidth]{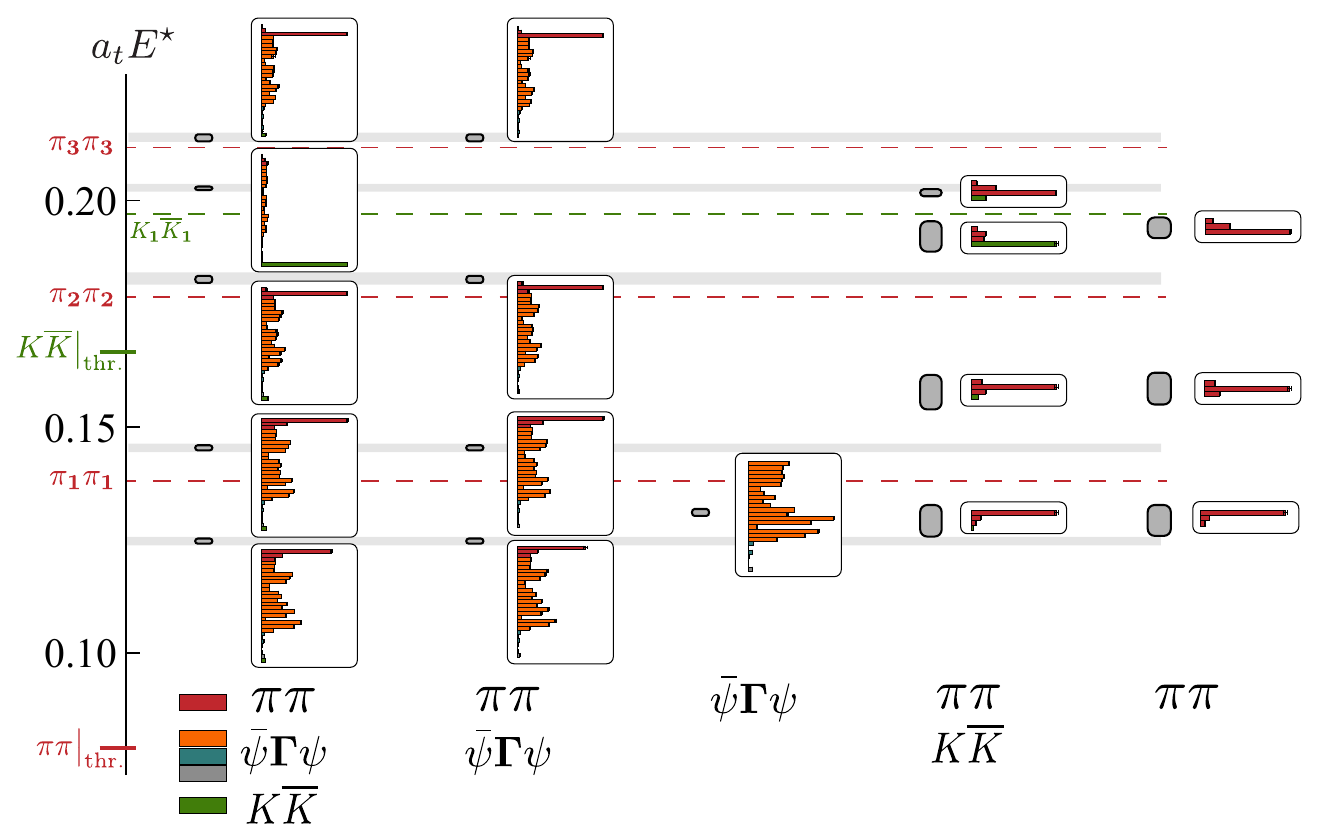}
\caption{Low-lying finite-volume energy spectrum (gray boxes) extracted from variational analysis of rest-frame correlations functions with varying operator basis. Histograms show, for each state, the relative size of overlap onto each operator in the basis. First column: large operator basis including many `single-hadron', $\bar{\psi} \mathbf{\Gamma} \psi$, operators, several `$\pi\pi$-like' operators, and one `$K\overline{K}$-like' operator. Second column: as before, excluding the $K\overline{K}$-like operator. Third column: using only `single-hadron' operators. Fourth, fifth columns: excluding the `single-hadron' operators. Dashed lines show the non-interacting $\pi\pi$ and $K\overline{K}$ energies on this $32^3$ lattice. Energy expressed in units of the temporal lattice spacing, $1/a_t \sim 6.0$~GeV. Figure adapted from one appearing in~\cite{Wilson:2015dqa}.
\label{fig:op_basis_comp} }
\end{figure*}

Figure~\ref{fig:op_basis_comp} shows the result of variational analysis of a large matrix of correlation functions performed on an $L^3 \sim (4\,\mathrm{fm})^3$ lattice. In the lattice units presented, the pion mass is $a_t m_\pi = 0.0393$ and the kaon mass is $a_t m_K = 0.0834$. The first column shows the spectrum extracted using 30 operators including several `$\pi\pi$-like' and `$K\overline{K}$-like' operators. The histograms show the relative size of overlaps, $\big\la 0 \big| \mathcal{A}(0) \big| E_n, \mathbf{P}; L\big\ra$, of each state onto each operator in the basis, and it is clear that the lowest two levels have strong overlap with both `single-meson' operators subduced from $J=1$ (orange) and $\pi\pi$ operators (red). The third column indicates the spectrum which is extracted if meson-meson operators are not included in the basis, and it is immediately clear that the spectrum is completely different, in particular in place of two low-lying states in the complete spectrum, only one state appears. It appears that without the meson-meson operators the variational system cannot find two orthogonal combinations to overlap with the two states. 

A rather simple explanation for this was presented in~\cite{Dudek:2012xn}. Imagine one could turn off the $\rho\to\pi\pi$ coupling and all $\pi\pi$ interactions, but retain the $\rho$ as a stable state in the spectrum --- in this limit, the finite-volume eigenstates will be a single $|\rho,L\rangle_0$ state of definite mass, and free $|\pi(\mathbf{k})\pi(\mathbf{k}),L\rangle_0$ states at energies $2\sqrt{m_\pi^2 + \mathbf{k}^2}$. These states would be interpolated by `single-meson' operators and `meson-meson' operators respectively. With a small but not zero coupling $\rho \to \pi\pi$, these basis states will admix to form finite-volume eigenstates, and in the case that the $\rho$ lies near just one $\pi\pi$ state we can treat this as a two-state system,
\begin{eqnarray*}
|E_1,L\rangle&=&\cos\theta\,|\rho,L\rangle_0+\sin\theta\,|\pi\pi,L\rangle_0, \\
|E_2,L\rangle&=&-\sin\theta\,|\rho,L\rangle_0+\cos\theta\,|\pi\pi,L\rangle_0,
\end{eqnarray*}
and in the expected case that a `single-meson' operator $\mathcal{O}$ overlaps significantly only with $|\rho,L\rangle_0$, the corresponding correlator would behave like
\begin{align}
  \sum_n e^{- E_{n}t}   \,&
\langle 0 \vert  \mathcal{O}(0)   \vert E_n, L \rangle \,  \langle E_n, L \vert \mathcal{O}^\dag(0) \vert 0 \rangle 
\nn \\
& \propto      \cos^2\theta \, e^{- E_{1}t} +    \sin^2\theta \, e^{- E_{2}t}
\,.
\label{eq:two-point_rho}
\end{align}
We note that as soon as any dynamical coupling $\rho\to\pi\pi$ is turned on, the mixing angle $\theta$ will be particularly sensitive to $L$ as the $\pi\pi$ (non-interacting) energy passes the $\rho$ ``mass''. If the free particles' energy is in the near vicinity (loosely, within the width) of a resonance, both $\cos^2\theta$ and $\sin^2\theta$ will be appreciable. It is therefore physically not possible for a local $\rho$-like operator to have significant preference to couple to either $E_1$ or $E_2$.
If we work with an operator basis that only has appreciable couplings to the compact $\rho$-like state it would prove impractical with finite statistics and a limited time-extent to resolve the presence of two exponentials.
As a consequence we would end up concluding the presence of a single state whose mass is somewhere between $E_1$ and $E_2$.

It can be justified that the `single-meson' operator has suppressed overlap onto the $\pi\pi$ state by appealing to the difference in spatial distributions of the operator and the state. The operator is a local object, which samples only a small region of space, while the two-meson state occupies the entire lattice volume, and hence the overlap is suppressed by the volume. We emphasize here that operators interpreted as being of  `single-hadron' type need not be just fermion bilinears (or three quark constructions in the case of baryons), but may contain any number of fermion fields (for example, \emph{tetraquark} meson constructions, $\sim \overline{\psi} \overline{\psi} \psi \psi$). What distinguishes them from multi-hadron operators is how they sample the spatial volume, in particular that they are local to a restricted region of space rather that the full spatial volume.

It is clear that we need to include `meson-meson' operators into our basis, but how many do we need? As an example, do we need to include $K\overline{K}$-like operators when studying $\pi\pi$ scattering, since these two channels can, in principle, couple. This is illustrated in the second column of Figure~\ref{fig:op_basis_comp} where we observe that excluding $K\overline{K}$-like operators does not change the spectrum below $K\overline{K}$-threshold. On the other hand, above $K\overline{K}$-threshold, where we expect there to be states which contain an admixture of $K\overline{K}$, we observe that one state disappears when the relevant operators are not included.

We conclude this discussion by providing a `rule of thumb' for the required operator basis: {\it in addition to `single hadron' operators capable of overlapping onto any resonances you might anticipate, be sure to include multihadron operators for all expected non-interacting levels in the energy region you wish to study  }.

In light of this, how should we view the spectra presented in Fig.~\ref{fig:isoscalar}, which were extracted without including any multihadron operators, but using a large basis of `single-hadron' operators? We might guess that we are overlapping only onto the `single-hadron-like' parts (pieces analogous to $|\rho, L\rangle_0$, but for other resonances) and hence the spectrum is indicating the \emph{presence} of a resonance (which is probably narrow) in some energy region, without precisely determining its properties. We will return to this question in Section~\ref{sec:narrow}, where we derive a relevant result for how narrow resonances might manifest themselves in finite-volume spectra.

\section{Examples of resonance determination \label{Sec:examples}}

Recent years have seen an increasing number of lattice QCD calculations making use of the L\"uscher approach in order to determine hadron resonance properties. In this section we will summarize this progress. We begin by reminding the reader that the procedure laid out in the previous two sections is not only useful in the extraction of resonances, scattering amplitudes in which no resonance appears can also be determined. $\pi\pi$ scattering with isospin=2 is the classic example, and this case has been studied in a number of lattice calculations
~\cite{Sharpe:1992pp,
Gupta:1993rn,
Kuramashi:1993ka,
Fukugita:1994ve,
Aoki:2002in,
Du:2004ib,
Yamazaki:2004qb,
Beane:2005rj,
Beane:2007xs,
Li:2007ey,
Feng:2009ij,
Yagi:2011jn,
Sasaki:2013vxa,
Fu:2013ffa,
Helmes:2015gla,
Dudek:2010ew,
Beane:2011sc,
Dudek:2012gj,
Bulava:2016mks}. 
Several recent calculations~\cite{
Dudek:2010ew,
Beane:2011sc,
Dudek:2012gj,
Bulava:2016mks} determine multiple energy levels and use these to determine the energy-dependence of the elastic phase-shift.

\begin{figure}
\centering
\includegraphics[width = \columnwidth]{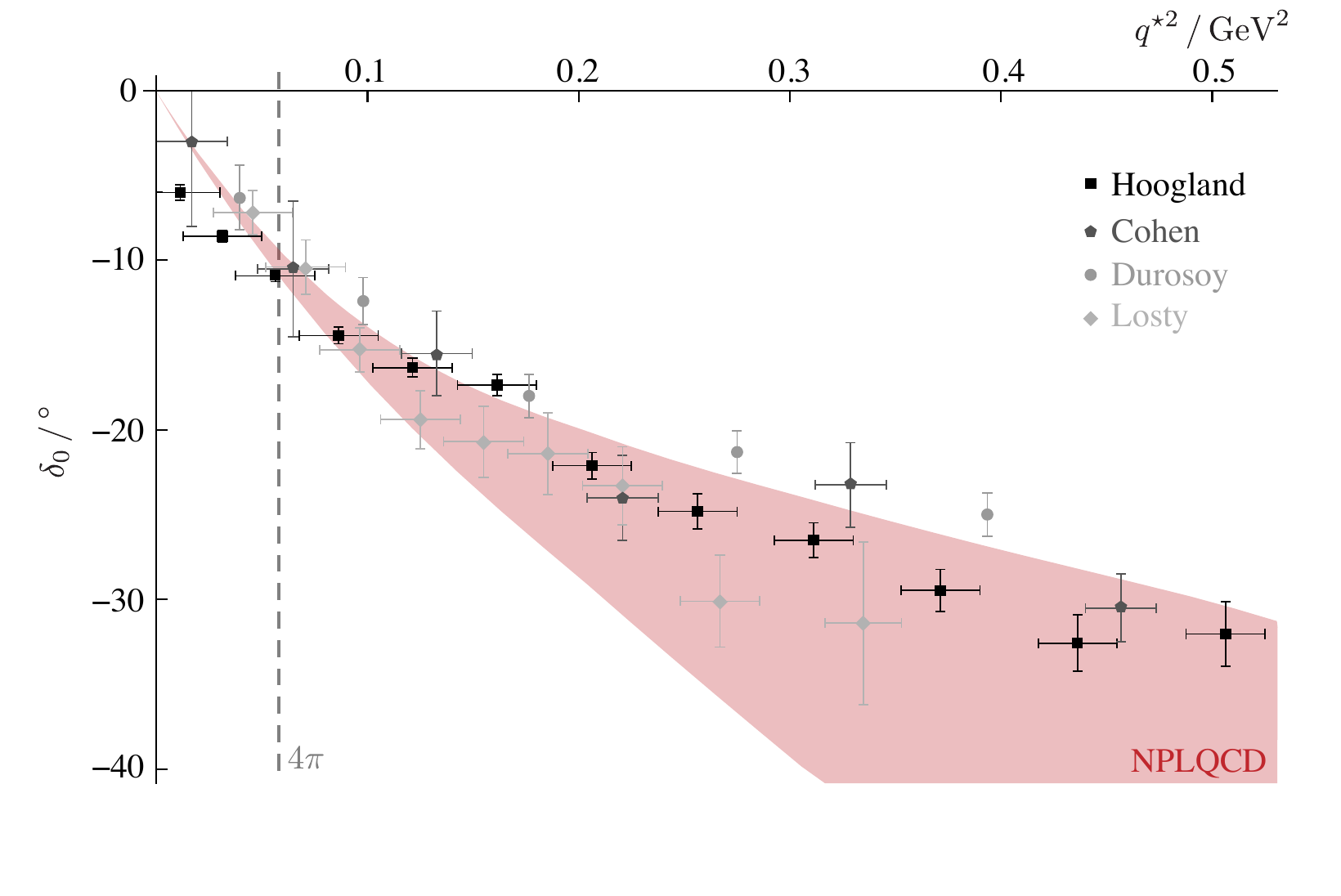}
\caption{
Elastic $I=2$ $\pi\pi$ scattering phase shift using NLO chiral perturbation theory constrained by lattice QCD finite-volume spectra computed at $m_\pi\sim 391$~MeV.
Figure adapted from~\cite{Beane:2011sc}.
}\label{fig:pipiI2NPLQCD}
\end{figure}

One application of these calculations has been to study the pion mass dependence of the scattering length~\cite{Yamazaki:2004qb, 
Beane:2005rj,
Beane:2007xs, Feng:2009ij,
Beane:2011sc, Yagi:2011jn, Sasaki:2013vxa,
Fu:2013ffa, Helmes:2015gla}, which may be compared to the expectations of chiral perturbation theory ~\cite{Weinberg:1966kf, Colangelo:2001df}, 
and these studies find that chiral perturbation theory describes the scattering length up to surprisingly large values of the pion mass, $m_\pi\sim400$~MeV.
The energy dependence of the $S$-wave $\pi\pi$ interaction has been studied by the NPLQCD Collaboration by considering low-lying rest and boosted lattice spectra across multiple volumes. These results, at a pion mass of $m_\pi\sim 390\,{\rm MeV}$, sufficiently constrain the NLO chiral expansion such that the effective range parameters and be extrapolated to the physical pion mass. The resulting phase shift is shown in Figure~\ref{fig:pipiI2NPLQCD}, where it is seen to be in reasonable agreement with experimental data.
%

\begin{figure}
\centering
\includegraphics[width = \columnwidth]{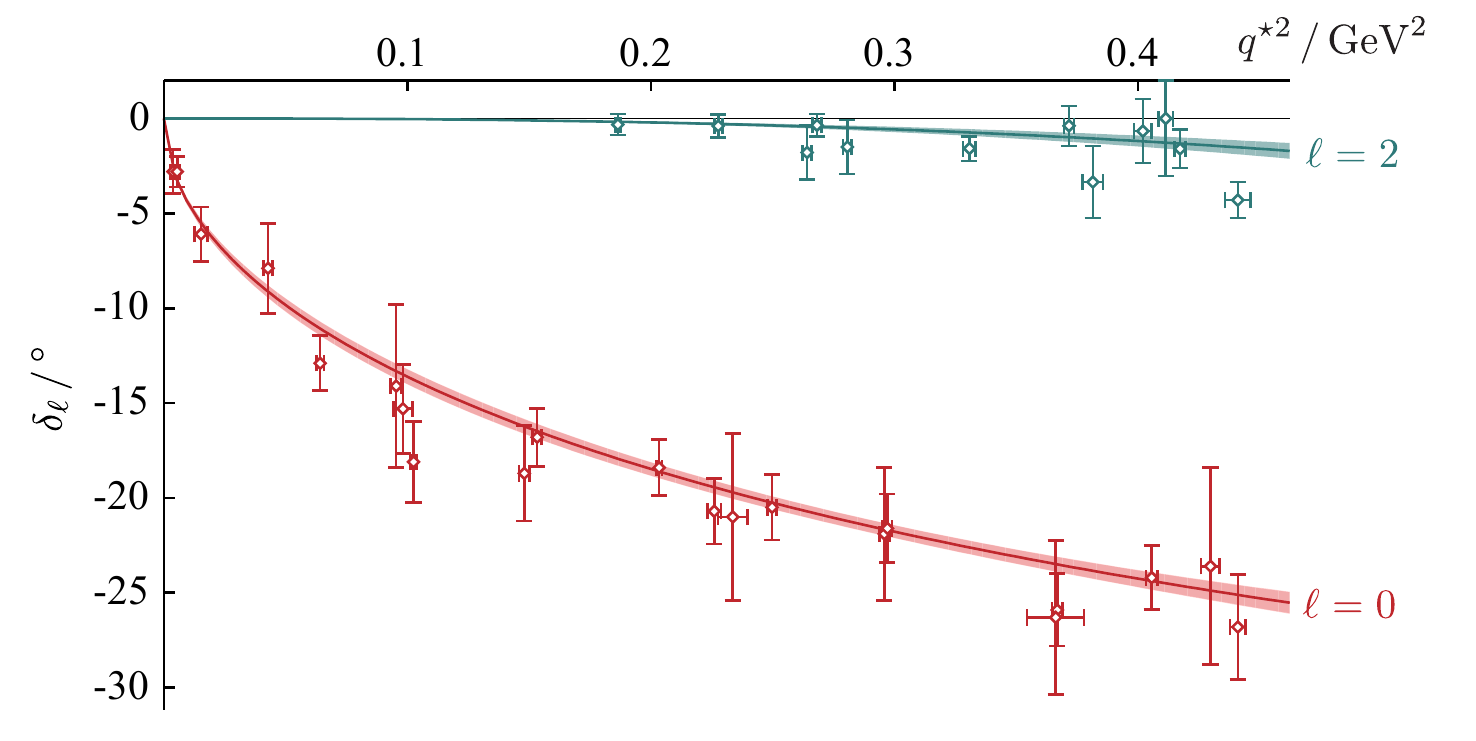}
\caption{Elastic $I=2$ $\pi\pi$ scattering phase-shifts in two partial waves determined from finite-volume spectra in three volumes with $m_\pi \sim 391$~MeV. The curves indicate scattering length descriptions of the finite-volume spectra. Figure adapted from one appearing in~\cite{Dudek:2012gj}.
}\label{fig:pipiI2}
\end{figure}

Using the same gauge field configurations as the NPLQCD study, the Hadron Spectrum Collaboration~\cite{Dudek:2012gj}
considered multiple irrep spectra in various moving frames to determine the energy dependence of the $S$ and $D$-wave elastic scattering phase-shifts up to the $4\pi$ threshold, as shown in Figure~\ref{fig:pipiI2}. The effect of the angular momentum barrier at threshold is clearly observed, with the $D$-wave being significantly reduced in magnitude with respect to the $S$-wave.

\subsection{Elastic resonances in $\pi\pi$ scattering \label{sec:elasticpipi}}

The most widely studied example of resonance extraction in lattice QCD is the $\rho$ resonance in $P$-wave isospin=1 $\pi\pi$ scattering
~\cite{
Aoki:2007rd,
Feng:2010es,
Lang:2011mn,
Aoki:2011yj,
Dudek:2012xn,
Pelissier:2012pi,
Wilson:2015dqa,
Bali:2015gji,
Bulava:2016mks}. 
At all but the smallest pion masses, the $\rho$ is an \emph{elastic} resonance, kinematically able to decay only to $\pi\pi$, and even at the physical pion mass, although the $\pi\pi\pi\pi$ channel is kinematically open, it has negligible coupling to the $\rho$~\cite{Olive:2016xmw}.

The pioneering application of the L\"uscher approach to this case was reported on in~\cite{Aoki:2007rd}. This early work considered a relatively small volume, determining two energy levels in the rest frame using a basis of operators featuring a $\pi\pi$-like construction and a $\overline{\psi} \Gamma \psi$ construction with vector quantum numbers. The two energy levels were utilized to give two points on the phase-shift curve using Eq.(\ref{eq:cotd}), which is the minimum required to determine the two free parameters appearing in a Breit-Wigner description of the scattering amplitude, while of course not providing a measure of goodness-of-fit.

Later works, in particular~\cite{Feng:2010es}, computing for several values of the pion mass, extended the approach by making use of some moving frames to obtain more energy levels and hence more points on the phase-shift curve. With one volume per pion mass,~\cite{Feng:2010es} had typically three points in the energy region corresponding to the resonance (and more outside this region), giving slightly more constraint on the resonance parameters. At about the same time,~\cite{Aoki:2011yj} also considered moving frames. 

\begin{figure}[!t]
\includegraphics[width = \columnwidth]{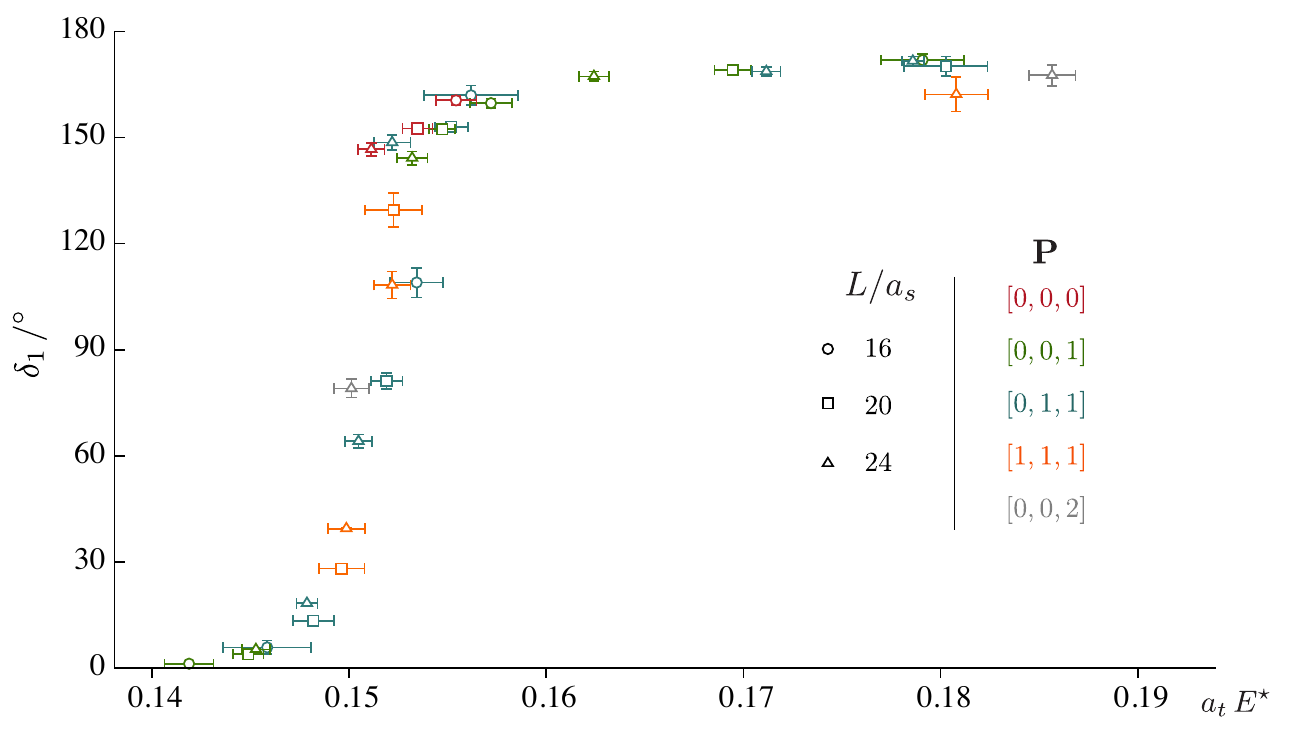}
\caption{Elastic $I=1$ $\pi\pi$ scattering phase-shifts in $P$-wave determined from finite-volume spectra in three volumes with $m_\pi \sim 391$~MeV. Figure adapted from one in~\cite{Dudek:2012xn}. Energy expressed in units of the temporal lattice spacing, $1/a_t \sim 5.7$~GeV.
}\label{fig:pipiI1840}
\end{figure}

\begin{figure*}[t]
\centering
\includegraphics[width = .6\textwidth]{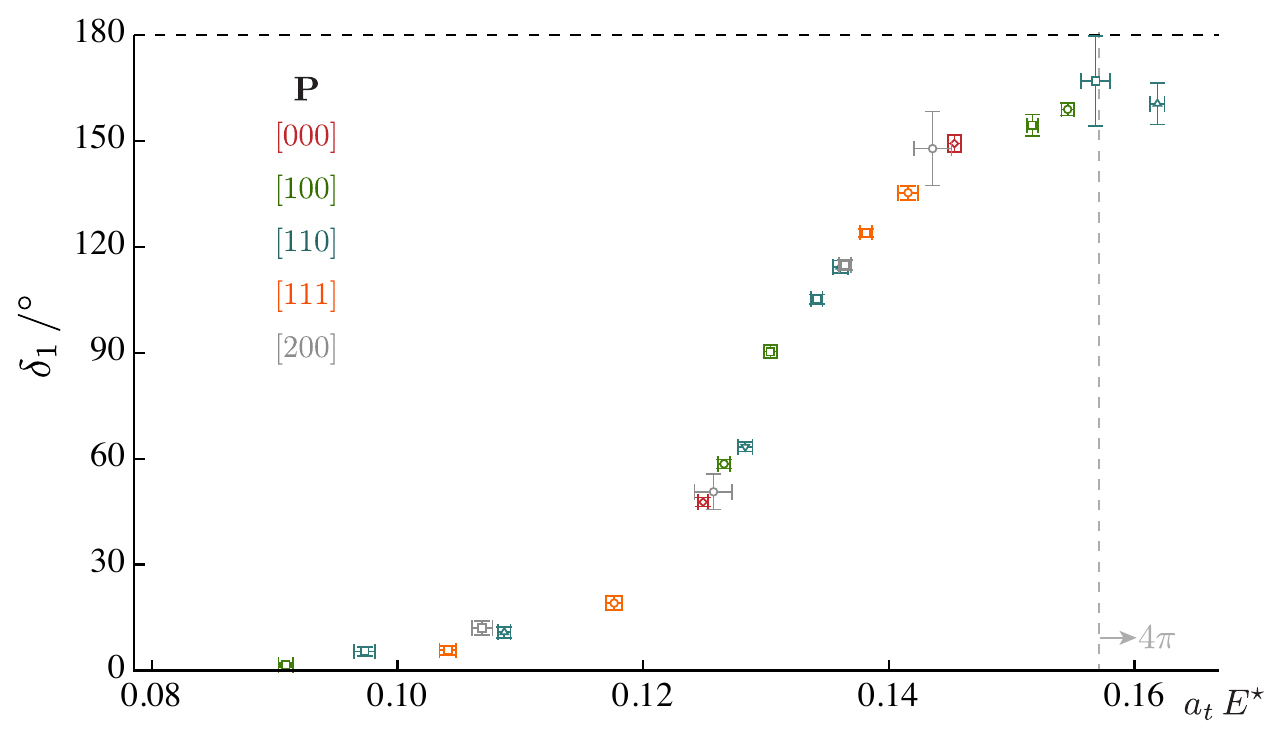}
\includegraphics[width = .35\textwidth]{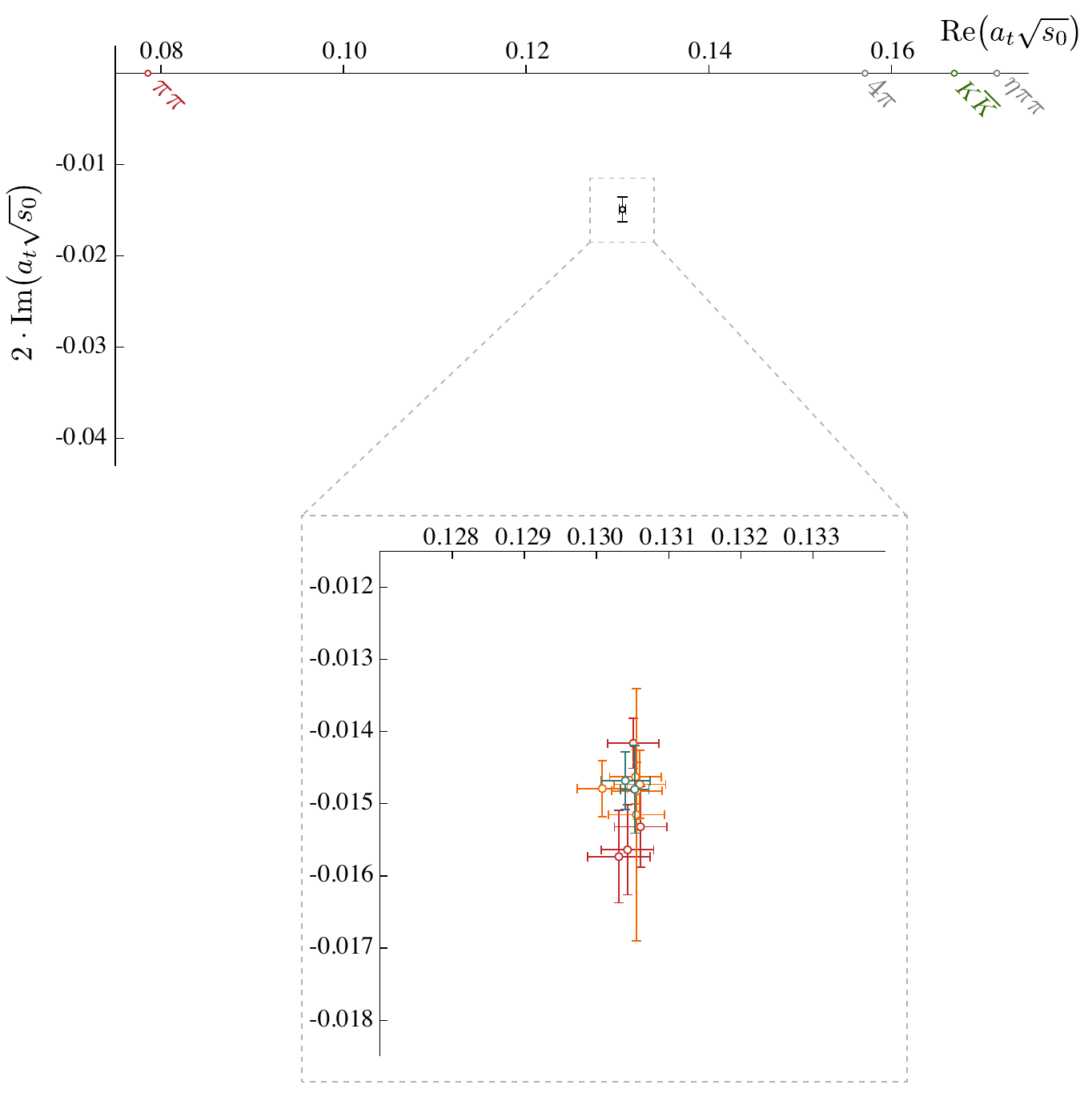}
\caption{Left: Elastic $I=1$ $\pi\pi$ scattering phase-shifts in $P$-wave determined from finite-volume spectra computed in a single $32^3$ volume with $m_\pi \sim 236$~MeV. Right: Resonance pole position for a wide range of amplitude parameterizations constrained to describe the finite-volume spectra. Energies expressed in units of the temporal lattice spacing, $1/a_t \sim 6.0$~GeV.
Figures adapted from those in~\cite{Wilson:2015dqa}. 
}\label{fig:pipiI1860}
\end{figure*}

\cite{Dudek:2012xn} and \cite{Wilson:2015dqa} perhaps best illustrate the power of extracting multiple energy levels using multiple volumes and/or several moving frames. Figure~\ref{fig:pipiI1840}, taken from~\cite{Dudek:2012xn}, shows the elastic phase-shift determined at $m_\pi \sim 391$~MeV using energy spectra in three volumes, $16^3, 20^3, 24^3$ ($L\sim 2.0, 2.4, 2.9$~fm), in all relevant irreps, in all frames up to $\mathbf{P}=[200]$. There can be no doubt from these data points that there is a resonance present with the energy dependence of the phase-shift being mapped out in detail across the entire elastic region, and the resonance parameters being tightly constrained. Figure~\ref{fig:pipiI1860}, taken from \cite{Wilson:2015dqa} illustrates that the use of multiple moving frames on a \emph{single larger volume} (which has a denser energy spectrum) can lead to the same level of detail in the mapping out of the elastic phase-shift. These calculations also computed spectra in irreps whose lowest angular momentum is not $\ell=1$, but rather $\ell = 3$, and in this way they were able to place some constraint on the size of the $F$-wave phase-shift, confirming that it is indeed negligibly small in the energy region where the $\rho$ appears, justifying the neglect of higher partial-waves in the finite-volume analysis.

Also illustrated in Figure~\ref{fig:pipiI1860} are the resonance pole positions found for a variety of amplitude parameterizations constrained to describe the finite volume spectra. It is clear from the tiny degree of scatter in the pole position, that the presence of a pole at that particular complex energy value is required, regardless of the other details of the parameterization, to describe this scattering system. 

\begin{figure}[!th]
\includegraphics[width = \columnwidth]{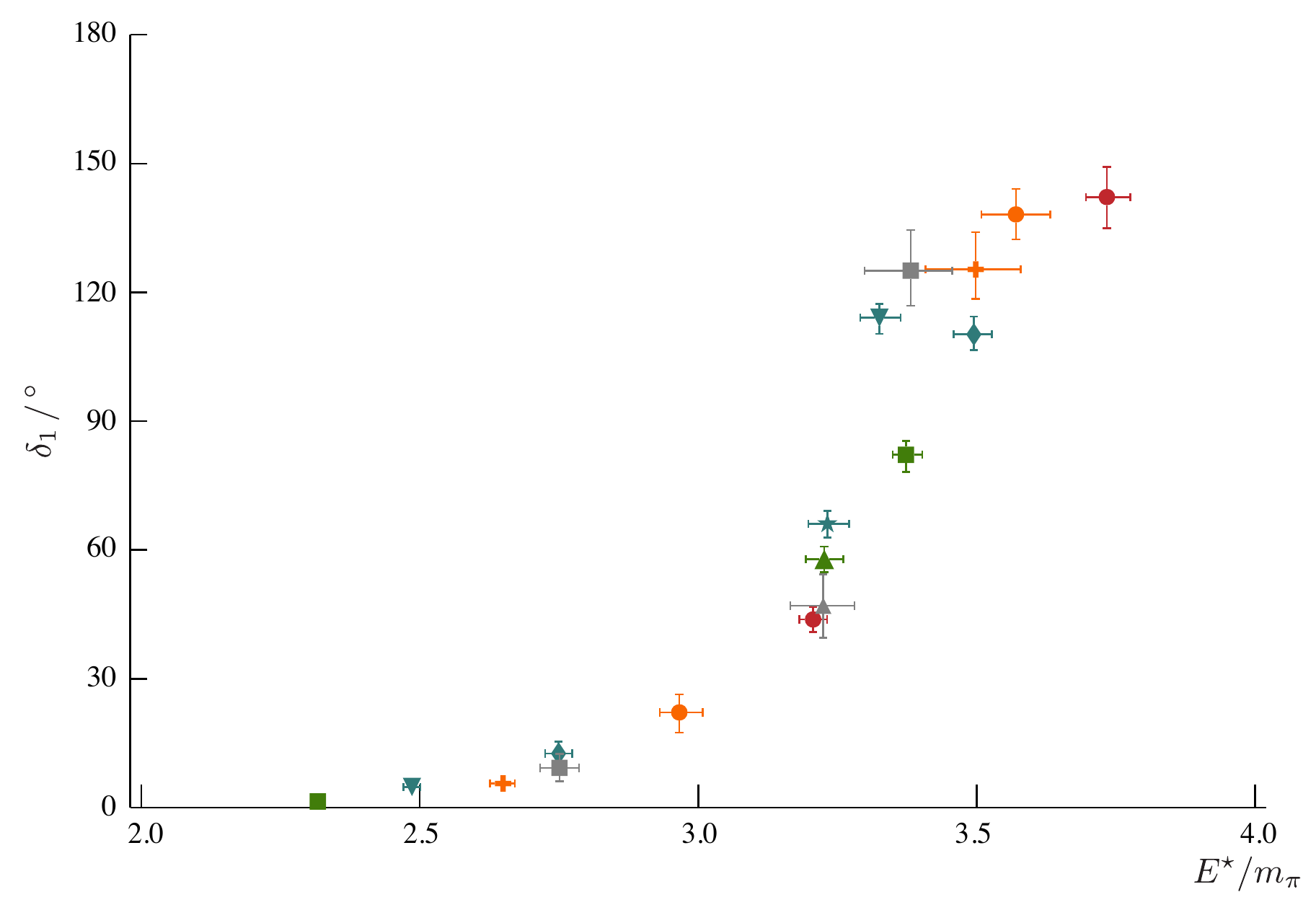}
\caption{Elastic $I=1$ $\pi\pi$ scattering phase-shifts in $P$-wave determined from finite-volume spectra computed the same $m_\pi \sim 236$~MeV configurations as used in the calculation presented in Figure~\ref{fig:pipiI1860}, but using a different correlator construction technique, operator basis and variational analysis method. Color coding as in Figure~\ref{fig:pipiI1860}. Figure adapted from one appearing in \cite{Bulava:2016mks}.}\label{fig:pipiI1Bulava}
\end{figure}

A calculation on the same lattice configurations as \cite{Wilson:2015dqa}, but using a different correlator construction technique, operator basis and variational analysis method is presented in~\cite{Bulava:2016mks}. As shown in Figure~\ref{fig:pipiI1Bulava}, the phase-shift extracted is compatible with that presented in Figure~\ref{fig:pipiI1860} within the larger statistical errors.

The $\rho$ is an isospin=1 resonance in $\pi\pi$ scattering in $P$-wave, and with this isospin, this is lowest allowed partial wave --- $\pi\pi$ scattering in $S$-wave cannot occur with isospin=1, but can with isospin=0. Experimentally the $S$-wave is very different to the $P$-wave, featuring not a narrow resonance ``bump'', associated with a rapid rise in phase-shift through $90^\circ$, but rather a gradual increase in phase-shift from $\pi\pi$ threshold up to the $K\overline{K}$ threshold, where some more rapid variation (associated with the $f_0(980)$ resonance) occurs. It has long been suspected that the slow increase in phase-shift is associated with a \emph{very broad} resonance, the $\sigma$, whose pole is located far into the complex energy plane. Recently, a highly constrained analysis of experimental $\pi\pi$ scattering data has confirmed that such a pole is present and determined its position with some precision (see~\cite{Pelaez:2015qba} for a review of the situation).

Very recently we have seen the first serious lattice QCD determination of elastic $\pi\pi$ scattering in the isospin=0 \mbox{$S$-wave}~\cite{Briceno:2016mjc}. Lattice QCD calculation of this channel had long been considered extremely challenging owing to the need to compute diagrams in which all the quarks and antiquarks annihilate, leading to something which is \emph{completely disconnected}. By computing a large number of propagation objects in the \emph{distillation} framework~\cite{Peardon:2009gh}, the Hadron Spectrum Collaboration were able to compute the required correlation functions and obtain finite-volume spectra at two pion masses, $m_\pi \sim 236, 391$~MeV. The lattices are the same ones used in the $\rho$ extractions described above, with three volumes at the heavier mass and a single larger volume at the lighter mass.

\begin{figure}[!th]
\includegraphics[width = \columnwidth]{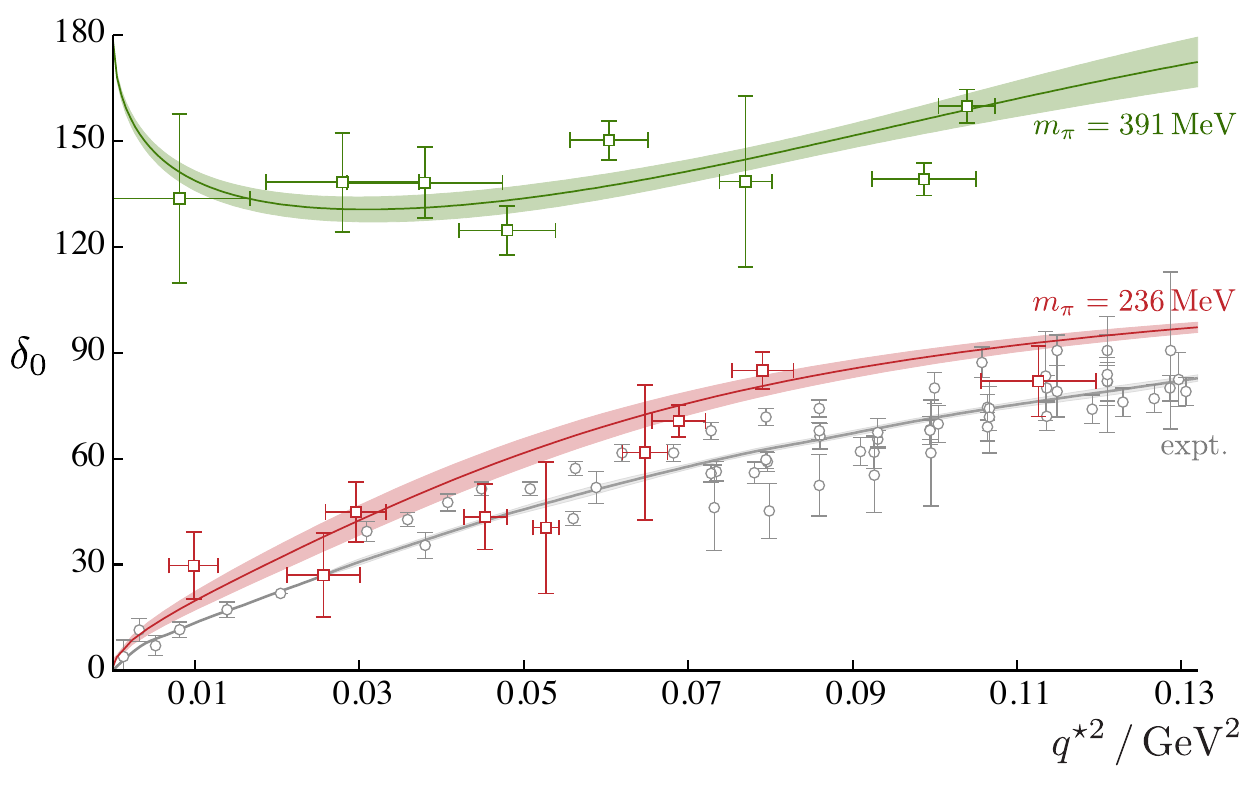}
\includegraphics[width = \columnwidth]{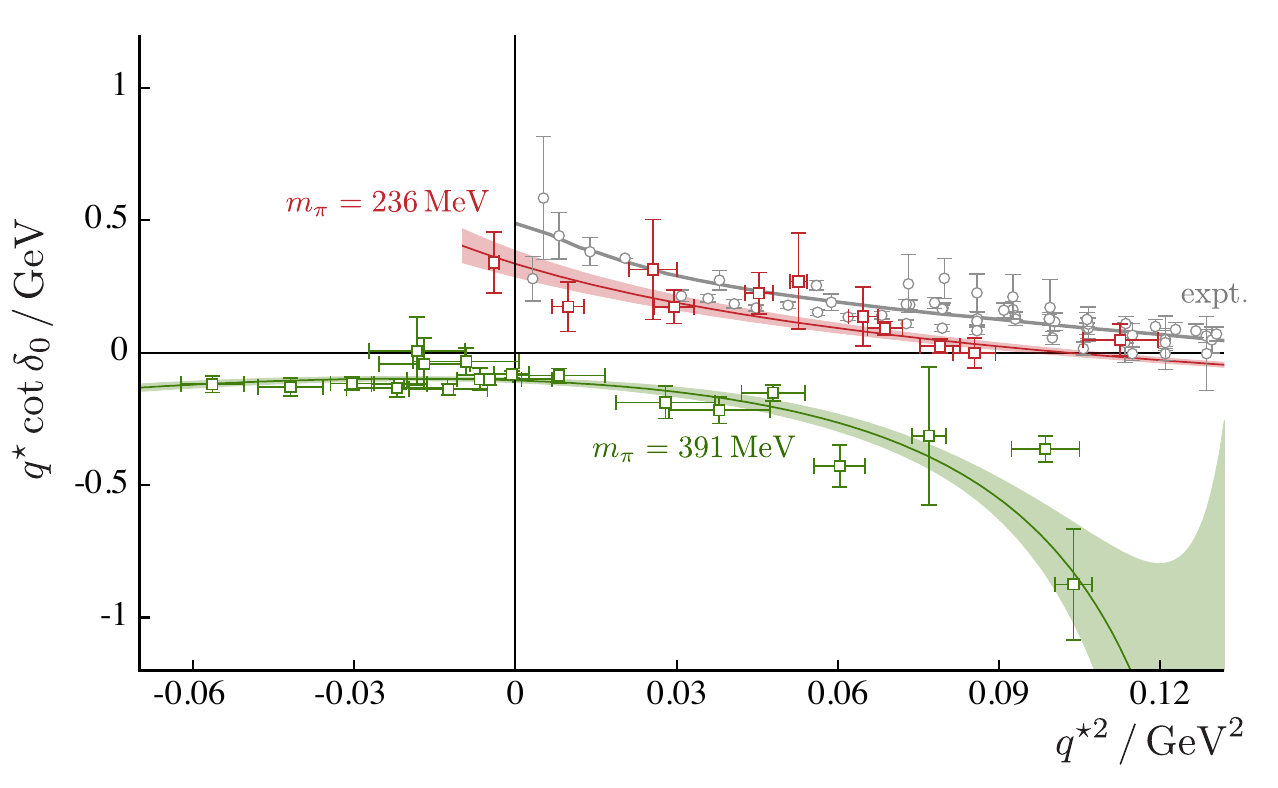}
\caption{Elastic $I=0$ $\pi\pi$ scattering phase-shifts in $S$-wave at two pion masses. Phase-shifts extracted from experimental data shown in grey along with a curve following from highly contrained analysis of that data. Figure adapted from one in~\cite{Briceno:2016mjc}. }\label{fig:pipiI0}
\end{figure}

Figure~\ref{fig:pipiI0} shows the elastic scattering phase-shift determined from spectra on these lattices for the two pion masses, and a clear change is observed between the two. At the heavier quark mass, the behavior is that of a bound-state lying just below threshold, while at the lighter mass we observe something much closer to the experimental situation, with a slow increase in phase-shift over the elastic region.

At the heavier quark mass, all analytic parameterizations of the scattering amplitude capable of describing the finite-volume spectra feature a pole located on the real energy axis, on the physical sheet, at ${E^\star = 758(4)}$~MeV, which is interpreted as a bound-state $\sigma$ (lying below the $\pi\pi$ threshold at $2\,m_\pi = 782$~MeV). At the lighter quark mass, the situation is somewhat less clear --- many different parameterizations are capable of describing the spectra, and while they do feature a pole far into the complex plane on the unphysical sheet, the position of that pole is not precisely determined, with considerable scatter observed as the amplitude parameterization is varied. This observation is not unique to the finite-volume situation --- the same scatter in pole position is observed when a variety of amplitudes forms are constrained using only the experimental elastic phase-shift data. It is only when amplitude forms which build in the required constraints of analyticity and crossing-symmetry are utilized that the pole position can be pinned down with precision~\cite{Pelaez:2015qba}.

This brings us back to a point previously raised in the context of analyzing the spectra that lie above multiple two-body open channels, discussed in Sec.~\ref{sec:fv_coupled}. We found that the quantization condition describing such systems, Eq.(\ref{eq:2chan}), relates a single energy level to multiple components of the scattering amplitude. We argued this issue can be circumvented by using flexible parametrizations of the energy dependence of the scattering amplitude. If a large density of states are determined, the systematic error in the resulting scattering amplitude on the real axis is expected to be small, because the quantization conditions constrains different linear combinations of the scattering amplitude in close proximity. That being said, it is quite possible that consistent results on the real energy axis might lead to significantly different values for the position of any resonant pole if it lies far into the complex plane. This motivates the need for further constraints on the scattering amplitude -- as an example, for light quark masses, it might be important to implement low-energy constraints imposed by chiral perturbation theory, such as the Adler zero present in the isoscalar $\pi\pi$ channel, which plays an important phenomenological role.

A system rather similar to the $\rho$ in $\pi\pi$ is the narrow $K^\star$ meson in $P$-wave $\pi K$ scattering with isospin=1/2.  There is an important difference, however, when this system is considered in a finite volume --- the $\pi K$ $S$-wave system does not decouple from the $P$-wave in moving frames, which significantly complicates the analysis since the \mbox{$S$-wave}, like the $\pi\pi$ case, appears to house a broad resonance. A few lattice calculations have considered elastic scattering in the $\pi K$ channel \cite{Bali:2015gji, Lang:2012sv, Fu:2012tj, Prelovsek:2013ela}.

\begin{figure}[!t]
\includegraphics[width = \columnwidth]{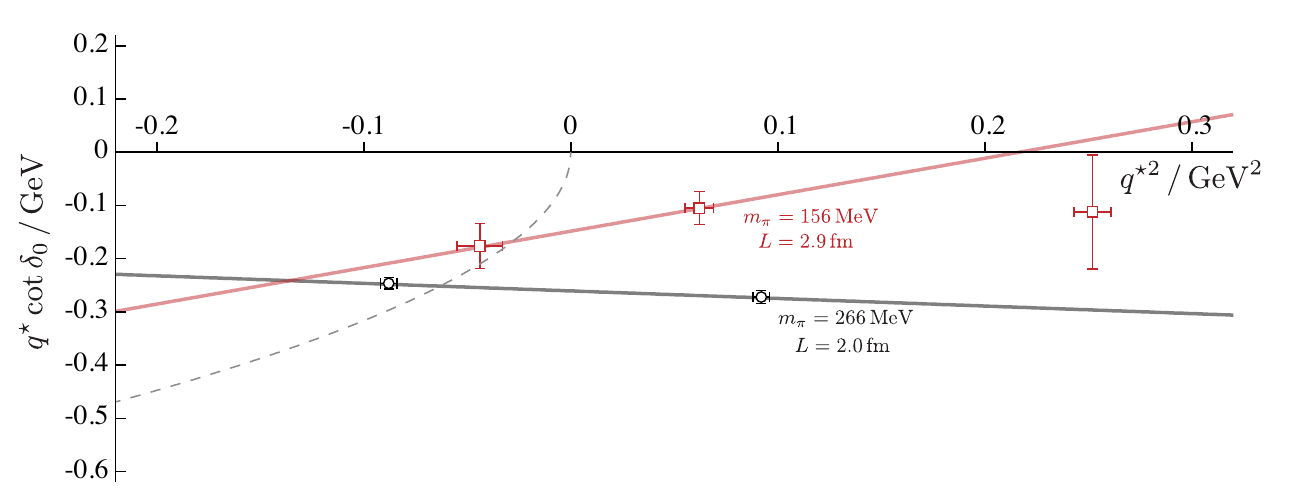}
\caption{$DK$ $S$-wave elastic scattering phase-shift at two values of the light quark mass. Straight lines indicate effective range fits to the lowest two points at each mass, with the point at which these lines intercept the dashed curve ($-|q^\star|$) being the position of a bound-state $D_{s0}^\star$ meson.
Figure adapted from one presented in~\cite{Lang:2014yfa}.
}\label{fig:DK}
\end{figure}

A number of papers have reported on the determination of elastic scattering amplitudes involving mesons containing a heavy quark~\cite{Mohler:2012na, Prelovsek:2013cra, 
Lang:2014yfa,
Lang:2015sba, Lang:2016jpk} -- these studies have typically considered only the rest frame in a single volume, giving a limited constraint on the energy dependence of the phase-shift. An example is shown in Figure~\ref{fig:DK} which displays the $DK$ $S$-wave elastic phase-shift at two pion masses computed in~\cite{Lang:2014yfa}. The use of a single volume and only the rest frame limits the number of relevant energy levels to two, with an effective range formula used to interpolate between the two in order to locate the position of a bound-state $D_{s0}^\star$ meson.

The need to determine multiple energy levels in a limited energy region in order to map out the scattering amplitude has lead to suggestions for approaches other than the use of moving frames. One example is to consider lattices in which one spatial direction is longer than the others, i.e. $L\times L \times (L' >L)$. In this way the unit of momentum in the $z$-direction is smaller than that in the $x,y$ directions, leading to a changed spectrum of non-interacting levels. In the interacting case, the spectrum is also changed, and the breaking of the cubic symmetry causes there to be more levels in the rest-frame spectrum. The complications are that, similar to the moving-frames case, operators in the irreps of a smaller symmetry group must be constructed, and unlike the moving frames case, new gauge fields are required for each new value of $L'$. This approach was applied in explicit calculation of the $\rho$ resonance in ~\cite{Pelissier:2012pi}. In principle one could also consider moving frames in these asymmetric boxes~\cite{Lee:2017igf}.  

Another proposed approach is to change the boundary conditions felt by the quark fields, by introducing a ``twist'' that can effectively induce a non-zero value of momentum~\cite{Bedaque:2004kc, Sachrajda:2004mi}. Typically the proposal, called \emph{partial twisting}, is to only apply the twist to ``valence'' quarks, i.e. to the propagators that go into correlator construction, so that new gauge-field generation is not required for each new twist value. Complications include the need to show that the application of a different boundary conditions to the valence versus the sea quarks does not lead to an unacceptable breakdown in the unitarity of the theory\footnote{This concern may be somewhat less if only heavy quarks are twisted, since their annihilation and `sea' behavior is likely to be less relevant}. \cite{Ozaki:2012ce} used partial twisting to obtain more points on the phase-shift curve in a calculation of $J/\psi \, \phi$ elastic scattering and similarly~\cite{Chen:2014afa} in the elastic scattering of isospin=1 $D\overline{D}$.

\subsection{Resonances in coupled-channel meson-meson scattering \label{sec:coupled}}

To date there have been four lattice QCD calculations which have used the L\"uscher approach to determine coupled-channel scattering amplitudes, all in the meson sector. The first was a study of coupled $\pi K$, $\eta K$ scattering, reported on in~\cite{Dudek:2014qha,Wilson:2014cna}, in which spectra were computed on three lattice volumes, including a large number of moving frames. The resulting energy levels were used to tightly constrain the two-channel scattering matrix, and this led to the conclusion that there is very little coupling between the channels, with the $\eta K$ channel being weakly repulsive. The $\pi K$ channel was found to feature a broad resonance and a virtual bound-state in $S$-wave, a bound-state $K^\star$ in $P$-wave, and hints of a narrow resonance in $D$-wave. 

The second reported coupled-channel study appeared in~\cite{Wilson:2015dqa}, where the extension of the spectra used to determine elastic $\pi\pi$ scattering in $P$-wave, to an energy region above $K\overline{K}$ threshold, was used to constrain coupled-channel $\pi\pi, K\overline{K}$ scattering. As in the previous case, relatively little coupling between the two channels was observed.

The third case is the only one to-date involving heavy quarks, and the first to consider three-channel scattering. Using the same distillation technology, operator constructions and analysis approach as the studies above,~\cite{Moir:2016srx} studied the coupled $D\pi, D\eta, D_s\overline{K}$ system, finding a near-threshold bound-state in $S$-wave, a deeply bound $D^\star$ in $P$-wave, and evidence for a narrow $D$-wave resonance coupled dominantly to $D\pi$.

\begin{figure}[t]
\includegraphics[width = \columnwidth]{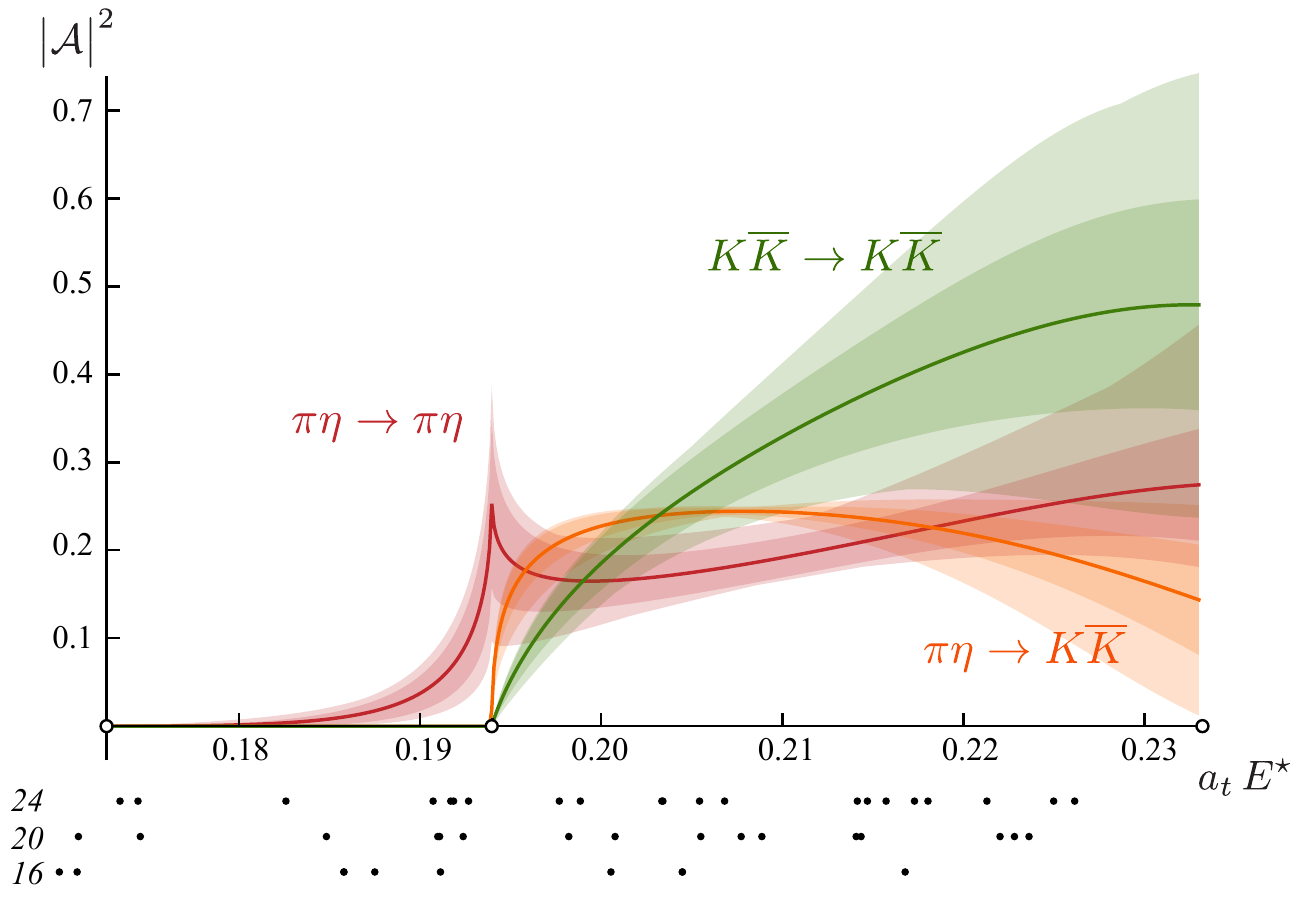}
\caption{$\pi \eta$, $K\overline{K}$ $S$-wave coupled amplitudes, expressed via $\big| \mathcal{A} \big| = \frac{1}{16\pi} \frac{2 \, q^\star}{E^\star} \big| \mathcal{M}_{\ell = 0} \big|$, plotted from $\pi \eta$ threshold up to $\pi \eta'$ threshold. Central values and inner bands indicate amplitudes and statistical errors taken from one particular successful parameterization, while the outer band indicates the degree of scatter observed (in the $1\sigma$ error bands) over a large number of successful parameterizations. Points below the graph show the positions of energy levels on each of three volumes used to constrain the amplitudes. Calculation performed at $m_\pi \sim 391$~MeV, with energy expressed in units of the temporal lattice spacing, $1/a_t \sim 5.7$~GeV. Figure adapted from one appearing in~\cite{Dudek:2016cru}. }\label{fig:a0amps}
\end{figure}

The fourth and most recent calculation,~\cite{Dudek:2016cru}, considered the coupled $\pi \eta, K\overline{K}$ system (with additional limited consideration of the coupling with the $\pi \eta'$ channel). Experimentally, the $S$-wave in this system features a strong enhancement at $K\overline{K}$ threshold that is usually ascribed to the presence of a resonance, the $a_0(980)$. In~\cite{Dudek:2016cru}, the Hadron Spectrum Collaboration presented spectra from three volumes in a range of frames with $m_\pi \sim 391$~MeV. In total, below $\pi \eta'$ threshold, they found 47 energy levels which were used to constrain the two-channel $\pi \eta, K\overline{K}$ scattering matrix. A large number of $K$-matrix parameterizations were found to be capable of describing the finite volume spectra -- Figure~\ref{fig:a0amps} shows all such amplitudes determined in this way which have an acceptable goodness-of-fit.

\begin{figure*}
\includegraphics[width = .45\textwidth]{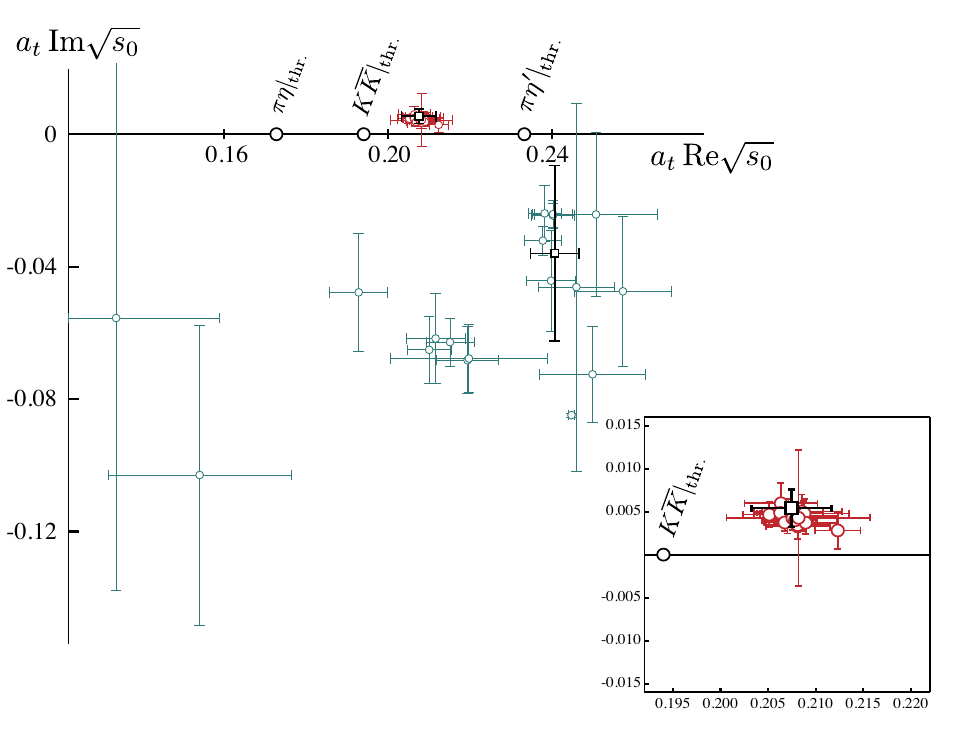}
\includegraphics[width = .45\textwidth]{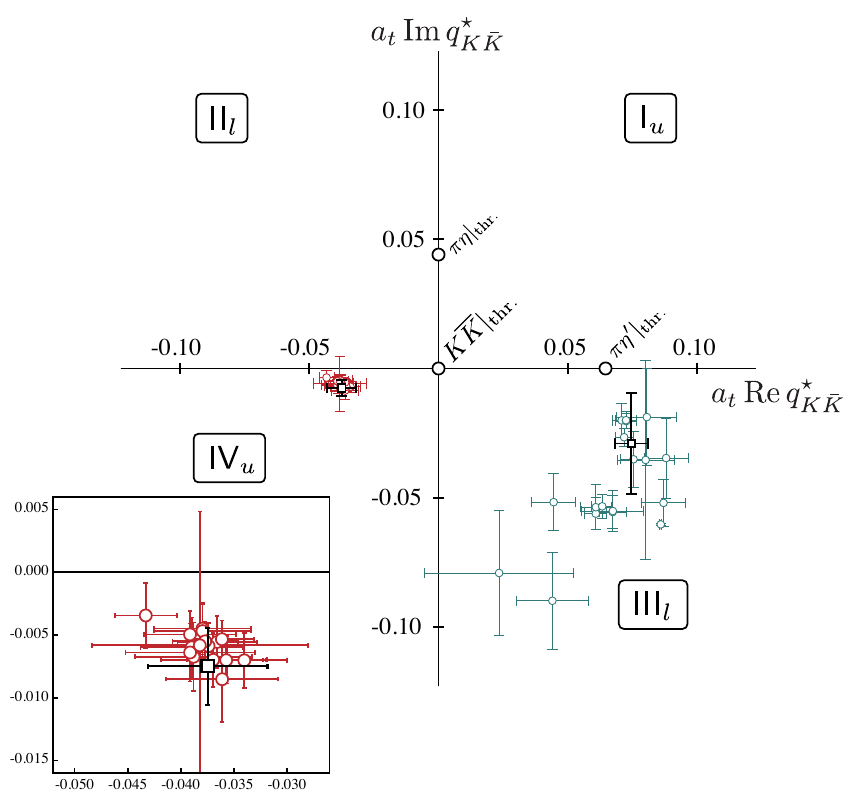}
\caption{
Pole singularities in the $S$-wave coupled-channel amplitudes plotted in Figure~\ref{fig:a0amps}. Left(right) panel: poles in complex energy plane (poles in complex $q^\star_{K\overline{K}}$ momentum plane). In both cases the red(blue) points show the sheet IV(III) pole found for a large number of amplitude parameterizations. Energy expressed in units of the temporal lattice spacing, $1/a_t \sim 5.7$~GeV. Figure adapted from one appearing in~\cite{Dudek:2016cru}. }\label{fig:a0poles}
\end{figure*}

Examining Figure~\ref{fig:a0amps}, we see that the $\pi \eta \to \pi \eta$ amplitude shows a `cusp-like' behavior at the $K\overline{K}$ threshold, and indeed a cusp is always allowed when a new kinematic threshold opens. However in this case, the strength of the effect, and the rapid turn-on of amplitudes leading to $K\overline{K}$ suggest there might be a resonance nearby. This can be examined by analytically continuing the amplitude parameterizations into the complex energy plane, and on doing so, a clear outcome emerges -- all successful parameterizations feature a pole on \mbox{sheet IV} ($\mathrm{Im}\, q^\star_{\pi \eta} > 0,\, \mathrm{Im}\, q^\star_{K\bar{K}} < 0$), and as can be seen in Figure~\ref{fig:a0poles}, there is very little scatter with variation in parameterization form. The residues of $\mathcal{M}$ at the pole can be factorized leading to couplings of the resonance to the $\pi \eta$ and $K\overline{K}$ channels which prove to be comparable in magnitude
\footnote{A description of these energy levels using amplitudes motivated by unitarized chiral perturbation theory~\cite{Guo:2016zep} gives a quite similar result for the pole position and couplings, although the validity of using the effective theory at this pion mass is not clear.}. 

Poles are also found on sheet III, which is closest to physical scattering above the $K\overline{K}$ threshold, but they are either far into the complex plane, or lie outside the energy region in which there are energy levels to constrain the amplitude. The position of these poles, which we emphasize are irrelevant for the behavior near the $K\overline{K}$ threshold, shows significant scatter with parameterization change, suggesting that they might be  artifacts.

The same study found that the $P$-wave amplitude below $\pi \eta'$ threshold is compatible with zero, and the \mbox{$D$-wave} scattering matrix features a narrow resonance with comparable couplings to both $\pi \eta$ and $K\overline{K}$ final states. The extraction of the amplitudes in $D$-wave is not at the same level of rigor as the $S$-wave: the three-body $\pi\pi\pi$ channel can couple to the $J^P = 2^+$ partial wave, but not to $0^+$, and the $D$-wave resonance lies above the $\pi\pi\pi$ threshold. The calculation reported in~\cite{Dudek:2016cru} did not include operators resembling $\pi\pi\pi$, nor was any attempt made to account for such a scattering channel, indeed the formalism to do so does not yet exist in a complete form (this will be discussed further in Section~\ref{Sec:extensions}).

This study illustrates some important points that are likely to be generally applicable to future studies of coupled-channel resonances using the L\"uscher approach. The extracted amplitudes do not have a simple `canonical' resonance behavior (as illustrated in Figure~\ref{fig:fv_flatte}), and it is only by obtaining a dense spectrum of energy levels across the relevant energy region (illustrated by the dots at the bottom of Figure~\ref{fig:a0amps}) that it is possible to constrain amplitude parameterizations sufficiently to determine the energy-dependence. The rest-frame spectrum in one volume alone (unless it were extremely large) would not be enough to understand this system. The use of a range of analytic parameterizations to describe the spectra, and the ability to continue these into the complex plane to search for pole singularities, leads to a description of the resonant physics in terms of a pole position and couplings, which one can argue is the least model-dependent approach possible. 

In summary,~\cite{Dudek:2014qha,Wilson:2014cna,Wilson:2015dqa, Moir:2016srx, Dudek:2016cru} present an approach to studying coupled-channel scattering in lattice QCD, based upon calculating the spectrum in many moving frames using a large basis of `single-hadron' and `hadron-hadron' operators, that can be extended to other scattering channels. The relevant finite-volume formalism is in place for systems featuring any number of coupled two-hadron channels, including those where the scattering hadrons have non-zero spin~\cite{Briceno:2014oea}. Nucleon-meson scattering such as $\pi N,\, \eta N \ldots$, in which excited nucleon resonances, $N^\star$s, are expected to appear, is one obvious example. Here the challenge is largely computational -- with the increase in the number of quarks comes an increase in cost -- this has restricted the scope of calculations~\cite{Lang:2016hnn}. A larger challenge, which becomes more acute at the mass of the pion is reduced, is the presence of three-hadron and higher multiplicity channels, for which a complete finite-volume formalism is not yet in place. We will return to this issue in Section~\ref{Sec:extensions}.

\section{Other approaches to resonance determination \label{Sec:other}}
We have presented an approach for resonance determination in lattice QCD which follows from the L\"uscher quantization condition, Eqn.~\ref{eq:QC_2body}, and illustrated its successful application in a number of explicit lattice QCD calculations. Alternative approaches have been proposed and in some cases applied, and in this section we will briefly review them, beginning with a consideration of the simplest possible approach: directly associating a particular finite-volume energy level with a resonance.
  
\subsection{Resonances in the L\"uscher formalism in the narrow-width approximation \label{sec:narrow}}

We might wonder if, in the case of a narrow resonance, we can use the width of the resonance as a small parameter and obtain a simple result for the appearance of the resonance in a finite volume. We can explore this in the simplest case of elastic scattering with a single partial-wave dominating. In this case the L\"uscher quantization condition, Eq.(\ref{eq:QC_2body}), takes the form,
\begin{align*}
&0= q^\star \cot \delta_\ell(E^\star)\\
 &+ 8 \pi E^\star \left[ \frac{1}{L^3} \!\sum_{\mathbf{k}} -\,  \mathrm{PV}\!\!\int\!\!\frac{d\mathbf{k}}{(2\pi)^3}  \right] \frac{1}{ 2\omega_k \, 2\omega_{Pk} \, (E - \omega_k - \omega_{Pk}) } ,
\end{align*}
where the removal of the imaginary part of $F$ and $\mathcal{M}^{-1}$ which is completely specified by unitarity has lead to the introduction of the \emph{principal-value} integration. 

For an amplitude dominated by a narrow elastic resonance, we may use the Breit-Wigner form, ${\cot \delta_\ell(E^\star) = \frac{m_0^2 - E^{\star 2}}{E^\star \, \Gamma(E^\star) }}$, with an energy-dependent width ${ \Gamma(E^\star) = \frac{g^2}{6\pi} \frac{(q^\star)^{2\ell +1}}{E^{\star 2}}  }$ that ensures the correct threshold behavior. The corresponding quantization condition will then be,
\begin{align*}
E^{\star 2}& = m_0^2 + 8 \pi \frac{E^{\star 2} \, \Gamma(E^\star) }{q^\star}& \\
 &\times \left[ \frac{1}{L^3} \sum_{\mathbf{k}} -\,  \mathrm{PV}\!\int\!\!\frac{d\mathbf{k}}{(2\pi)^3}  \right] \frac{1}{ 2\omega_k \, 2\omega_{Pk} \, (E - \omega_k - \omega_{Pk}) },
\end{align*}
and in the limit of the width being small (either due to a small coupling, or a small phase-space for decay), we expect a solution very close to $E^\star = m_0$, and further solutions very close to the non-interacting two-body energies, $E = \omega_k + \omega_{Pk}$. Specifically these ``nearly non-interacting'' levels are the energies which solve
\begin{equation*}
E - \omega_k - \omega_{Pk} = - \frac{\nu}{m_0^2} \frac{1}{2\omega_k \, 2 \omega_{Pk}  }\, 8 \pi \frac{E^{\star 2} \, \Gamma(E^\star) }{q^\star}, 
\end{equation*}
where $\nu$ is the degeneracy of the non-interacting state. An explicit equation for the energy of the level near $m_0$ can be found at lowest non-trivial order in the small width of the resonance,
\begin{widetext}
\begin{equation*}
E^\star_R(L) = m_R \left( 1 + \frac{\Gamma_R}{m_R} + 8 \pi \frac{\Gamma_R}{q_R^\star}   \left[ \frac{1}{L^3} \sum_{\mathbf{k}} -\,  \mathrm{PV}\!\int\!\!\frac{d\mathbf{k}}{(2\pi)^3}  \right] \frac{1}{ 2\omega_k \, 2\omega_{Pk} \, (E_R - \omega_k - \omega_{Pk}) }   + \mathcal{O}(\Gamma_R^2) \right)^{1/2},
\end{equation*}
\end{widetext}
where the pole-position mass ($m_R$) and width ($\Gamma_R$) have been introduced. 

Clearly, for a small width, this is a small volume-dependent shift away from $m_R$, unless it happens that $E_R = \sqrt{m_R^2 + \mathbf{P}^2}$ is approximately equal to a non-interacting energy level, in which case the shift can be enhanced due to the effect of the pole. This is precisely the `avoided level crossing' behavior observed in Figure~\ref{fig:fv_elastic_res} for a Breit-Wigner resonance.

Because the sum over discrete momenta is not invariant under boosts, the volume-dependent correction term will differ in different frames, and we would not expect the energy of the resonance to perfectly satisfy the dispersion relation expected for a single particle, i.e. $E_R(\mathbf{P}, L) \neq \sqrt{ \big( E_R^\star(L) \big)^2 + \mathbf{P}^2 }$. This was explicitly observed in the results presented in~\cite{Wilson:2015dqa}, which we show in Figure~\ref{fig:pipiI1860} --- we see that when we boost the moving frame energy levels that lie near the resonance position back to the cm-frame, they do not all lie at the same energy, and it is this effect which allows us to map out the energy-dependence of the phase-shift within the narrow width of a resonance.

It is worth pointing out that the corresponding analysis for a \emph{bound-state} leads to the finite-volume corrections being not power law as above, but rather exponentially suppressed like $e^{-\kappa L}$ with $\kappa$ the binding momentum. Because $E$ for a bound-state is always below threshold, the function in the sum above is smooth and the difference with the integral is exponentially small~\cite{Luscher:1985dn, Konig:2017krd, Hansen:2016ync,  Korber:2015rce, Meissner:2014dea,  Briceno:2013hya, Briceno:2013bda,  Konig:2011nz, Konig:2011ti, Bour:2011ef, Davoudi:2011md, Beane:2003da}

It appears that we can conclude (at least in the elastic case) that if there is a narrow resonance of mass $m_R$, there will be a finite-volume energy level close to $m_R$, with a volume-dependent shift that is likely to be small unless a non-interacting level happens to lie very close by. In the limit of zero width due to there being no coupling to the decay channel, we would interpret this state as being like the $| \rho, L\ra_0$ basis state we discussed in Section~\ref{sec:meson-meson}, that will be interpolated well by `single-hadron-like' operators. We can thus combine our observations in this section and those in Section~\ref{sec:meson-meson} to propose an explanation of the spectra extracted in calculations like~\cite{Dudek:2013yja}, which use only a large basis of `single-hadron-like' operators and no multihadron operators, one example of which we previously presented in Figure~\ref{fig:isoscalar}. It is likely that these calculations are resolving the \emph{presence} of relatively narrow resonances in the relevant energy regions, and the masses plotted should be viewed as being \emph{imprecise} guides to the $m_R$ values of these resonances. What the decay properties of these states are, or whether there are broader resonances present, cannot be determined from such calculations.

\subsection{Resonances and `naive' level counting \label{sec:counting}}

The previous section suggested that the presence of a narrow resonance might be inferred by there being an `extra' energy level beyond those expected in the non-interacting spectrum. A number of recent calculations
~\cite{
Prelovsek:2013xba, Prelovsek:2014swa, Lee:2014uta, Padmanath:2015era, Lang:2015sba, Lang:2016jpk, Lang:2016hnn} 
make use of bases of operators that include multihadron operators and extract spectra that, in some cases, are likely to be reasonably close to the complete spectrum in a given energy region. Typically only the rest frame is considered, and the use of relatively small volumes limits the density of states extracted. Since these spectra commonly extend into energy regions where more than one scattering channel is kinematically open, a coupled-channel L\"uscher analysis of the type described in Section~\ref{sec:coupled} would be desired, yet there are insufficient numbers of levels to fully constrain multichannel parameterizations. Instead an approach is followed that appeals to narrow resonance arguments of the type we presented in the previous section, and as we have argued, at best this approach can suggest the presence of a narrow resonance, but cannot determine its properties with any precision.

The calculated finite-volume spectrum is compared to the known multichannel non-interacting spectrum, with each calculated level being associated with a non-interacting level, until all non-interacting levels in the relevant energy region have been exhausted. Any calculated level that appears \emph{in excess} of this counting is thus a candidate to be due to the presence of a narrow resonance. Further confirmation of these assignments are sought from the values of the overlaps $\langle E_n | \mathcal{O} | 0 \rangle$ extracted from the variational solution, with large overlaps onto `single-hadron' operators used to suggest a resonance assignment. 

An example, albeit not an ideal one, owing to the complicating presence of scattering hadrons of non-zero spin, and three-body channels which are not considered, is the analysis presented in~\cite{Padmanath:2015era}. Using a basis of meson-meson operators and single-hadron operators of both quark-bilinear and tetraquark constructions, they extract a rest-frame spectrum in the $T_1^+$ irrep (where the $X(3872)$ experimental state would be expected to appear if it is a $J^P=1^+$ resonance) in a single volume. Observing an `extra' level beyond the $\eta_c$`$\sigma$', $D \overline{D}^*$, $J/\psi \, \omega$ and $\chi_{c1}$`$\sigma$' levels expected in a narrow energy window, they choose to associate this state with the $X(3872)$.

The converse argument has also been made, that if \emph{no level appears in excess} of those expected in the non-interacting spectrum, \emph{there is likely to be no resonance}. An example is the calculation presented in~\cite{Prelovsek:2014swa} of the hidden-charm $I=1$ sector, where a basis of meson-meson operators is supplemented with tetraquark operators, and the rest-frame spectrum of the $T_1^+$ irrep is determined in a single volume. Each calculated lattice level is matched with an expected nearby non-interacting level, with no lattice levels left over, and it is concluded that there is no signal for a $Z_c$ resonance.

\begin{figure}[t]
\centering
\includegraphics[width = .3\textwidth]{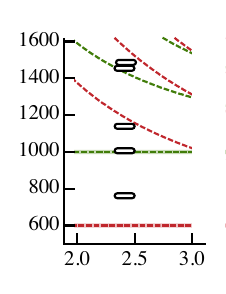}
\caption{Rest-frame spectrum in a $(2.4 \,\mathrm{fm})^3$ lattice for the amplitude plotted in Fig.~\ref{fig:f0980}. \label{fig:f0980_fv_one_volume} }
\end{figure}

We conclude this section by raising a note of caution regarding the use of the \emph{absence} of an `extra' level to conclude the absence of resonant behavior. In Section~\ref{sec:examples} we presented an example of a two-channel amplitude which features a relatively narrow resonance corresponding to a pole singularity on the second sheet, but whose finite-volume spectrum does not obviously feature an isolated energy level that we can associate with the resonance (Fig.~\ref{fig:f0980}). In Figure~\ref{fig:f0980_fv_one_volume} we present an example of how the approach described in the current section could be misleading when applied to this amplitude --- a calculation of the rest-frame spectrum in a quite reasonable $(2.4\,\mathrm{fm})^3$ volume would give the spectrum shown, which would certainly not lead one to conclude the presence of a relatively narrow resonance close to the $K\overline{K}$ threshold and might, unfortunately, lead one to incorrectly conclude the \emph{absence} of any resonance in this energy region.

In summary, the level counting approach is based largely on expectations from elastic scattering, and in the case of coupled-channels, while we might get simple narrow resonances which do behave like elastic resonances (e.g. Fig.~\ref{fig:fv_flatte}), and which come with an `extra' level, we can also have more subtle situations where a resonance is present, but no `extra' level appears. The approach described in Sec.~\ref{sec:coupled}, where the complete spectrum in many moving frames and/or volumes is used to tightly constrain the energy dependence of the scattering matrix, does not suffer from the weaknesses outlined above.

\subsection{Finite volume EFT Hamiltonian approach \label{sec:hamiltonian}}

Following the diagrammatic approach presented in Section~\ref{sec:luscher}, given an explicit quantum field theory Lagrangian describing the interactions between hadron fields, one can proceed to determine and parametrize the propagators of multi-particle states in a finite-volume, and the location of the poles of these objects would give the prediction of the finite-volume spectrum for that given QFT.

The procedure we have been describing thus far is one where a particular hadron-level Lagrangian is never specified, but instead we use relations that hold for any unitary quantum field theory, which relate the finite-volume spectrum to the infinite-volume $S$-matrix.

Alternatively, one can specify a particular choice of hadron-level lagrangian, consider the finite-volume propagator, say of two-particle states, and tune the parameters of the theory in order to reproduce the observed finite-volume spectrum~\cite{Bernard:2007cm,Beane:2012ey,Hall:2013qba,Wu:2014vma}. Because this procedure assumes a suitable low-energy effective field theory that describes the interacting system in the region of interest being evaluated in a finite-volume, it is commonly referred as the \emph{finite volume EFT Hamiltonian approach}. 

Although the L\"uscher quantization condition is never explicitly used in this approach, we can show that the finite-volume spectra of the EFT Hamiltonian have the same relationship with the infinite volume scattering amplitudes as given by the L\"uscher quantization condition (up to exponentially suppressed corrections) --- provided one works within the validity of the EFT and incorporates enough non-negligible partial waves
\footnote{
A similar technique to those presented in the references above was used in~\cite{Briceno:2013lba} to generalize the L\"uscher condition to two-nucleon systems
}. We can illustrate the equivalence of the two approaches in the simple case of elastic scattering of equal mass spinless bosons in a single dominant partial-wave, in the overall rest-frame~\cite{Wu:2014vma}. The development of the L\"uscher approach presented in Section~\ref{sec:narrow} leads to 
\begin{equation}
q^\star \cot \delta = \left[ \frac{1}{L^3} \sum_\mathbf{k} - \mathrm{PV}\!\int\!\frac{d\mathbf{k}}{(2\pi)^3} \right] \frac{4\pi}{k^2 - q^{\star2}}, \label{eq:fv_lusc}
\end{equation}
while the equivalent expression in the finite-volume EFT approach~\cite{Hall:2013qba} follows from the relation determining the relative momentum of the two-body system,
\begin{equation*}
\omega_q = \sqrt{m^2 + q^{\star2}} = \Delta_0 - \frac{2 \pi^2}{L^3} \sum_\mathbf{k} \frac{g^2(k)}{\omega_k - \omega_q}, 
\end{equation*}
via
\begin{equation*}
\frac{g^2(k)}{\omega_q - \Delta_0 - \Sigma(k)} = - \frac{1}{\pi \omega_q} \frac{1}{q^\star \cot \delta - i q^\star}.
\end{equation*}
In these expressions, $\Delta_0$ and $g$ are bare parameters or functions in the EFT being considered and the self-energy, $\Sigma(q^\star) = 2\pi^2 \int\!\frac{d\mathbf{k}}{(2\pi)^3} \, \frac{4\pi}{k^2 - q^{\star 2}} \frac{g^2(k)}{\omega_q - \omega_k + i \epsilon}$. Eliminating the imaginary part that is fixed by unitarity, we are left with
\begin{equation*}
q^\star \cot \delta = \frac{2\pi}{\omega_q} \frac{1}{g^2(q^\star)} \left[ \frac{1}{L^3} \sum_\mathbf{k} - \mathrm{PV}\!\int\!\frac{d\mathbf{k}}{(2\pi)^3} \right] \frac{g^2(k)}{\omega_k - \omega_q},
\end{equation*}
and if $g(k)$ is a smooth function of the momentum, then the difference between the sum and the integral will be an exponentially small quantity such that
\begin{equation*}
q^\star \cot \delta = \left[ \frac{1}{L^3} \sum_\mathbf{k} - \mathrm{PV}\!\int\!\frac{d\mathbf{k}}{(2\pi)^3} \right] \frac{4\pi}{k^2 - q^{\star 2}} + \mathcal{O}(e^{-m L}),
\end{equation*}
which is clearly equivalent to Eq.(\ref{eq:fv_lusc}) for reasonably large volumes.

Given that the finite volume spectra of the Hamiltonian and the L\"uscher formalism have the same relationship to on-shell scattering, neither method offers any more or less information about the physical scattering amplitude. The EFT Hamiltonian can be viewed as a particular amplitude parameterization. When constrained by the same lattice QCD results, all sufficiently flexible parameterizations should agree on the on-shell scattering amplitude. The alternative parameterizations will in general give different analytic continuation into the complex plane and this may manifest itself in variation in the position and couplings of any resonance poles.

The Hamiltonian formulation has been used in the analysis of recent numerical lattice QCD studies, with the study of the $\Lambda$(1405) generating significant attention \cite{Hall:2014uca,Liu:2016wxq}. 
Owing to the computational cost, the numerical studies of this state have been limited to the use of only local interpolating operators \cite{Menadue:2011pd,Hall:2014uca}. At heavy quark masses ($m_\pi\gtrsim 400\,{\rm MeV}$) the $\Lambda$ is clearly isolated as a deeply-bound state, far below the $\pi\Sigma$ threshold. As the quark masses are lowered, this state becomes resonant in $\pi\Sigma$ and lies somewhere near the vicinity of the $\overline{K}N$ threshold. 
Without having access to a dense finite-volume spectrum, it has not been possible to constrain a general parameterization of the coupled-channel scattering as described in Section~\ref{sec:coupled}.
A more phenomenological approach is taken by introducing a Hamiltonian that describes the meson-baryon interactions through a ``bare'' state. The parameters of this Hamiltonian are then constrained by the quark-mass evolution from the bound-state region at larger quark masses, supplemented with a coupling strength informed by the physically-observed width.

This work has gone further to investigate the electromagnetic coupling to this finite-volume state. The claimed result is that the strangeness magnetic moment of the $\Lambda$(1405) vanishes in the vicinity of the physical point \cite{Hall:2014uca}. Such a result has highlighted the tremendous discriminating power of external probes to resolve the structure of resonant states. In this case, a vanishing strangeness moment would suggest that the strange quark is bound to a $\overline{K}$ meson, which cannot carry any magnetization in an $S$-wave. As discussed above, even in the narrow-width approximation, the quantization condition leads to finite-volume eigenstates which boost nontrivially. Finite-volume eigenstates in the $\Lambda$ system could then differ significantly in composition in boosted frames. The influence of this modified structure on three-point correlation functions remains to be quantified. A systematic approach to matrix elements will be discussed further in Section~\ref{Sec:formfactors}.

The Hamiltonian formulation has also been used in further studies of the $N^*$(1535) \cite{Liu:2015ktc} and Roper \cite{Liu:2016uzk,Kiratidis:2016hda,Wu:2017qve} resonances. In these exploratory investigations, two-body operators have not been used in the extraction of the spectrum. As discussed in Sec.~\ref{sec:meson-meson}, this implies that the spectrum obtained is incomplete and one cannot trust that the spectra obtained are indeed accurate, which would subsequently the resulting scattering amplitudes and resonant poles.

\subsection{Unitarized chiral perturbation theory and chiral extrapolations\label{sec:uchipt}}

An approach closely related to that described in the previous section considers the particular EFT of chiral perturbation theory, unitarized using some prescription~\cite{Oller:1997ng, Dobado:1996ps, Oller:1998hw, GomezNicola:2001as, Pelaez:2006nj}, in a finite-volume~\cite{Agadjanov:2016mao, Chen:2012rp, Doring:2012eu, Doring:2011nd, Molina:2015uqp, Doring:2011vk, Garzon:2013uwa, Bernard:2010fp, Nebreda:2011di, Rios:2008zr, Guo:2008nc, Hanhart:2008mx}. In practice, these works rederive or generalize the L\"uscher spectral condition using $\chi$PT as a starting point. The final result always has the same functional form as Eq.(\ref{eq:QC_2body}) (up to exponentially suppressed corrections), with the scattering amplitude being that of $\chi$PT or some suitable unitarization (U$\chi$PT ). The advantage of this procedure is that it typically builds into the amplitudes some low-energy properties of QCD (relevant for light quarks), and it allows for the relation of observables obtained using different values of the quark masses, a ``chiral extrapolation''.

This methodology has been applied in the study of non-resonant scattering lengths in the heavy-light meson sector~\cite{Liu:2012zya}, as well as in chiral extrapolations of the $\rho$~\cite{Bolton:2015psa, Hu:2016shf}. First exploratory studies of the chiral extrapolations of the $a_0(980)$~\cite{Guo:2016zep} and $\sigma$~\cite{Doring:2016bdr} resonances have appeared, analyzing the finite-volume energy levels calculated by the Hadron Spectrum Collaboration~\cite{Dudek:2016cru,Briceno:2016mjc}.

From a technical standpoint there is no new insight coming from considering U$\chi$PT in a finite volume as compared to simply using Eq.(\ref{eq:QC_2body}) with amplitude forms given by infinite-volume U$\chi$PT. In the case of multichannel scattering there is not a unique procedure for unitarizing $\chi$PT, so one should be careful not to assume that the corresponding amplitudes are a direct consequence of QCD. Pioneering lattice QCD calculations of the excited state spectrum are currently being performed with relatively heavy quark masses, $m_\pi \gtrsim 300$~MeV, where the reliability of $SU(2)$ flavor $\chi$PT would be questionable, and furthermore, the application of $\chi$PT for channels with explicit strange-quark degrees of freedom requires making an expansion about the $SU(3)$ flavor symmetric point of QCD, and obtaining reliable estimates associated with this poor approximation of QCD is notoriously challenging.

The use of chiral effective field theories to extrapolate hadron results down in quark mass is not a new approach. In the past, when states which appear as resonances at the physical quark mass were only realized as bound states at higher quark masses, this approach was used to extrapolated lattice results for the $\rho$ \cite{Leinweber:2001ac,Bruns:2004tj,Allton:2005fb,Armour:2005mk} and $\Delta$ \cite{Leinweber:1999ig,Young:2002cj,Bernard:2005fy,Pascalutsa:2005nd} resonances, for instance. In those works, it was necessary to put in the decay thresholds artificially, using the experimental width to constrain the relevant coupling parameters. The significant new feature moving into the true resonance regime is the ability to directly determine the width (and hence coupling) within the lattice calculation and thereby directly study the quark mass dependence of the complex pole location --- such as highlighted in Figure~\ref{fig:rho_mpi_0}.

\subsection{Other approaches\label{sec:various}}

There have been several other approaches proposed to determine resonance properties in lattice QCD calculations, here we will briefly summarize them. 

\vspace{0.5cm}
In what we might call the `tuned threshold' or \emph{UKQCD method} (see e.g.~\cite{McNeile:2006bz}), the approach requires us to find a value of quark mass such that the resonance to be studied happens to lie rather close to the kinematic threshold of its two-body decay channel. That is, if the resonance decay is $R\to a b$, $m_R \approx m_a + m_b$. Under these conditions it is argued that the resonance to decay-channel coupling, $g_{R \to ab}$, can be determined by considering the correlation function with a `single-hadron' operator with the quantum numbers of $R$ at the source, and an $ab$-like `meson-meson' operator at the sink. The finite volume of the lattice appears to play no role in this approach. It was applied to the decays $b_1 \to \pi \omega$ and $\pi_1 \to \pi b_1$ in~\cite{McNeile:2006bz}, and has recently been applied to baryon decays in~\cite{Alexandrou:2013ata, Alexandrou:2015hxa}.

Underlying this framework are two necessary assumptions which might lead to important systematic errors. First, it is assumed that states interpolated using single-particle and multi-particle operators correspond solely to a resonance, $| R\rangle$, and a scattering state, $| ab\rangle$, respectively. As we saw in Sec.~\ref{sec:meson-meson}, this in general not true. Second, one must assume that the resonance state is an asymptotic state, and the overlap between the multi-particle and resonance state, $\langle R| ab\rangle$, is equal to the scattering amplitude coupling these two ``channels". This, of course, is not true for actual resonances, but the process of tuning the threshold such that resonance cannot actually decay, making it a bound-state at threshold, increases the possible validity. Of course in practice it may not be possible to find a quark mass where the required near-degeneracy occurs, or that quark mass may be very far from the physical quark mass, making the extrapolation questionable. 

\vspace{0.5cm}
The \emph{histogram method} is based on the construction of a probability distribution which mimics the scattering cross section~\cite{Bernard:2008ax}. This method utilizes the low-lying eigenstate spectrum across a uniform distribution of finite volumes, binned as a function of energy, one then generates a histogram which counts the number of states. In the vicinity of a resonance there will be a pronounced enhancement in this ``density of states'' distribution. While offering a clear visualization of the resonance structure, a direct comparison has demonstrated that it is not as straightforward to implement as the conventional L\"uscher technique~\cite{Giudice:2012tg}, and it is not clear that the approach would be useful for the study of resonances which do not appear as a clear `bump' in the scattering amplitude.

\vspace{0.5cm}
A recently proposed approach to coupled-channel scattering in a finite-volume makes use of an \emph{optical potential} to convert the problem into one featuring just a single channel with the effect of the other channels absorbed into the optical potential~\cite{Agadjanov:2016mao}. The intention is to extend the power of the L\"uscher method to high energies where many-particle states can go on-shell without having to explicitly take into account such effects (by hiding them in the optical potential), and although this appears to be a formally sound idea, putting this into practice even for the simplest of systems introduces some model-dependence. At this early stage it is not clear how much information must be supplied by the finite-volume spectrum to make this approach practical -- in the coupled $\pi \eta$, $K\overline{K}$ case considered for illustration in \cite{Agadjanov:2016mao}, the density of states required was extremely high, at the level where, if one actually had this set of energy levels, one could bin in energy, and in each bin solve coupled versions of Eq.(\ref{eq:QC_2body}), to find \emph{all} elements of the scattering matrix without needing to introduce any auxiliary optical potential.

\vspace{0.5cm}

Lastly we consider the  \emph{potential method}, which has been championed by the HAL QCD collaboration~\cite{Ishii:2006ec, Aoki:2009ji, HALQCD:2012aa, Aoki:2012tk, Aoki:2013cra}. This is the only approach which does not make direct use of the discrete spectrum in a finite-volume. Instead, from correlation functions constructed using operators featuring spatially displaced hadron operators, one accesses a quantity which the authors associate with a non-relativistic inter-hadron potential. By solving a Schr\"odinger equation including this potential, one obtains scattering amplitudes and/or bound-state energies.

This formalism has largely been implemented in the study of baryon-baryon systems, where unlike the results obtained by other groups using fairly standard determinations of sub-threshold energy levels~\cite{Berkowitz:2015eaa, Beane:2011iw, Beane:2012vq, Yamazaki:2012hi, Yamazaki:2015asa}, the HAL QCD collaboration sees no evidence for a bound deuteron or di-neutron for heavier than physical quark masses. At this stage it is not clear what is the source of this discrepancy, whether it is that the long-distance part of the potential, to which shallow bound-states are most sensitive is not being accurately determined by HAL QCD, or whether, as HAL QCD have suggested~\cite{Iritani:2016jie, Iritani:2017rlk}, that the spectra obtained in~\cite{Berkowitz:2015eaa, Beane:2011iw, Beane:2012vq, Yamazaki:2012hi, Yamazaki:2015asa}, which do not make full use of the variational technique presented in Section~\ref{sec:variational}, are not reliable.~\footnote{We point the reader to a response by the NPLQCD collaboration~\cite{Beane:2017edf} in regards to the remarks made by the HAL QCD collaboration~\cite{Iritani:2017rlk}.} Another application of the potential method by the HAL QCD collaboration has been in a multi-channel study of the $Z_c(3900)$~\cite{Ikeda:2016zwx}.

An important observation is that while systems potentially featuring shallow bound-states (baryon-baryon), and complicated coupled-channel meson systems ($Z_c$) have been studied in the potential approach, the simplest resonant systems have not. In particular, there has been no presentation of $\pi\pi$ $I=1$ scattering and the $\rho$ resonance, which would seem to be an ideal benchmark calculation. The sensitivity of the tightly-bound $\rho$ to the very short-distance part of a $\pi\pi$ `potential' might be a problem for the method, as it is not clear that the short-distance region can be precisely determined.

\section{Coupling resonances to external currents \label{Sec:formfactors}}

As we have seen already in this review, the field has witnessed a tremendous burst of activity providing progress in the ability to determine masses, widths and couplings to two-body hadronic decay channels of hadronic resonances. In particular, a general quantization condition for two-particle systems in a finite volume has been derived, Eq.(\ref{eq:QC_2body}), which relates the discrete spectrum in a finite volume to scattering amplitudes in infinite volume, and approaches based upon Eq.(\ref{eq:QC_2body}) have been successfully implemented in a number of numerical calculations.

Going beyond this, we can consider if there are further rigorous tools we can bring to bear to aid development of an understanding of hadron resonances. The response of resonances to external probes, such as electromagnetic or weak currents, has the potential to reveal substantial new information about the internal structure of resonant systems. For instance, as discussed above, it has been demonstrated that a vanishing strange-quark magnetization of the $\Lambda(1405)$ would suggest that the strange quark is most-probably bound in a kaon and hence a clear signature that the resonance has the structure of a $\bar{K}N$ molecule \cite{Hall:2014uca}. Similarly, the nature of the Roper resonance has challenged theorists for many years, and resolving the relevant electromagnetic transition form factors~\cite{Wilson:2011aa,Segovia:2015hra}, which can also be measured~\cite{Aznauryan:2008pe,Dugger:2009pn,Aznauryan:2009mx, Aznauryan:2011qj}, could help to distinguish between different proposed scenarios, such as a quark orbital excitation, explicit gluonic excitation, or hadronic molecule. The gluonic content of resonances could also be studied directly by evaluation of gluonic operators which have been previously considered for stable hadrons~\cite{Meyer:2007tm,Horsley:2012pz,Detmold:2016gpy}. Such matrix elements would provide a more rigorous framework with which to build on existing analyses where gluonic content has been inferred from operator overlaps \cite{Dudek:2013yja, Liu:2012ze, Dudek:2012ag, Dudek:2011bn, Edwards:2011jj, Dudek:2011tt, Dudek:2010wm, Dudek:2009qf}.

To isolate matrix elements of resonant states in lattice QCD, we once again must apply a formalism which maps between the quantities directly obtained from finite-volume correlation functions and the desired infinite-volume observables. 

Firstly we will present a diagrammatic representation for an infinite-volume scattering amplitude featuring external currents, which we will refer to as a \emph{transition amplitude}. In close analogy to scattering amplitudes, the diagrammatic representation of the transition amplitude describing $n$ incoming and $n'$ outgoing hadrons, with $j$ insertions of external currents, $\mathcal{T}_j(n\to n')$, can be expressed as 

\vspace{2mm}
\noindent\emph{the sum over all diagrams with $n$ incoming and $n'$ outgoing hadronic legs (that have been amputated and put on-shell) and $j$ insertions of the external current. The momenta flowing through the external currents are constrained to satisfy conservation of momentum, but are otherwise free. All intermediate hadron propagators are evaluated with the $i\epsilon$-prescription and all intermediate loop momenta are integrated}.

\begin{figure}
\begin{center}
\centering
\subfigure[]{\includegraphics[width = \columnwidth]{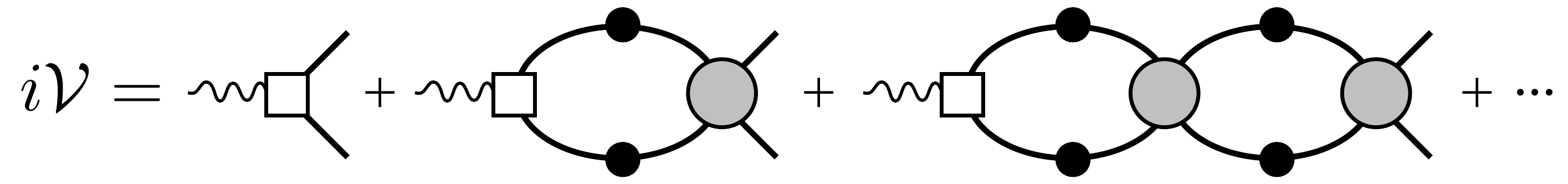} \label{fig:iV}}
\subfigure[]{\includegraphics[width = \columnwidth]{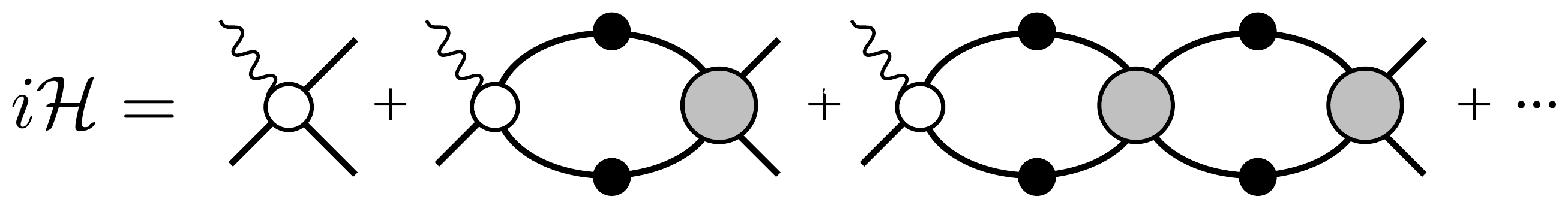} \label{fig:iH}} 
\caption{  Shown are the diagrammatic definitions of the \mbox{(a) $0\to2$} and (b) $1\to2$ transition amplitudes.
The fully dressed propagators and kernels are defined in Fig.~\ref{fig:scat_amp}. The empty squares and circles denote electroweak kernels, which are purely real below multiparticle thresholds.  
}\label{fig:current_insertion_1}
\end{center}
\end{figure}

\vspace{2mm}
Figure~\ref{fig:current_insertion_1} illustrates this in the case of (a) a $0 \to 2$ transition, relevant to e.g. the vector decay constant of the $\rho$ resonance decaying to $\pi\pi$ and (b) a $1 \to 2$ transition, relevant to e.g. the transition form factor of the $\Delta$ baryon resonance in $\gamma^\star N \to \pi N$. The case of $2 \to 2$ transitions, relevant to e.g. deuteron photo-disintegration $\gamma d \to n p$ where the deuteron is a bound-state in $n p$ scattering, can be found in~\cite{Briceno:2015tza}. 

Focussing on the $1 \to 2$ transition, it is clear that $i\mathcal{H}$ should depend on the virtuality of the external current, $Q^2$. For a fixed value of $Q^2$, with the three external hadrons on-shell, the only remaining freedom is in the direction of the relative momentum of the two outgoing hadrons, and it follows that in analogy to the partial-wave expansion of the scattering amplitude, we can partial-wave expand the transition amplitude,
\begin{equation}
\mathcal{H} = \sqrt{4\pi} \, \mathcal{H}_{\ell m}(Q^2, E^\star)\, Y_{\ell m}(\hat{\mathbf{q}}^\star),
\label{eq:Hellm}
\end{equation}
where for simplicity we have assumed the external hadrons are all spinless, although the generalization is straightforward. The energy dependence of the partial-wave transition amplitude, $\mathcal{H}_{\ell m}(Q^2, E^\star)$, must be closely related to that of the corresponding scattering amplitude, $\mathcal{M}_\ell(E^\star)$. For example if the transition process is $\gamma^\star \pi \to \pi \pi$ in $P$-wave, we would expect a close relationship with the scattering process $\pi \pi \to \pi \pi$ in $P$-wave. Indeed this relation can be seen by comparing the diagrams in Figure~\ref{fig:iH} with those in Figure~\ref{fig:scat_ampa}, where we observe that the iterated `rescattering' is the same in both cases, with the only difference being the presence of an additional kernel characterizing the coupling to the external current. In particular we would expect that if the strong rescattering in $\mathcal{M}_\ell$ gives rise to a pole in $E^\star$ corresponding to a composite particle (e.g. a bound-state or a resonance), that same pole singularity must also be present in $\mathcal{H}_{\ell m}$. 

In fact we can rigorously define what we mean by a \emph{resonance transition form-factor} by considering the behavior of $\mathcal{H}$ near the pole. Recall that for the scattering amplitude, we could factorize at the pole to define couplings to the incoming and outgoing channels, e.g.
\begin{equation}
	\lim_{s\to s_0} \mathcal{M}(ab \to cd) \sim \frac{ g_{R \to ab} \; g_{R \to  cd} }{s_0 -s},
\end{equation}
similarly we can factorize $\mathcal{H}$,
\begin{equation}
	\lim_{s\to s_0} \mathcal{H}(\gamma e \to cd) \sim \frac{ F_{\gamma e \to R}(Q^2) \; g_{R\to cd} }{s_0 -s},
	\label{eq:res_ff}
\end{equation}
where $F_{\gamma e \to R}(Q^2)$ is the transition form-factor for $\gamma e \to R$ as a function of the virtuality of $\gamma$. In these expressions $R$ may be a bound-state, in which case $s_0$ is real and below kinematic threshold, and $g, F$ are (in a suitable convention) purely real, or $R$ may be a resonance, in which case $s_0$ is complex, and $g, F$ may also be complex.

\vspace{30mm}
\subsection{Determining matrix elements in lattice QCD\label{sec:ff_lattice}}

Retaining our focus on amplitudes like $\mathcal{H}$ which feature hadrons in both incoming and outgoing states, with insertion of a single current, we are led to consider \emph{three-point functions} of the following form,
\begin{widetext}
\begin{equation}
\hspace{-0.5cm}C^{(3)}_L(x_4, y_4, z_4, \mathbf{P}, \mathbf{P}') \equiv \int_L\!d\mathbf{x} \, d\mathbf{y}\, d\mathbf{z}\; 
e^{-i \mathbf{P}' \cdot( \mathbf{z} - \mathbf{y} )} \, 
e^{-i \mathbf{P} \cdot( \mathbf{y} - \mathbf{x} )} \,
\Big[ \big\la 0 \big|
		T \, \mathcal{A}(z)\, \mathcal{J}(y) \, \mathcal{B}^\dag(x)
      \big| 0 \big\ra \Big]_L, \label{eq:three-point_func}
\end{equation}
where $\mathcal{B}^\dag$ has the quantum numbers of the desired incoming state, $\mathcal{A}$ has the quantum numbers of the desired outgoing state, and $\mathcal{J}$ is the current operator. As in the two-point function case, there is a dispersive representation in terms of the discrete eigenstates of QCD in a finite-volume,
\begin{align}
C^{(3)}_L(x_4, y_4, z_4, \mathbf{P}, \mathbf{P}')
=  L^9 \, \sum_{n,n'}& \, 
e^{- E_{n'} (z_4 - y_4)} \, 
e^{- E_n  (y_4 - x_4)} \nn \\
&\times \big\la 0 \big| \mathcal{A}(0) \big| E_{n'}, \mathbf{P}', L \big\ra \;
\big\la E_{n'}, \mathbf{P}', L \big| \mathcal{J}(0) \big| E_{n}, \mathbf{P}, L \big\ra  \;
\big\la E_{n}, \mathbf{P}, L \big| \mathcal{B}^\dag(0) \big| 0\big\ra, 
\end{align}
where the finite-volume \emph{transition matrix elements}, $\big\la E_{n'}, \mathbf{P}', L \big| \mathcal{J}(0) \big| E_{n}, \mathbf{P}, L \big\ra$ are what we would like to determine in explicit calculations.
\end{widetext}

In the simplest case where the initial and final states are both QCD-stable hadrons, the finite-volume transition matrix elements have an immediate interpretation -- they can be related to the infinite volume transition amplitudes with only exponentially suppressed finite-volume corrections. For example if $\mathcal{A}$ and $\mathcal{B}$ are both isospin-1 pseudoscalar operators and the current has vector quantum numbers
\footnote{strictly speaking, operators in cubic or little-group irreps that contain subductions of the relevant helicities
}
, the lowest energy contributions to the correlator comes from the \emph{pion form-factor}, with the matrix-element having a Lorentz covariant decomposition (in infinite volume
\footnote{in finite volume there is an extra factor of $\big[ (2 \omega L^3) (2 \omega' L^3) \big]^{-1/2}$ due to the different normalization of finite-volume states.}
),
\begin{equation}
\big\la \pi(\mathbf{P'}) \big| \mathcal{J}^\mu \big|  \pi(\mathbf{P}) \big\ra = (P' + P)^\mu  \, F_\pi(Q^2),
\label{eq:pionff}
\end{equation}
where the virtuality $Q^2 = - (P'-P)^2$. Current conservation constrains the value of $F_\pi(0)$, but the form-factor is otherwise unconstrained, and non-perturbative calculations are required to determine the $Q^2$ dependence.
Provided volumes such that $m_\pi L \gg 1$ are used, the exponentially suppressed finite-volume corrections can be neglected, and such stable-hadron form-factors can be computed in a relatively straightforward manner in lattice QCD. Well studied cases include the $\pi$ and nucleon form factors \cite{Green:2015wqa, Shultz:2015pfa, Chambers:2017tuf, Collins:2011mk,Alexandrou:2011db,Capitani:2015sba,Lin:2010fv,Yoon:2016dij}.

Our interest here though is in composite particles, and in this case the relation between the finite-volume transition matrix element and the infinite-volume transition amplitude is not so simple, and the formalism to relate the two in general cases has only very recently been laid out.

\subsection{Lellouch-L\"uscher formalism and its generalizations\label{sec:LL_formalism}}

The first suggestion for how to relate transitions in infinite volume to the matrix elements calculated in a finite volume came from Lellouch and L\"uscher, who described the case of the weak decay $K \to \pi\pi$~\cite{Lellouch:2000pv} in the rest frame of the kaon. This idea has subsequently been generalized to systems in moving frames
~\cite{Kim:2005gf, Christ:2005gi}, 
coupled channels~\cite{Hansen:2012tf}, 
$\gamma^\star\to\pi\pi$~\cite{Meyer:2011um, Feng:2014gba}, 
bound state photodisintegration~\cite{Meyer:2012wk}, 
$\gamma^\star N \to \Delta$~\cite{Agadjanov:2014kha}, 
and elastic resonance form factors~\cite{Bernard:2012bi} among others. In this subsection we will sketch the framework as presented in~\cite{Briceno:2015csa, Briceno:2014uqa, Briceno:2015tza} which holds for generic two-body systems.~\footnote{~\cite{Agadjanov:2016fbd} presented a rederivation of the general result presented in~\cite{Briceno:2014uqa} applied to the specific $B\to K^\star\to K\pi$ weak decays.}

~\cite{Briceno:2014uqa} showed that the relationship between finite-volume matrix elements and infinite-volume transition amplitudes can be obtained using a diagrammatic representation of three-point functions, following a procedure similar to the one described in Sec.~\ref{sec:diagram_rep} for two-point functions\footnote{an earlier study using similar methods,~\cite{Bernard:2012bi}, was restricted to the non-relativistic case}.

\begin{figure}[t]
\begin{center}
\centering
\includegraphics[width=\columnwidth]{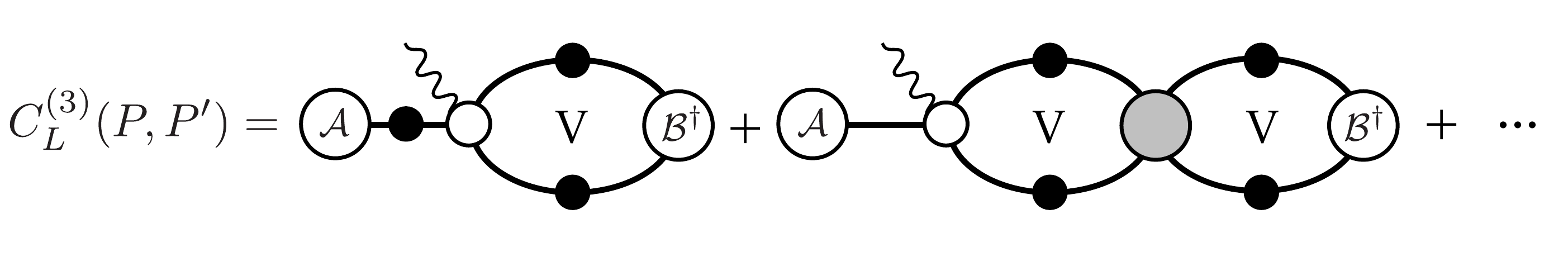}
\caption{ The diagrammatic representation of the momentum-space three-point function coupling one- and two-particle states. All objects are defined in Figs.~\ref{fig:scat_amp}, \ref{fig:FVcorr_F_function}, and \ref{fig:current_insertion_1}.}\label{fig:FV_1to2_corr}
\end{center}
\end{figure}

As an example of this formalism, let us consider the three-point correlation functions with a single QCD-stable hadron interpolated by $\mathcal{A}$ and a two-hadron state interpolated by $\mathcal{B}^\dag$ (relevant to e.g. $\gamma \pi \to \pi \pi$). In momentum space, this is diagrammatically depicted in Fig.~\ref{fig:FV_1to2_corr}. As was the case for two-point functions, we can express the correlation function in terms of on-shell infinite-volume quantities (in particular, the transition amplitudes $\mathcal{H}^{\rm in}(Q^2,P)$), and finite-volume quantities, 
\begin{align}
\label{eq:CLtwoterms}
C^{(3)}_{L}(P,P') 
 =&
A^\star(P')   \,   \Delta(P')  \,
\mathcal{H}^{\rm in}(Q^2,P) \nonumber \\
 & \times \frac{1}{F^{-1}(P,L) + \mathcal M(P)}   \,
  B^\star(P) 
  +\cdots 
\end{align}
where $\Delta$ is the single-hadron propagator. One can then Fourier transform to Euclidean space-time, and match the resulting expression to Eq.(\ref{eq:three-point_func}). After a substantial amount of algebra, one arrives at the following expression relating the finite-volume matrix element of the external current to infinite-volume transition amplitudes,
\begin{align}
\label{eq:onetotwomain}
 L^3
 \Big|  \big\langle E'_n,& \textbf P', L  \big|   {\mathcal J}(0) \big| E^{(1)}_{0}, \textbf P,L \big\rangle \Big|
  \nonumber\\ 
   &= \frac{1}{\sqrt{2E^{(1)}_{0}(\mathbf{P})}}   \sqrt{ \, \mathcal H^{\mathrm{in}} \, \Big [ \mathcal R(E_{n}', \textbf P')\Big ] \,  \mathcal H^{\mathrm{out}} } \,,
\end{align}
where the matrix $\mathcal R$ was defined in Eq.(\ref{eq:Rdef}). $\mathcal H^{\mathrm{in}}$ and $\mathcal H^{\mathrm{out}}$ are row and column vectors in the same space in which $\mathcal R$ resides, namely the space of angular momenta and open channels, {and each element corresponds to the angular momentum component of the amplitude associated with the given channel}. For example, when only one channel is kinematically open, we can explicitly write the product appearing in the equation above as,
\begin{align}
\mathcal H^{\mathrm{in}} \, \Big [ \mathcal R(E_{n}', \textbf P')\Big ] \,  \mathcal H^{\mathrm{out}} 
 \equiv
\mathcal H^{\mathrm{in}}_{\ell m} \, \Big [ \mathcal R(E_{n}', \textbf P')\Big ] _{\ell m,\ell' m'}\,  \mathcal H^{\mathrm{out}}_{\ell' m'}, \nonumber \\
\label{eq:HRH}
\end{align}
where prior to partial wave projection, $\mathcal{H}$ is defined diagrammatically in Fig.~\ref{fig:iH}, and its partial wave projection was defined in Eq.(\ref{eq:Hellm}).

This result demands some commentary: firstly, we observe that there is not a one-to-one mapping between the finite-volume matrix element and the infinite-volume transition amplitudes. In general, the reduction of rotational symmetry tells us that multiple angular momentum transition amplitudes will appear in a matrix element computed for a given lattice irrep (see e.g. Table~\ref{tab:subduction} for the $\mathbf{P} =\mathbf{0}$ case), and this is implicit in Eq.(\ref{eq:onetotwomain}). 
Secondly, we see from Eq.(\ref{eq:Rdef}) that $\mathcal R$ depends on the volume, on two-body kinematics, and importantly, also the \emph{two-body scattering dynamics} in $\mathcal{M}$. As a result, in order to correctly extract transition amplitudes from finite-volume matrix elements, one must first determine the two-body spectrum and then use Eq.(\ref{eq:QC_2body}) to constrain the relevant scattering amplitudes. Furthermore, since $\mathcal R$ depends not just on $\mathcal M$, but also its derivative, we require a parameterization of the energy-dependence of the scattering amplitude, which can in principle introduce a source of systematic uncertainty.

Although we have sketched a case in which the asymptotic hadrons are spinless, the extension to hadrons with spin is straightforward~\cite{Briceno:2015csa}. The related formalism for $0 \to 2$ transitions was first introduced by Meyer in the context of extracting the timelike $\pi$ form factor via $\gamma^\star\to\pi\pi$~\cite{Meyer:2011um}, and the generalization to make it applicable to generic $0\to 2$ processes is presented in~\cite{Briceno:2015csa}.

\subsection{Applications \label{sec:examples_LL}}

The primary application of the finite-volume formalism to date has been to the process $K \to \pi \pi$ where the weak interaction induces a hadronic decay of the otherwise QCD-stable kaon. The weak interaction does not conserve isospin, so both isospin=0 and isospin=2 $\pi\pi$ states can appear in this decay. Experimentally the $I=0$ channel is observed to be an order of magnitude larger than the $I=2$ channel, and this can be qualitatively explained by the presence of the resonant $\sigma$ in $I=0$ and the absence of resonant behavior in $I=2$. The role these amplitudes play in constraining sources of CP violation in the Standard Model has ignited a significant research program to determine the $K \to \pi \pi$ amplitudes precisely using lattice QCD~\cite{Bai:2015nea, Blum:2012uk, Boyle:2012ys, Blum:2011pu, Blum:2011ng}.

Kinematically, the decay $K\to\pi\pi$ decay is rather simple -- the cm-energy of the two-pion state must coincide with the mass of the kaon, which is possible since the pion momenta lie in a continuum in infinite volume. However in lattice QCD calculations one is not at liberty to specify the energies of $\pi\pi$ states in a given volume, since these are governed by the dynamics of the system, as we have previously discussed. What one \emph{can} determine is the transition amplitude at nonzero virtualities, here defined as $Q^2=-(P_K-P_{\pi\pi})^2$, and interpolate to the physically relevant point, $Q^2=0$ and $E_{\pi\pi}=E_{K}$. Technical challenges have to-date prevented this approach from being applied fully, instead what has been done mostly is to implement carefully chosen boundary conditions (as described at the end of Section~\ref{sec:elasticpipi}) to tune a $\pi\pi$ energy level such that it coincides with the kaon mass. Furthermore, in all of these studies a single operator is used to extract the finite-volume $\pi\pi$ state, which we have extensively discussed can lead to large uncontrolled systematic errors in the spectrum, matrix elements, and amplitudes.

\begin{figure*}
\centering
\includegraphics[width = \textwidth]{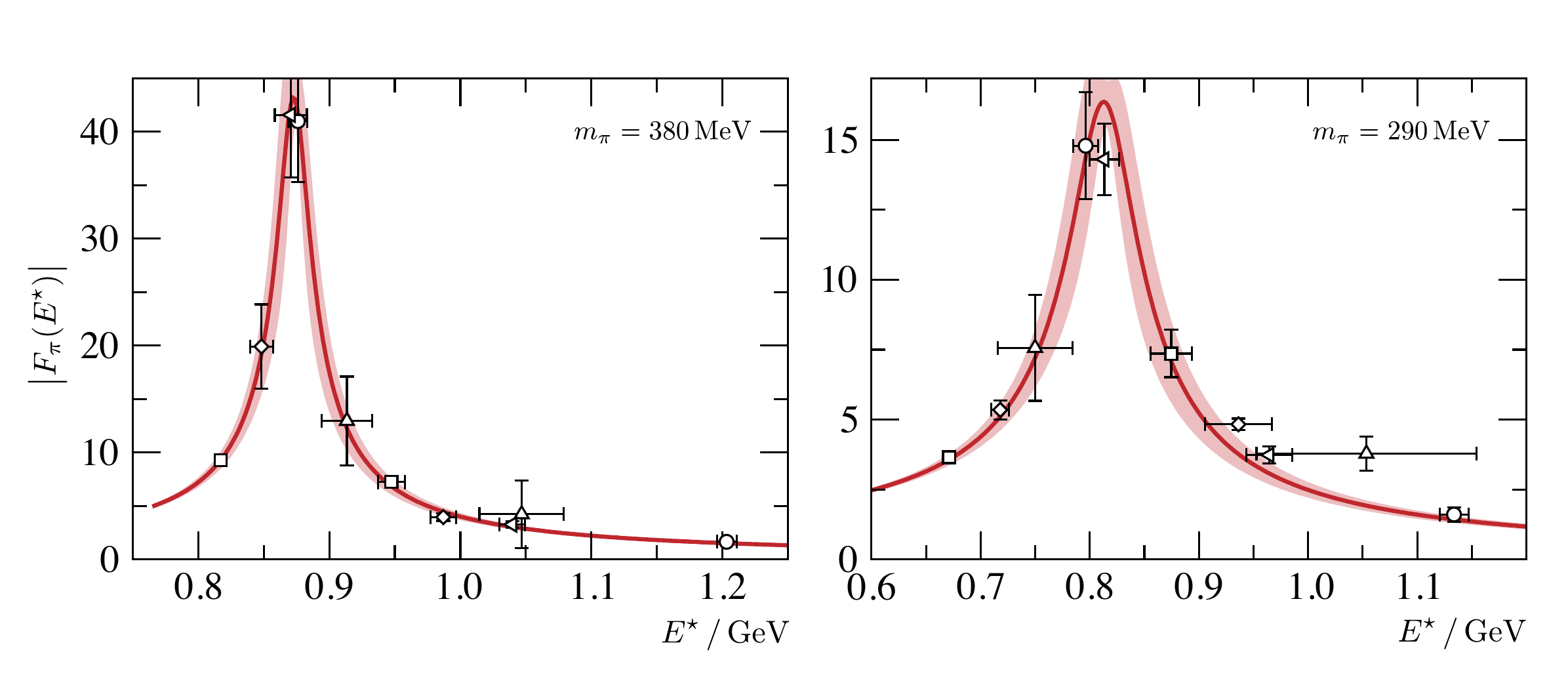}
\caption{ Shown is the $\pi$ form-factor in the timelike region obtained in~\cite{Feng:2014gba} using two different values of the light-quark masses corresponding to $m_\pi=290,\,380$~MeV. }\label{fig:form_factor} 
\end{figure*}

For $0 \to 2$ transitions, the case of $\gamma^\star\to\pi\pi$ has been recently explored~\cite{Feng:2014gba}, partially motivated by the important role this amplitude plays in constraining the dominant hadronic contribution to anomalous magnetic moment of the muon. The infinite-volume transition amplitude is a Lorentz vector whose dynamics can be parametrized by a single dimensionless function which corresponds to the pion form-factor, Eq.(\ref{eq:pionff}), for time-like virtualities, $Q^2 = - (E^\star)^2 < 0$. By restricting attention to energies below the $K\overline{K}$ threshold, Feng et al.~\cite{Feng:2014gba} determined $F_\pi(E^\star)$ at two values of the quark masses corresponding to $m_\pi=290,\,380$~MeV, shown in Figure~\ref{fig:form_factor}. One clearly observes the presence of the $\rho$ resonance in the energy dependence of the form-factor, as we would expect since $\gamma^\star \to \pi \pi$ should have the same pole singularities as $\pi \pi \to \pi \pi$ in $P$-wave.

\vspace{5mm}
The prototype amplitude to illustrate $1 \to 2$ scattering is $\pi\gamma^\star\to\pi\pi$, which plays a role in a wide range of phenomenology, principally through the isospin=1 \mbox{$P$-wave} amplitude, which features the $\rho$ resonance (see, for example,~\cite{Colangelo:2014dfa, Colangelo:2014pva, Wess:1971yu, Witten:1983tw, Huston:1986wi, Capraro:1987rp}). This amplitude was recently determined for the first time using lattice QCD in~\cite{Briceno:2015dca, Briceno:2016kkp} using the correlator construction technology laid out in~\cite{Shultz:2015pfa}. With a pion mass of $391$~MeV, a large number of three-point correlations functions were computed corresponding to a range of momenta for the $\pi$ operator, the $\pi\pi$ operator, and the current insertion. The $\pi\pi$ operators used were those found to optimally interpolate finite-volume energy levels through variational analysis in~\cite{Dudek:2012xn} -- they contain linear combinations of $\pi\pi$-like constructions and $\rho$-like `single-meson' operators. From these correlation functions, finite-volume matrix elements of the type appearing on the left-hand-side of Eq.(\ref{eq:onetotwomain}) were determined, and the corresponding infinite-volume amplitude $\mathcal{H}$ could then be obtained using parameterizations of the scattering amplitude capable of describing the finite-volume spectrum. A global analysis, making use of parameterization of the $E^\star$ and $Q^2$ dependence of the transition amplitude was performed.

\begin{figure*}
\centering
\includegraphics[width = \textwidth]{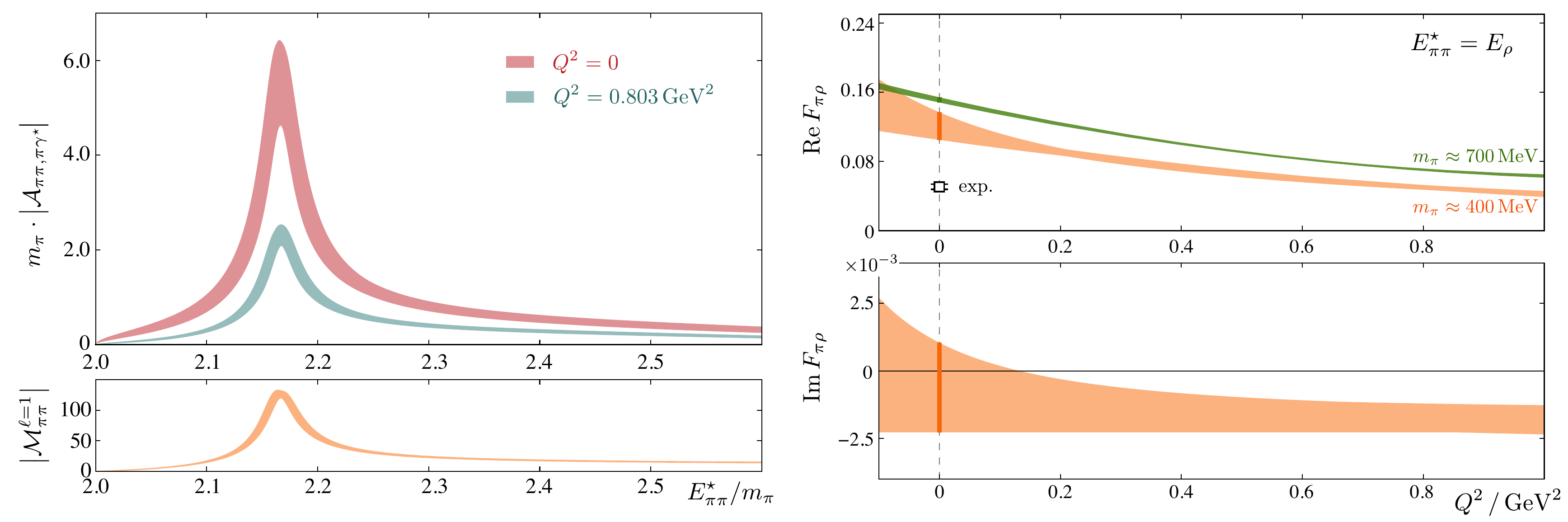}
\caption{ Left panel: shown is the $\pi\gamma^\star\to\pi\pi$ transition amplitude as a function of the c.m. energy of the system for two values of the photon virtuality obtained~\cite{Briceno:2015dca, Briceno:2016kkp}. This is compared to the elastic $\pi\pi$ scattering amplitude. Right panel: Shown are the real and imaginary components of the $\pi\to\rho$ form factor obtained for quark masses where the $\rho$ is stable~\cite{Shultz:2015pfa} and when the $\rho$ is unstable~\cite{Briceno:2015dca, Briceno:2016kkp}. This is compared to the experimental photocoupling~\cite{Huston:1986wi, Capraro:1987rp}.
}\label{fig:pi_to_rho} 
\end{figure*}

The results for the transition amplitude as a function of $E^\star$ at two different values of $Q^2$ are shown in the left panel of Figure~\ref{fig:pi_to_rho}, where the enhancement due to the $\rho$ resonance is clearly visible. In the right panel of Figure~\ref{fig:pi_to_rho} we show the $\rho \to \pi \gamma$ transition form-factor rigorously extracted from the residue of the $\rho$-pole in our parameterized transition amplitude as indicated in Eq.(\ref{eq:res_ff}). The smallness of the imaginary part of $F$ is correlated with the narrow width of the $\rho$ at $m_\pi \sim 391$~MeV.

It is worth remarking the rapid progress there has been in this area of the field. In the last couple of years, we have seen the ideas first set in place by Lellouch and L\"uscher~\cite{Lellouch:2000pv} generalized to increasingly complex reactions~\cite{Briceno:2015csa, Briceno:2014uqa, Briceno:2015tza}. Furthermore, we are entering an era where the sophistication of these ideas are matched by that of the numerical tools. Two illustrated examples are the two discussed in this section. First, the idea of studying the $\gamma^\star\to\pi\pi$ amplitude was proposed by Meyer by considering a system of $\pi\pi$ at rest~\cite{Meyer:2011um}. This was shortly after generalized by Feng et al.~\cite{Feng:2014gba} for $\pi\pi$ systems with nonzero momenta, and in the same study the authors performed the first numerical implementation. Similarly, the idea of studying $1\to2$ reactions was first proposed in 2014 by two groups \cite{Agadjanov:2014kha, Briceno:2014uqa}, and within a year these ideas were first put into practice in the context of the $\gamma^\star \pi \to \pi \pi$ amplitude~\cite{Briceno:2015dca, Briceno:2016kkp}.

With the formalism now laid out and tested in the simplest of systems, one can expect to see application to a variety of more interesting systems. A contemporary example comes in the charmonium sector, where $XYZ$ resonances defy explanation in simple models. The vector $Y$ resonances can have their decay constants studied rigorously using the approach presented above, and the relative size of these can be compared to the production rates in $e^+e^-$ and may also inform suggestions as to the internal structure of these states. Radiative transitions between various $XYZ$ resonances and conventional charmonium states can also be studied rigorously, and the corresponding form-factors can similarly be used to infer details of internal structure.

\section{Contemporary Extensions  \label{Sec:extensions}}

In this document we have focused our attention on ideas that have attained a certain maturity and that have been tested in at least one numerical study. In this section we will discuss a selection of extensions that have not yet quite reached this level. 

\subsection{Particles with nonzero intrinsic spin}
 
The bulk of serious calculations to-date have considered meson-meson scattering systems where the scattering particles are spinless. The formalism to deal with stable scattering particles of non-zero spin is in place~\cite{Briceno:2014oea}, and with it one could consider systems featuring mesons like the $\omega$ or the $D^\star$, which are stable for pion masses larger than physical, or the $J/\psi$ which only has a small hadronic width and which can be rendered stable by neglecting $c\bar{c}$ annihilation. The formalism would also have an application in meson-baryon and baryon-baryon scattering.

Scattering systems where the hadrons are not spinless are typically more complicated, featuring coupled amplitudes even if only one kinematic channel is open. For example, in the case of pseudoscalar-vector scattering, the $S$-wave has $J^P=1^+$ quantum numbers (\mbox{$^3\!S_1$ in the $^{2S+1}L_J$ notation}), but so does the $D$-wave ($^3\!D_1$) and these channels will mix in a $2 \times 2$ scattering matrix. 
The limited set of calculations that have been performed here so far have typically ignored this fact, for example in the study of $\pi \rho$ and $\pi \omega$ scattering in $J^P=1^+$ in~\cite{Lang:2014tia}, the `spinless' L\"uscher formalism was applied, effectively setting the $D$-wave amplitude to zero by fiat. In addition, not all possible meson-meson operator constructions corresponding to non-interacting levels in the energy region of interest were included, so it is not clear that the complete finite-volume spectrum was determined. 

Similarly, to date, calculations of two-baryon systems~\cite{Berkowitz:2015eaa, Beane:2011iw, Beane:2012vq, Yamazaki:2012hi, Yamazaki:2015asa} typically ignore the fact that different partial waves mix, either dynamically  due to the tensor force, or due to the mismatch between the cubic boundary and rotational symmetry (in finite volume).  The first attempt to constrain the contribution of the tensor force in the deuteron channel was carried out in~\cite{Orginos:2015aya}, using ideas proposed in~\cite{Briceno:2013lba, Briceno:2013bda}, but the signal obtained, with heavier than physical light quarks, was consistent with zero. 

The approach outlined in Sections~\ref{Sec:spectrum}, in which the finite-volume spectrum is determined in a range of volumes and/or moving frames, providing a large number of energy levels to constrain the various channels which are coupled by spin, is likely to be the way forward to study such systems.

\subsection{Three-particle systems \label{sec:three_part}}

A larger challenge is presented by the fact that most resonances in QCD with physical mass quarks are actually able to decay to \emph{three-body} final states, and the formalism we have been primarily discussing in this document, which is expressed in Eq.(\ref{eq:QC_2body}), applies only to systems \emph{below three-body thresholds}. This has motivated extensions of this formalism for energies where three-particle states can go on-shell~\cite{Polejaeva:2012ut, Briceno:2012rv, Hansen:2015zga, Hansen:2014eka, Briceno:2017tce}~\footnote{Interesting progress has also been made for 1+1 dimensional systems~\cite{Guo:2016fgl, Guo:2017ism}.}.

The current state of the art formalism is outlined in~\cite{Briceno:2017tce}, where a quantization condition was derived relating the finite-volume spectrum to scattering amplitudes for systems where two- and three-particle channels are coupled, for energies below the four-particle threshold. We do not present the result here, but it can be compactly written in form analogous to that of coupled two-particle systems, expressed in Eq.(\ref{eq:2chan}). In arriving at this result, two restrictions have been imposed, causing this to be not the most general equation for three-particle systems. First, all individual particles involved are assumed to be identical and carry no intrinsic spin, and as a result, there are only two channels that are kinematically open. This is a fairly mild assumption that is put in place only to simplify the kinematics. Second, the energies of the two-body subsystem (inside the three-particle system) must be below any pole in the two-particle $K$-matrix. In~\cite{Hansen:2015zga, Hansen:2014eka} it was shown that poles in the $K$-matrix can lead to power-law finite-volume effects which have thus far not been accounted for. The removal of this restriction is not so simple, but it is not expected to be insurmountable.  
   
Although this formalism has yet to be implemented in numerical studies, there have been two important formal checks performed for the simpler system considered in~\cite{Hansen:2015zga, Hansen:2014eka}, where two- and three-particle states decouple due to an additional symmetry in the system (e.g. $G$-parity in multi-pion systems). First, in~\cite{Hansen:2015zta, Hansen:2016fzj} the authors demonstrated that this quantization condition reduces to previously determined perturbative expressions for the finite-volume corrections to the energy of the lowest lying state in a system composed of three weakly repulsive scalar bosons~\cite{Beane:2007qr, Tan:2007bg}. Another important check was performed in~\cite{Hansen:2016ync}, where the authors showed that this quantization condition also reproduces the finite-volume corrections of Efimov bound states considered in~\cite{Meissner:2014dea}.

\subsection{Elastic form factors of resonances~\label{sec:elastic_FF}}

There are rigorously defined properties of resonances that can be computed in lattice QCD, which are not necessarily accessible in experiments. A good example are the form-factors of unstable particles, which may be extracted from the residue at the resonance pole of a  $2 \to 2$ transition amplitude featuring a current insertion of whatever construction is of interest. The required finite-volume formalism, whose construction is analogous to those presented in Section~\ref{Sec:formfactors} is given in~\cite{Briceno:2015tza, Bernard:2012bi}. From the form-factors, one can proceed to rigorously determine radii or other moments for these states, and this could give a concrete tool for distinguishing compact and molecular states, for example. Furthermore, it was recently demonstrated by Ji~\cite{Ji:2013dva} that quark distribution functions of QCD-stable states can be accessed non-perturbatively via lattice QCD. This idea, in conjunction with the formalism laid out in~\cite{Briceno:2015tza}, could allow for the determination of the quark distribution functions of \emph{unstable states}, allowing a view of the bound structure of resonance states.

\section{Outlook \label{Sec:outlook}}

The study of scattering processes and resonance properties within lattice QCD is entering an exciting period. In the last few years we have witnessed tremendous progress both in the development of the relevant formalism, and its application in explicit calculation. In this review we have presented some of the highlights of this program.

At the core of the formalism is the observation that infinite-volume scattering amplitudes control the discrete spectrum of eigenstates in a finite periodic box. We have discussed techniques for solving the inverse problem, where the discrete spectrum is extracted from a lattice QCD calculation and the scattering amplitudes are initially unknown. In order to get to the point where these techniques can be applied, it is necessary to reliably determine the excited state spectrum within a lattice QCD computation. We have illustrated the power of the variational method applied to matrices of two-point correlation functions computed using a large basis of operators, and shown the vital importance of including relevant multi-hadron operators in this basis.

Recent calculations have successfully determined elastic scattering amplitudes in several channels, and the extension into the coupled-channel sector has now been demonstrated in a number of cases.

Going beyond the simplest hadron-hadron scattering processes, we have seen progress in the development of formalism that handles reactions in which an electroweak current couples to a scattering system. As well as allowing rigorous first-principles QCD study of a range of phenomenologically interesting reactions, this formalism can also be used to compute processes which are experimentally inaccessible, such as the form-factors of hadron resonances.

Exploratory studies are being performed using unphysically large values of the quark masses, with reasoning beyond just the usual argument of decrease in cost: as the pion mass decreases, the threshold for three-hadron channels opening also decreases. Eq.(\ref{eq:QC_2body}) as presented is restricted in validity to energies below the lowest coupled three-hadron threshold. This unwanted constraint has motivated studies that are attempting to generalize Eq.(\ref{eq:QC_2body}) for systems where three or more particles can go on-shell. These works, which we have briefly reviewed, are advancing rapidly, but remain in the developmental stage. There is hope that the most general quantization condition involving two- and three-particle states will be derived in the next few years and its implementation will follow.

\begin{acknowledgments}
We thank the authors of papers whose figures we reproduce in this article for their permission to do so, and S.~Sharpe, M.~Hansen and C.~Thomas for their comments on a early draft of the manuscript. RAB and JJD acknowledge support from U.S. Department of Energy contract DE-AC05-06OR23177, under which Jefferson Science Associates, LLC, manages and operates Jefferson Lab. JJD acknowledges support from the U.S. Department of Energy Early Career award contract \mbox{DE-SC0006765}. RDY acknowledges support from Australian Research Council grants CE110001004, FT120100821 and DP14010306.

\end{acknowledgments}

\bibliography{bibi_inspire}

\vspace{1cm}
\section{Numerical evaluation of $F$ \label{Sec:F_funct}}

The matrix $F(E, \mathbf{P}; L)$, whose inverse appears in Eq.(\ref{eq:QC_2body}), is, in the case of single-channel scattering, an object whose matrix elements are given by Eq.(\ref{eq:Fscdef}). The appropriate generalization to multiple scattering channels, including the case where the scattering particles are not spinless is presented in~\cite{Briceno:2014oea} -- here we will focus on the numerical implementation of the simplest case of elastic scattering of spinless particles. In this case we can show that, for a frame momentum $\mathbf{P} = \frac{2\pi}{L} \mathbf{d}$, 
\begin{widetext}
\begin{equation}
\hspace{-2cm}
F_{\ell m; \ell' m'} = i \frac{1}{16 \pi} \frac{2\, q^\star}{E^\star} 
\left[ \delta_{\ell,\ell'} \delta_{m,m'} 
+ i \frac{1}{\gamma} \frac{1}{\pi^{3/2}} \sum_{\bar{\ell}, \overline{m}} 
b(\ell m; \bar{\ell} \overline{m} \,| \ell' m') \;
\left(\frac{q^\star L}{2\pi}\right)^{-(\bar{\ell} +1)} 
\mathcal{Z}^{\mathbf{d}}_{\bar{\ell} \overline{m}}\left(1;\left(\frac{q^\star L}{2\pi}\right)^2 \right) 
\right]
\label{eq:Fnum}
\end{equation}
where $\gamma = E/E^\star$, $b(\ell m; \bar{\ell} \overline{m} \,| \ell' m')  = \sqrt{\frac{(2\ell+1)(2\bar{\ell} +1)}{2 \ell' +1} } 
\big\langle \ell m; \bar{\ell} \overline{m} \big| \ell' m'\big\rangle
\big\langle \ell 0; \bar{\ell} 0 \big| \ell' 0\big\rangle
$, and where the \emph{L\"uscher zeta functions} are defined by
\end{widetext}
\begin{equation}
\mathcal{Z}^{\mathbf{d}}_{\ell m}\left(s;x \right) = \sum_{\mathbf{r} \in \mathcal{P}_\mathbf{d}} \frac{|\mathbf{r}|^\ell \, Y_{\ell m}(\hat{\mathbf{r}})}{(|\mathbf{r}|^2 - x^2)^s}.
\end{equation}
In this expression the sum is performed over $\mathcal{P}_\mathbf{d} = \big\{ \mathbf{r} \in \mathbf{R}^3 \big|\, \mathbf{r} = \hat{\gamma}^{-1}( \mathbf{n} - \alpha \mathbf{d} ) \big\}$ where $\mathbf{n}$ is a triplet of integers and $\alpha = \frac{1}{2}\left( 1 + \frac{m_1^2 - m_2^2}{E^{\star 2}} \right)$ for scattering particle masses $m_1, m_2$. The operation $\hat{\gamma}^{-1}$ is defined as ${\hat{\gamma}^{-1}\mathbf{x} = \frac{1}{\gamma} \mathbf{x}_{||} + \mathbf{x}_\perp}$ with $\mathbf{x}_{||}$, $\mathbf{x}_\perp$ being the components of $\mathbf{x}$ parallel to and perpendicular to the direction of $\mathbf{d}$. A longer discussion can be found in~\cite{Rummukainen:1995vs} and \cite{Leskovec:2012gb}.

The zeta function in Eq.(\ref{eq:Fnum}) contains the factor $\left( |\mathbf{r}|^2 - \left(\frac{q^\star L}{2\pi}\right)^2 \right)^{-1}$, which is singular if ${q^{\star } = \frac{2\pi}{L} |\mathbf{r}|}$, which corresponds to a two-particle state taking an allowed non-interacting energy in the $L\times L \times L$ volume. We can easily illustrate this in the rest-frame ($\mathbf{d} = \mathbf{0}$) where $q^{\star } = \frac{2\pi}{L} |\mathbf{n}|$ and hence ${E^\star = \sqrt{m_1^2 + \left(\frac{2\pi}{L}\right)^2 |\mathbf{n}|^2} + \sqrt{m_2^2 + \left(\frac{2\pi}{L}\right)^2 |\mathbf{n}|^2}}$, which are precisely the non-interacting energy levels. Strategies for numerically evaluating the zeta function can be found in~\cite{Rummukainen:1995vs, Kim:2005gf, Fu:2011xz, Leskovec:2012gb, Yamazaki:2004qb}.  

The matrix $F$ becomes block-diagonal if we consider the remaining symmetries of the cube (or the boosted cube), using the operation known as \emph{subduction} into irreducible representations (irreps) of the relevant symmetry group. Subduction from $\ell, m$ into the $n^\mathrm{th}$ embedding of $\ell$ into the irrep $\Lambda$ (row $\rho$) is achieved by applying
\begin{equation*}
\mathbb{S}^{\Lambda \rho n}_{\ell m}(R)  = \sum_\lambda \mathcal{S}^{\Lambda \rho n}_{\ell \lambda} \, D^{(\ell)}_{m \lambda}(R)
\end{equation*}
where $R$ is the rotation which carries a vector along $[001]$ into the direction of $\mathbf{d}$, and where we are summing over a quantity $\lambda$ that can be associated with helicity. The helicity-based subductions, $\mathcal{S}^{\Lambda \rho n}_{\ell \lambda}$ can be found in~\cite{Thomas:2011rh, Dudek:2012gj}. It follows that we can subduce $F$ as follows:
\begin{equation*}
\delta_{\Lambda, \Lambda'} \delta_{\rho, \rho'}\,  F^{\Lambda}_{\ell n; \ell' n'} =
\sum_{m, m'} \Big( \mathbb{S}^{\Lambda \rho n}_{\ell m}(R) \Big)\,  F_{\ell m; \ell' m'} \, \Big( \mathbb{S}^{\Lambda' \rho' n'}_{\ell' m'}(R) \Big)^*,
\end{equation*}
where the orthogonality of different irreps and different rows within an irrep is expressed explictly on the left-hand-side. We can write the subduced $F$ in the form,
\begin{equation*}
 F^{\Lambda}_{\ell n; \ell' n'} = i \frac{1}{16 \pi} \frac{2\, q^\star}{E^\star} 
\left[ \delta_{\ell,\ell'} \delta_{n,n'} + i f^\Lambda_{\ell n; \ell' n'} \left( \left( q^\star L / 2\pi \right)^2 \right)  \right], \label{eq:f}
\end{equation*}
and above threshold ($q^\star > 0$), $f$ can be defined to be real, so that the imaginary part of $F$ is $\frac{1}{16 \pi} \frac{2\, q^\star}{E^\star} \delta_{\ell,\ell'} \delta_{n,n'}$. Since elastic unitarity ensures that $\mathrm{Im}\, \frac{1}{\mathcal{M}_\ell} = - \frac{1}{16 \pi} \frac{2\, q^\star}{E^\star}$, the L\"uscher quantization condition, Eq.(\ref{eq:QC_2body}), which can be recast into the form $\det \big[ \mathcal{M}^{-1} + F \big] =0$, is effectively then a real equation. Importantly if $\mathcal{M}$ is parameterized in a manner which does not respect unitarity, we can see that this equation will have an imaginary part which is volume-independent and which cannot be solved. These arguments generalize naturally to the coupled-channel case, and it is there that the condition that $\mathcal{M}$ satisfy unitarity is a practical constraint.

\begin{figure}
\centering
\includegraphics[width = \columnwidth]{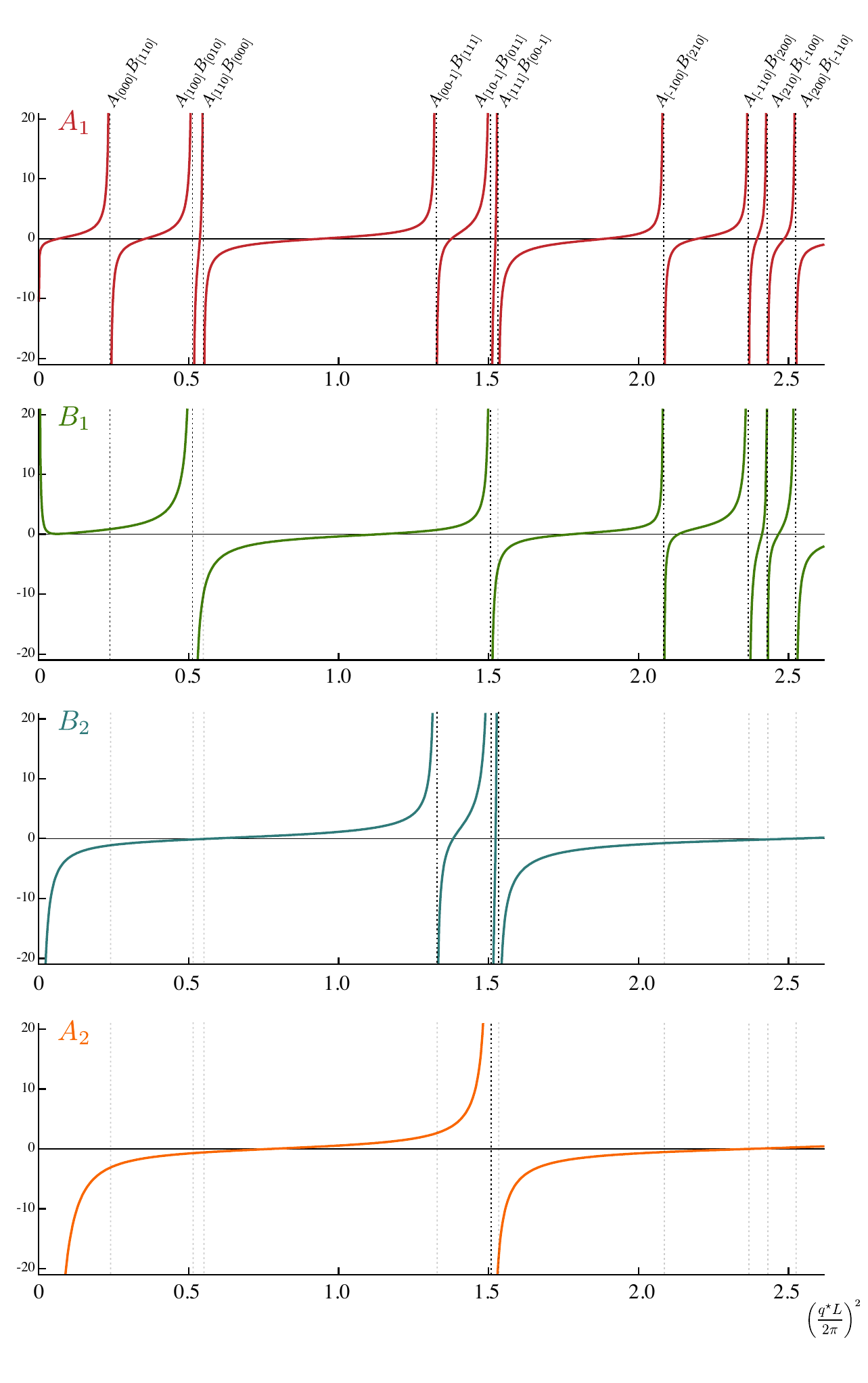}
\caption{$f^{\Lambda}_{\ell n ; \ell' n'}$ for $\mathbf{d}=[110]$. From top to bottom: $f^{A_1}_{01;01}$, $f^{B_1}_{11;11}$, $f^{B_2}_{11;11}$, $f^{A_2}_{21;21}$, which are in each case the lowest $\ell$ subduced into that irrep. The scattering particles `$A$', `$B$', are chosen to have masses 300 MeV, 500 MeV, respectively and $L$ is chosen to be 3.6 fm. The divergences can be seen to correspond to the energies of non-interacting $AB$ pairs having momenta of the type shown at the top of the diagram.
}\label{fig:Fplot}
\end{figure}

In Figure~\ref{fig:Fplot} we illustrate a typical behavior for $f^\Lambda$ in the case of the four one dimensional irreps, $A_1, B_1, B_2, A_2$ of $\mathbf{d} = [110]$. Note that there are many divergences in these functions, and these divergences correspond to the possible non-interacting particle pairs in these irreps. Since the non-interacting spectrum is different in different irreps, the set of divergences differs depending upon $\Lambda$. Note also that there can be very tight regions in $(q^\star L / 2\pi)^2$ between neighboring divergences -- when we come to find solutions to Eq.(\ref{eq:QC_2body}) using numerical rootfinding techniques it will be important to ensure that we pay attention to this.

\begin{figure}
\centering
\includegraphics[width = 0.9\columnwidth]{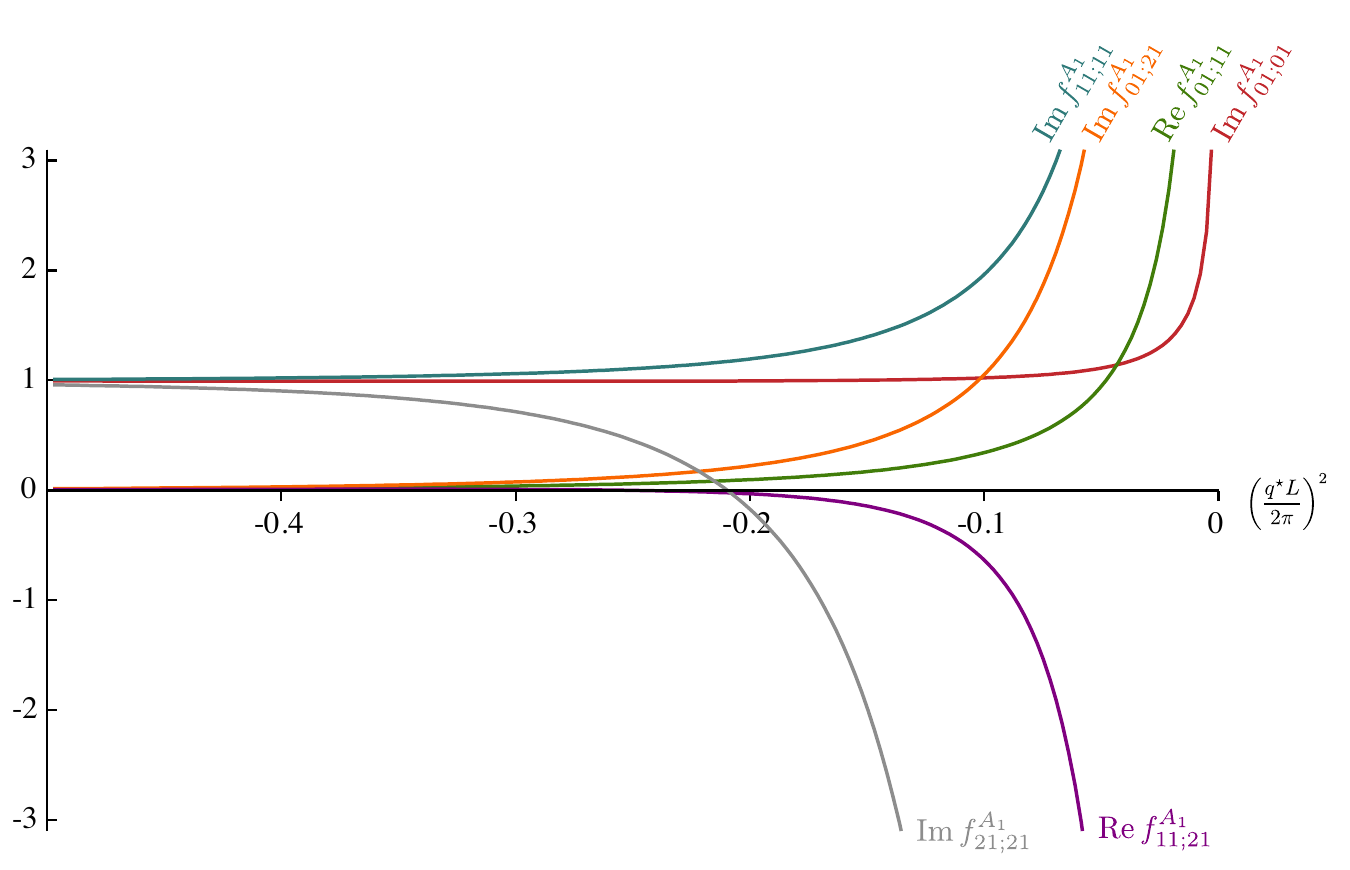}
\caption{$f^{A_1}_{\ell n ; \ell' n'}$ for $\mathbf{d}=[110]$ plotted below the kinematic threshold. Masses and volume as in Figure~\ref{fig:Fplot}.
}\label{fig:Fplot_subthresh}
\end{figure}

In Figure~\ref{fig:Fplot_subthresh} we illustrate the behavior of some elements of a typical $f^\Lambda$ \emph{below threshold}. We observe that the diagonal elements $f^\Lambda_{\ell n; \ell n}$, as we go further below threshold, we tend to value of $i$, and the off diagonal elements tend to $0$. This ensures that $F^\Lambda \to 0$ as we go some way below threshold. It follows that bound-states lying far below threshold appear in finite-volume at an energy that is very close to their infinite-volume energy. Another consequence is that in the case of multi-channel scattering, a kinematically closed channel does not have a significant impact on the quantization condition at energies far below its threshold, so it is not necessary to include in Eq.(\ref{eq:QC_2body}) the continuation of distant closed channels. Note that there is a narrow region below threshold where the finite-volume functions are not zero, and this indicates that the effect of a closed channel can `leak' down a little way below its threshold -- the effect is exponentially suppressed with increasing volume.

\end{document}